\documentclass[12pt]{iopart}
\usepackage{epsfig,graphicx}

\newcommand{\bra}{\langle}
\newcommand{\ket}{\rangle}
\newcommand{\bigbra}{\left\langle}
\newcommand{\bigket}{\right\rangle}
\newcommand{\order}{{\mathcal O}}

\newcommand{\sgn}{\textrm{sgn}}

\newcommand{\one}{{\rm 1\!\!I}}

\newcommand{\be}{\begin{equation}}
\newcommand{\ee}{\end{equation}}
\newcommand{\bd}{\begin{displaymath}}
\newcommand{\ed}{\end{displaymath}}
\newcommand{\vsp}{\vspace*{3mm}}
\newcommand{\room}{\rule[-0.1cm]{0cm}{0.6cm}}

\newcommand{\F}{{\mathcal F}}
\newcommand{\R}{{\rm I\!R}}

\newcommand{\bB}{\ensuremath{\mathbf{B}}}

\newcommand{\bI}{\ensuremath{\mathbf{I}}}

\newcommand{\bR}{\ensuremath{\mathbf{R}}}

\newcommand{\blambda}{{\mbox{\boldmath $\lambda$}}}

\newcommand{\bxi}{{\mbox{\boldmath $\xi$}}}

\newcommand{\bomega}{{\mbox{\boldmath $\omega$}}}

\newcommand{\bOmega}{{\mbox{\boldmath $\Omega$}}}

\newcommand{\here}{\makebox(0,0)}

\newcommand{\vertex}{\here{\circle*{2}}}

\newcommand{\rate}{\tilde{\eta}}

\newcommand{\del}{{\delta_{\!N}}}

\newcommand{\chiR}{\chi_{\rm\!_R}}

\begin{document}\small

\title[Minority games with inner
product strategy definitions]{Generating functional analysis of
minority games with inner product strategy definitions}

\author{ACC Coolen and N Shayeghi}
\address{
Department of Mathematics, King's College London\\ The Strand,
London WC2R 2LS, United Kingdom}

\begin{abstract}
\noindent
We use generating functional methods to solve the so-called inner product versions of the
minority game (MG), with fake and/or real market histories, by generalizing the theory
developed recently for look-up table MGs with real histories.
The phase diagrams of the lookup table and inner product MG versions
are generally found to be identical, with the exception of inner product MGs where
histories are sampled linearly, which are found to be structurally critical. However,
we encounter interesting differences both in the theory (where the role of the
history frequency distribution in lookup table MGs is taken over by the eigenvalue spectrum of a history covariance matrix in inner product MGs)
and in the static and dynamic phenomenology of the models.
Our theoretical predictions are supported by numerical simulations.
\end{abstract}

\ead{ton.coolen@kcl.ac.uk,nima@mth.kcl.ac.uk}


\section{Introduction}

\noindent
 Minority Games (MG) \cite{ChalZhan97,ChalZhan98} are
relatively simple and transparent models that were designed to increase our understanding of the
complex collective processes which result from inductive decision
making by large numbers of interacting agents in simplified `markets'. They can be
 seen
as mathematical implementations of the so-called El Farol bar
problem \cite{Arth94}. Many versions of the MG have by now been
studied in the literature.
 They differ
in the type of microscopic dynamics used (e.g. batch versus
on-line, stochastic versus deterministic), in the definition of
the information provided to the agents (real-valued versus
discrete, true versus fake market history) and the agents'
decision making strategies, and also in the specific recipe
defined for the conversion of the external information into a
trading action (inner products versus look-up tables). In MG models
with `fake' market memory, proposed first in \cite{Cavagna99},
 at each point in time all agents are given random data upon which to base their
decisions, rather than the actual market
history. They have the advantage of being Markovian and were
therefore the first to be studied and solved in the theoretical
physics literature using techniques from equilibrium
\cite{Replicas1,Replicas2} and non-equilibrium
\cite{HeimelCoolen2001,CoolenHeimelSherr2001,CoolenHeimel2001}
statistical mechanics, as first developed for and applied to spin-glasses.
Models with true market history are non-Markovian and therefore more
demanding, but also here theoretical progress is being made
\cite{ChalMars00,Coolen2005}.
 For a more extensive
introduction to MGs and their statistical mechanical theory we
refer to the recent textbooks \cite{BookCMZ,BookAC}.

 To our knowledge, however, there has
not yet been any such exact solution for MG models in which the
decision making strategies are defined via inner products rather
than look-up tables \cite{CavaGarrGiarSher99,GMS2000}; not even
for the simplest Markovian case, where the market histories are
fake. Inner product MGs have so far tended to be studied either
numerically, or at the level of investigating properties of the
stochastic microscopic laws, in spite of the fact that one might
well argue that the definitions of the `inner product MGs' are
possibly closer to how financial time series tend to be predicted
by practitioners \cite{Bouchaud} (as these usually involve generalized
linear market prediction models). Moreover,
in fact not a single study  has so far been
published of inner product MGs with real as opposed to fake histories, there are not even
papers based on numerical simulations only.

In this paper we first generalize the existing generating functional theory \cite{Coolen2005} that was developed for look-up table MGs
 with real and/or fake histories
to a  larger class of MG versions, which includes the familiar look-up table MGs and inner product MGs as special cases.
We then focus on the
analysis of the dynamics and phase transitions of inner product MG versions with and without real histories,
and show that in the infinite system size limit:
(i) for inconsistent fake histories the inner product and lookup table MG
versions behave fully identically, including short time correlation and response functions and the volatility (apart from a trivial transformation
related to time-scales),
(ii) similar to what was found in \cite{Coolen2005} for look-up table MGs,
for inner product MGs the introduction of real histories affects the observables in the ergodic region,
but does not generally change the phase diagram, except when the histories are sampled by unbounded functions (see below), (iii) the key observable to measure the differences
between real and fake histories in our generalized MG models is the spectrum of the history covariance matrix
in the stationary state, which for
inner product MGs reduces to the history frequency distribution, (iv) for inner product MGs an approximate calculation of this spectrum leads to a
fully closed stationary state theory, the predictions of which are supported by numerical simulations, and
(v) inner product MGs with predominantly real histories where the inner product is taken between
strategy entries and the actual values of the past market bids (or, more generally, by an unbounded function of these bids rather than e.g. a saturating function) are permanently critical, giving
a volatility that is either zero or infinity (separated by a novel critical value for the control parameter $\alpha$, unrelated to the standard ergodicity
breaking transition point).

\section{Definitions}

\noindent We imagine having a system of $N$ agents, labeled by
$i=1,\ldots,N$. At each step $\ell\in\{0,1,2,\ldots\}$ of the game
each agent submits a `bid' $b_i(\ell)\in\R$ to the market. The
(re-scaled) cumulative bid at stage $\ell$ is defined as
\begin{equation} A(\ell)=\frac{1}{\sqrt{N}}\sum_i b_i(\ell)+A_e(\ell)
\label{eq:define_bid}
\end{equation}
Here $A_e(\ell)$ denotes an external contribution that represents e.g. random market perturbations or actions by
market regulators. Each agent $i$ determines his bid $b_i(\ell)$
at each step $\ell$ on the basis of external information, which is
to represent historic market data, using his preferred decision
making strategy at that point in time. Each agent has $S$ such
strategies, labeled by $a=1,\ldots,S$. Profit is made by those
agents who find themselves subsequently in the minority group,
i.e. in the case of $A(\ell)>0$ by those with $b_i(\ell)<0$ and in
the case of $A(\ell)<0$ by those with $b_i(\ell)>0$.

 MG versions differ in their definition of the information
provided to the agents
 and in how this information is converted into trading
decisions.
 In the original MG, the information consisted of the signs of the overall
 market bids $A(\ell)$ over a number of time steps in the past. In inner product MGs, in contrast, the external information at stage $\ell$ will
 consist of a $p$-dimensional vector $\bI(\ell)$ (generalizing the
`information vector' of \cite{CavaGarrGiarSher99,GMS2000}), where
 $\alpha=p/N$ remains finite as $N\to\infty$, with entries
\begin{equation}
I_{\lambda}(\ell) =\frac{1}{\sqrt{p}}~ f[ (1-\zeta) A(\ell-
\lambda) + \zeta Z(\ell,\lambda)]
 \label{eq:define_info_IP}
\end{equation}
and  with $\lambda\in\{1,\ldots,p\}$. The function $f[.]$ (which we choose to be anti-symmetric) allows
the information given to the agents to involve more general properties of the overall
bids than just their sign, on which the original model
\cite{ChalZhan97,ChalZhan98} was based.  The
$Z(\ell,\lambda)$ are zero-average Gaussian random variables,
which represent the `fake' alternative to the true market data
$A(\ell-\lambda)$, and $p$ is the number of iteration steps in the
past for which market information is made available. For reasons
that will become clear we write
$I_\lambda(\ell)=p^{-\frac{1}{2}}\F_\lambda[\ell,A,Z]$, so
\begin{eqnarray}
 \F_\lambda[\ell,A,Z]&=& f[(1-\zeta)  A(\ell-\lambda) + \zeta
Z(\ell,\lambda)] \label{eq:Flambda_IP}
\end{eqnarray}
If we ensure that $f[.]$ remains finite, one has
$\lim_{N\to\infty}\sum_{\lambda} [I_{\lambda}(\ell)]^2=\order(1)$
for all $\ell$. Definition (\ref{eq:define_info_IP}) implies that
for $\zeta=1$ the external information is fully random, whereas
for $\zeta=0$ it represents true historic market data. Following
\cite{Coolen2005} we
 distinguish between two types of definitions for the statistical
 properties of the `fake memory' variables:
\begin{eqnarray}
{\rm consistent\!:} &~~ Z(\ell,\lambda)=Z(\ell-\lambda), &~~ \bra
Z(\ell)Z(\ell^\prime)\ket=S^2\delta_{\ell \ell^\prime}
\label{eq:typeA}
\\
{\rm inconsistent\!:} &~~ Z(\ell,\lambda)~{\rm all~independent}, &~~
\bra
Z(\ell,\lambda)Z(\ell^\prime\!,\lambda^\prime)\ket=S^2\delta_{\ell
\ell^\prime}\delta_{\lambda\lambda^\prime} \label{eq:typeB}
\end{eqnarray}
We take the agents' strategies to be represented by vectors
$\bR^{ia}\in\R^p$ with components $R_{\lambda}^{ia}$.
 In accordance with standard definitions \cite{CavaGarrGiarSher99,GMS2000}
we choose these components to be independent zero-average and
unit-variance Gaussian random variables, assigned before the start
of the game, and remaining fixed throughout. They represent
quenched disorder.
 The arrival of information $\bI(\ell)$ at step $\ell$ prompts each agent $i$ to submit the
 bid $b_i(\ell)=\sum_{\lambda} R^{i a_i(\ell)}_{\lambda}
I_{\lambda}(\ell)$, where $a_{i}(\ell)$ denotes the
preferred strategy at stage $\ell$ of agent $i$. In order to
detect which of their $S$ private strategies to use, all agents
keep track of valuations $p_{ia}(\ell)$ of their strategies. These
 measure to what extent each strategy would have led to
minority decisions  if it had been used all the time:
 \begin{equation}
 p_{ia}(\ell+1)=p_{ia}(\ell)-\frac{\rate}{\sqrt{N}}
~ A(\ell)\sum_{\lambda}R^{ia}_{\lambda} I_{\lambda}(\ell)
 \end{equation}
 The
factor $\rate>0$ represents a learning rate. The active strategy
$a_{i}(\ell)$ of trader $i$ at stage $\ell$ is chosen to
be  the one with the highest valuation $p_{ia}(\ell)$ at that
moment, i.e. $a_{i}(\ell)= {\rm arg}~\max_a
\{p_{ia}(\ell)\}$.
 Our
full equations now become, in more explicit form and with the
notation (\ref{eq:Flambda_IP}):
\begin{eqnarray}
p_{ia}(\ell+1)&=&p_{ia}(\ell)-\frac{\rate}{N\sqrt{\alpha
}}\sum_{\lambda} R_{\lambda}^{ia}\F_\lambda[\ell,A,Z]
 A(\ell) \label{eq:valuation_dynamics}
\\ A(\ell)&=& \frac{1}{\sqrt{\alpha}N}\sum_i
\sum_{\lambda} R_{\lambda}^{i a_i(\ell)}
\F_\lambda[\ell,A,Z]
  \label{eq:totalbid_eqn}
\end{eqnarray}
 Henceforth we restrict ourselves to the case $S=2$, two
strategies per agent so $a\in\{1,2\}$, and introduce the
conventional definitions \bd q_i(\ell)=\frac{1}{2}[p_{i1}(\ell)-
p_{12}(\ell)],~~~~~~
\bomega^i=\frac{1}{2}(\bR^{i1}\!+\bR^{i2}),~~~~~~
\bxi^i=\frac{1}{2}(\bR^{i1}\!-\bR^{i2}) \ed as well as
$\bOmega=N^{-1/2}\sum_i \bomega^i$. This allows us to write
$\bR^{i a(\ell)}=\bomega^i+\sgn[q_i(\ell)]\bxi^i$. We then
proceed to include decision noise in the familiar way via the
substitution $\sgn[q_i(\ell)]\to \sigma[q_i(\ell),z_i(\ell)]$,
 in which the $\{z_{i}(\ell)\}$ are
independent zero average random numbers, described by a symmetric
and unit-variance distribution $P(z)$. The function $\sigma[q,z]$
is taken to be
 non-decreasing in $q$ for any $z$, and
 parametrized by a parameter $T\geq 0$
 such that $\sigma[q,z]\in \{-1,1\}$, with $\lim_{T\to
 0}\sigma[q,z]=\sgn[q]$ and
 $\lim_{T\to\infty}\int\!dz~P(z)\sigma[q,z]=0$.
The two main examples are
 additive and multiplicative noise:
\begin{eqnarray}
  {\rm additive:~~~~~~~~~~} && \sigma[q,z]=\sgn[q+Tz]
  \label{eq:additive}
\\
  {\rm multiplicative:~~~} &&
  \sigma[q,z]=\sgn[q]~\sgn[1+Tz]
  \label{eq:multiplicative}
\end{eqnarray}
 $T$ measures
  the degree of randomness in the agents' decision making,
  with $T=0$ bringing us back to  $a_{i}(\ell)={\rm arg}~\max_a \{p_{ia}(\ell)\}$,
  and with random strategy selection for $T=\infty$.
In on-line MG theories one will in practice only need the decision
noise average
\begin{equation}
\sigma[q]=\int\!dz~P(z)\sigma[q,z] \label{eq:define_sigma}
 \end{equation}
 Upon also adding the usual external perturbation fields
$\{\theta_i(\ell)\}$ to define response functions, our microscopic
equations (\ref{eq:valuation_dynamics},\ref{eq:totalbid_eqn})
are then replaced by
\begin{eqnarray}&&
q_{i}(\ell+1)=
q_{i}(\ell)+\theta_i(\ell)-\frac{\rate}{N\sqrt{\alpha
}}\sum_{\lambda=1}^p \xi_{\lambda}^{i}\F_\lambda[\ell,A,Z]
 A(\ell) \label{eq:q_eqn}
\\
&& A(\ell)= A_e(\ell)+\!\frac{1}{\sqrt{\alpha N}}
\sum_{\lambda=1}^p\Big\{\Omega_{\lambda}\!+\frac{1}{\sqrt{N}}\sum_i\sigma[q_i(\ell),z_i(\ell)]\xi_\lambda^i\Big\}
\F_\lambda[\ell,A,Z] ~~\label{eq:A_eqn}
\end{eqnarray}
The values of $A(\ell)$ for $\ell\leq 0$ and of the $q_i(0)$ play
the role of initial conditions.

Equations (\ref{eq:q_eqn},\ref{eq:A_eqn}) are identical to those
studied in \cite{Coolen2005} for look-up table MGs; the only
difference between look-up table and inner product definitions for
the information-to-decision conversion in the MG is in the
definition of the function $\F_\lambda[\ell,A,Z]$. We may thus
take over from \cite{Coolen2005} the derivation of the effective
single agent problem, and simply make the appropriate substitutions
in the final result\footnote{In fact, the formulation of the
derivation in \cite{Coolen2005} was chosen with
 the present generalization in mind.}:
 \begin{eqnarray}
 \hspace*{-20mm}
 {\rm look\!\!-\!\!up~table\!:}&~~~
\blambda\!\in\!\{-1,1\}^M\!,~~2^M\!=p, &~~  \F_\blambda[\ell,A,Z]= \sqrt{\alpha N}~\delta_{\blambda,\blambda(\ell,A,Z)}\\[1mm]
 \hspace*{-20mm} &&\hspace*{-10mm} \lambda_{\mu}(\ell,A,Z)=\sgn[(1\!-\!\zeta) A(\ell\!-\!\mu) + \zeta
Z(\ell,\mu)]\nonumber
\\[2mm]
\hspace*{-20mm}
{\rm inner~product\!:}&~~~
\lambda\!\in\!\{1,\ldots,p\}, &~ \hspace*{-10mm}
  \F_\lambda[\ell,A,Z]= f[(1\!-\!\zeta) A(\ell\!-\!\lambda) + \zeta
Z(\ell,\lambda)] \label{eq:link}
 \end{eqnarray}
The inner product models of \cite{CavaGarrGiarSher99,GMS2000} correspond to choosing  recipe (\ref{eq:link}), with
$\zeta=1$ (fake history) and $f[A]=A$, giving $\F_\lambda[\ell,A,Z]=
Z(\ell,\lambda)$.

\section{Generating functional analysis of generalized MGs}

\subsection{The effective single agent equation}

\noindent We choose the real-time duration of individual
iterations of the MG equations to be $\del=\rate/2p$.
This was proposed within the replica approach
\cite{Replicas1,Replicas2,ChalMars00}, and later confirmed to be
the canonical scaling via generating functional analysis studies
\cite{CoolenHeimel2001,Coolen2005}. The generating functional
analysis in \cite{Coolen2005}, which via (\ref{eq:link}) can be
made to apply also to inner-product models, leads
as always to closed deterministic dynamical order parameter
equations which are fully exact in the limit $N\to\infty$ and on
times which do not scale with $N$. The order parameters are the
disorder-averaged single-site correlation and response function,
which in terms of the original discrete iterations
$\ell=0,1,2,\ldots$ are defined as
\begin{eqnarray}
C(\ell,\ell^\prime)&=& \lim_{N\to\infty}
\frac{1}{N}\sum_i\overline{\bra
\sigma[q_i(\ell),z_i(\ell)]\sigma[q_i(\ell^\prime),z_i(\ell^\prime)]\ket}
\label{eq:identifyC}
\\
G(\ell,\ell^\prime)&=&\lim_{N\to\infty}\frac{1}{N}\sum_i\frac{\partial}{\partial\theta_i(\ell^\prime)}\overline{\bra
\sigma[q_i(\ell),z_i(\ell)]\ket}
 \label{eq:identifyG}
\end{eqnarray}
The brackets $\bra\ldots\ket$ refer to averaging over all
realizations of both the decision noise variables $\{z_i(\ell)\}$
and the fake bid variables $\{Z(\ell,\lambda)\}$, at all times. It
was shown in \cite{Coolen2005} (to which we refer for details)
that for $N\to\infty$ and in terms of new time variable
$t=\ell\del$ (which will become real-valued in this limit) the
order parameters (\ref{eq:identifyC},\ref{eq:identifyG}) are to be
extracted self-consistently from the following disorder-free
effective single agent process
 with a retarded
self-interaction and zero-average Gaussian noise $\eta(t)$ with
non-trivial temporal correlations $\bra
\eta(t)\eta(t^\prime)\ket=\Sigma(t,t^\prime)$:
\begin{eqnarray}
\frac{d}{dt}q(t)&=& \theta(t)-\alpha\int_0^t\!dt^\prime~
R(t,t^\prime)~\sigma[q(t^\prime)] +\sqrt{\alpha}~\eta(t)
\label{eq:single_trader}
\end{eqnarray}
 As a
result of the limit $N\to\infty$ we must also write
$C(\ell,\ell^\prime)\to C(t,t^\prime)$ and $G(\ell,\ell^\prime)\to
G(t,t^\prime)$. One always has $G(t,t^\prime)=0$ for $t\leq
t^\prime$ (due to causality), and $C(t,t^\prime)=C(t^\prime,t)$
with $C(t,t)=1$. If we write averages over realizations of the
non-Markovian process (\ref{eq:single_trader}) as
$\bra\ldots\ket_\star$, the kernels $\{C,G\}$ must satisfy for
$t>t^\prime$:
\begin{equation}
C(t,t^\prime)=\bra \sigma[q(t)]\sigma[q(t^\prime)]\ket_\star
~~~~~~~~G(t,t^\prime)=\frac{\delta}{\delta\theta(t^\prime)}\bra
\sigma[q(t)]\ket_\star \label{eq:finalCG}
\end{equation}
 The relation between the auxiliary
kernels $\{R,\Sigma\}$ and the order parameters
$\{C,G\}$ was found to be defined via an effective
process for the overall bid $A(\ell)$, namely by
\begin{eqnarray}
R(t,t^\prime) &=& \frac{\delta}{\delta A_e(t^\prime)}
~\lim_{\del\to 0}~\left. \bigbra\bigbra
\overline{W}[\ell^\prime,\ell;\{A,Z\}] A(\ell)
 \bigket\bigket_{\!\{A,Z\}}\right|_{\ell=t/\del,\ell^\prime=t^\prime/\del}
 \label{eq:finalR}
 \\
\Sigma(t,t^\prime) &=& \rate ~\lim_{\del\to 0} ~\del^{-1}
\left.\bigbra\bigbra \overline{W}[\ell,\ell^\prime;\{A,Z\}]
A(\ell)A(\ell^\prime) \bigket
\bigket_{\!\{A,Z\}}\right|_{\ell=t/\del,\ell^\prime=t^\prime/\del}
 \label{eq:finalSigma}
\end{eqnarray}
Here $\bra\bra \ldots\ket\ket_{A,Z}$
refers to an average over this effective stochastic process for
the bids $\{A\}$ and the pseudo-history $\{Z\}$, a process that was found \cite{Coolen2005} to take the form
\begin{eqnarray}
 A(\ell)&=& A_e(\ell)+\phi_\ell -
\frac{1}{2}\rate~\sum_{\ell^\prime<\ell}
 G(\ell,\ell^\prime)\overline{W}[\ell,\ell^\prime;\{A,Z\}]A(\ell^\prime)
\label{eq:effective_bid_process}
\end{eqnarray}
with zero-average Gaussian random fields $\{\phi\}$, characterized
by \begin{eqnarray} \bra
\phi_\ell\phi_{\ell^\prime}\ket_{\{\phi|A,Z\}}&=&\frac{1}{2}[1+C(\ell,\ell^\prime)]
\overline{W}[\ell,\ell^\prime;\{A,Z\}] \label{eq:finall_phistats}
\end{eqnarray}
The function $\overline{W}[.;.]$ in these formulae is defined as
\begin{equation}
\overline{W}[\ell,\ell^\prime;\{A,Z\}]=\frac{1}{p}\sum_\lambda\F_\lambda[\ell,A,Z]\F_\lambda[\ell^\prime,A,Z]
\label{eq:defineW}
 \end{equation}
  The macroscopic theory is closed. The
remaining problem is to (i) express the kernels $\{R,\Sigma\}$ in
terms of $\{C,G\}$ using (\ref{eq:finalR},\ref{eq:finalSigma}),
followed by (ii) calculating the order parameters $\{C,G\}$ from
(\ref{eq:finalCG}).

It is quite satisfactory to find that the theory of
\cite{Coolen2005} can be made to apply to a much broader class of
MG models, which saves us from having to redo the generating
functional analysis here.
 Mathematically, the differences
between the two main MG families (look-up table versus inner
product strategies) are within the macroscopic theory found to be
limited to which expression to substitute in equations
(\ref{eq:finalR},\ref{eq:finalSigma},\ref{eq:effective_bid_process})
for the function $\overline{W}[\ell,\ell^\prime;\{A,Z\}]$, which
measures the similarity between the market `histories' (whether
fake or real) as observed at stages $\ell$ and $\ell^\prime$ of
the process:
\begin{eqnarray}
{\rm look\!\!-\!\!up~table:}&&
\overline{W}[\ell,\ell^\prime;\{A,Z\}]=\delta_{\blambda(\ell,A,Z),\blambda(\ell^\prime,A,Z)}
 \label{eq:W_LU}
\\[4mm]
{\rm inner~product}:&&
\overline{W}[\ell,\ell^\prime;\{A,Z\}]=\label{eq:W_IP}
\\[1mm]
&&\hspace*{-20mm}{\small \frac{1}{\alpha N}\sum_\lambda
f[(1-\zeta) A(\ell-\lambda) + \zeta Z(\ell,\lambda)]
  ~f[(1-\zeta)
A(\ell^\prime\!-\lambda) + \zeta Z(\ell^\prime\!,\lambda)]
}\nonumber
\end{eqnarray}
Here $\blambda(\ell,A,Z)\in\{-1,1\}^M$ (with $2^M=p$) denotes the
`history string' as observed by agents in the lookup table MGs at
time $\ell$, with entries $\lambda_k(\ell,A,Z)=\sgn[(1-\zeta)
A(\ell-k)] + \zeta Z(\ell,k)]$. Similarly, the differences between the two
`fake memory' definitions (\ref{eq:typeA},\ref{eq:typeB}) are
limited to the details of the zero-average Gaussian variables
$\{Z\}$ in (\ref{eq:W_LU},\ref{eq:W_IP}), and the associated
averaging process $\bra \ldots\ket_{Z}$.

\subsection{Time translation invariant stationary states}

\noindent In the special case of fully ergodic and
time-translation invariant states (TTI) without anomalous
response, where $C(t,t^\prime)=C(t-t^\prime)$,
$G(t,t^\prime)=G(t-t^\prime)$, $R(t,t^\prime)=R(t-t^\prime)$ and
$\Sigma(t,t^\prime)=\Sigma(t-t^\prime)$, one can derive from the
effective single agent equation relatively simple and familiar
expressions for persistent order parameters \cite{Coolen2005}. The
relevant scalar quantities are
 $\chi=\int_0^\infty\!dt~G(t)$,  $\chi_R=\int_0^\infty\!dt~R(t)$,
 $c=\lim_{t\to\infty}C(t)$, and the fraction $\phi$ of `frozen'
 agents in the game. They were found \cite{Coolen2005} to obey:
\begin{eqnarray}
\phi&=& 1- {\rm Erf}[u] \label{eq:phi_eqn}
\\
c &=&\sigma^2[\infty]\Big\{ 1-{\rm Erf}[u]+\frac{1}{2u^2}{\rm
Erf}[u]-\frac{1}{u\sqrt{\pi}}\rme^{-u^2}\Big\} \label{eq:c_eqn}
\\
\chi&=& {\rm Erf}[u]/\alpha\chiR
 \label{eq:chi_eqn}
\end{eqnarray}
with the short-hands
$u=\sqrt{\alpha}\chiR\sigma[\infty]/S_0\sqrt{2}$ and
$S_0^2=\Sigma(\infty)$. In order to find the TTI stationary
solution $\{\phi,c,\chi\}$ and the phase transition point (defined
by $\chi\to\infty$), we therefore do not need to solve for our
order parameter kernels in full but just need to extract
expressions for $\chiR$ and $S_0$ from the stochastic overall bid
process (\ref{eq:effective_bid_process}). These latter two
quantities can be written as
\begin{eqnarray}
\chiR &=& \lim_{\del\to 0}\left\{\overline{W}[0,0;\{A,Z\}]+
\sum_{\ell=1}^\infty \frac{\partial}{\partial A(0)}
\bigbra\!\bigbra \overline{W}[\ell,0;\{A,Z\}] A(\ell)
 \bigket\!\bigket_{\!\{A,Z\}}\right\}~~~
 \\
S_0^2 &=&  \lim_{\del\to 0}
\lim_{L\to\infty}\frac{\rate}{L^2\del}\sum_{\ell,\ell^\prime=1}^L
\bigbra\!\bigbra \overline{W}[\ell,\ell^\prime;\{A,Z\}]
A(\ell)A(\ell^\prime) \bigket\!\bigket_{\!\{A,Z\}}
\end{eqnarray}
Note that $A(0)$ and $\overline{W}[0,0;\{A,Z\}]$ express the
initial conditions; for the purpose of evaluating the stationary
state we may now drop the global bid perturbations $\{A_e(t)\}$.

\subsection{Expressions for the kernels $R$ and $\Sigma$}

\noindent In order to work out the theory also away from TTI
stationary states and compare it to that of look-up table MGs,
knowledge of just $\chiR$ and $S_0$ is no longer sufficient; one
needs to calculate the kernels $R$ and $\Sigma$ in full. Here it
is no longer clear to what extent the analysis of
\cite{Coolen2005} can be adapted. The differences between the two
model families start to manifest themselves.
 We first define
\begin{eqnarray}
\hspace*{-25mm}
 \Delta_{r+1}(\ell_0,\ldots,\ell_{r})&=& p^r
\Big\bra\!\Big\bra
\overline{W}[\ell_0,\ell_r;\{A,Z\}]\prod_{i=1}^{r}
\overline{W}[\ell_{i-1},\ell_i;\{A,Z\}]
 \Big\ket\!\Big\ket_{\!\{A,Z\}}
\label{eq:general_Deltas}\\ \hspace*{-25mm}
 \tilde{\Delta}_{r+r^\prime+2}(\ell_0,\ldots,\ell_{r};\ell^\prime_0,\ldots,\ell^\prime_{r^\prime})&=&
 p^{r+r^\prime+1} \Big\bra\!\Big\bra \overline{W}[\ell_0,\ell^\prime_{0};\{A,Z\}]\overline{W}[\ell_r,\ell^\prime_{r^\prime};\{A,Z\}]
\nonumber \\
 \hspace*{-25mm} && \hspace*{-15mm}\times \Big[\prod_{i=1}^{r}
\overline{W}[\ell_{i-1},\ell_i;\{A,Z\}]\Big]\Big[\prod_{j=1}^{r^\prime}
\overline{W}[\ell^\prime_{j-1},\ell^\prime_j;\{A,Z\}]\Big]
 \Big\ket\!\Big\ket_{\!\{A,Z\}}~~
\label{eq:general_Deltahat}
\end{eqnarray}
As was done in in \cite{Coolen2005} we re-write the global bid
equation (\ref{eq:effective_bid_process}) as
\begin{eqnarray*}
\sum_{\ell^\prime\leq \ell}\Big\{ \delta_{\ell\ell^\prime} +
\frac{1}{2}\rate
 G(\ell,\ell^\prime)\overline{W}[\ell,\ell^\prime;\{A,Z\}]
 \Big\}A(\ell^\prime)
&=& A_e(\ell)+\phi_\ell
\end{eqnarray*}
and we invert the operator on the left-hand side, using
$\del=\rate/2p$:
\begin{eqnarray}
\hspace*{-20mm} A(\ell)&=& A_e(\ell)+\phi_\ell+\sum_{r>0}(-\del p
)^r\sum_{\ell_1\ldots \ell_r} G(\ell,\ell_1)G(\ell_1,\ell_2)\ldots
G(\ell_{r-1},\ell_r) \nonumber \\
 \hspace*{-20mm}
 &&\times
\overline{W}[\ell,\ell_1;\{A,Z\}]
\overline{W}[\ell_1,\ell_2;\{A,Z\}]\ldots
\overline{W}[\ell_{r-1},\ell_r;\{A,Z\}] ~
\big[A_e(\ell_r)+\phi_{\ell_r}\big] \label{eq:Aexplicit}
\end{eqnarray}
 We insert (\ref{eq:Aexplicit}) into
(\ref{eq:finalR}), and consider only infinitesimal external bid
perturbations $A_e$:
\begin{eqnarray}
\hspace*{-25mm}
 \room R(t,t^\prime) &=&
\delta(t-t^\prime)+\lim_{\del\to 0}\frac{1}{\del}
\left\{\sum_{r>0}(-\del)^{r} \sum_{\ell_1\ldots \ell_{r-1}}\!
G(\ell_0,\ell_1)G(\ell_1,\ell_2)\ldots G(\ell_{r-1},\ell_r)
\right.\nonumber
\\ \hspace*{-25mm} &&
\left. \hspace*{20mm} \times~ p^r \Big\bra\!\Big\bra
\overline{W}[\ell_0,\ell_r;\{A,Z\}] \prod_{i=1}^{r}
\overline{W}[\ell_{i-1},\ell_i;\{A,Z\}]
 \Big\ket\!\Big\ket_{\!\{A,Z\}}\right\}
 \Big|_{\ell_0=\frac{t}{\del},\ell_r=\frac{t^\prime}{\del}}
 \nonumber
 \\
 \hspace*{-25mm}
 &=&
\delta(t-t^\prime)+\lim_{\del\to 0}
\frac{1}{\del}\left\{\sum_{r>0}(-\del)^{r}\! \sum_{\ell_1\ldots
\ell_{r-1}} G(\ell_0,\ell_1)\ldots G(\ell_{r-1},\ell_r)\nonumber
\right.
\\ \hspace*{-25mm} &&
\hspace*{50mm}\left.\times~ \Delta_{r+1}(\ell_0,\ldots,\ell_{r})
\room\right\}
\Big|_{\ell_0=\frac{t}{\del},\ell_r=\frac{t^\prime}{\del}}
 \label{eq:workoutR}
\end{eqnarray}
This is clearly identical to the corresponding expression found in
\cite{Coolen2005} (although now the meaning of the kernels
$\Delta_{k}(\ldots)$ is allowed to be different, dependent on
which model family we choose to apply the generalized theory to).
Similarly we can insert
 (\ref{eq:Aexplicit})  into
(\ref{eq:finalSigma}), again with $A_e\to 0$, and find
\begin{eqnarray}
\hspace*{-15mm} \Sigma(t,t^\prime) &=& \rate \lim_{\del\to 0}
\frac{1}{\del} \left\{ \sum_{r,r^\prime\geq 0}(-\del)^{r+r^\prime}
\sum_{\ell_1\ldots \ell_r}G(\ell_0,\ell_1)\ldots
G(\ell_{r-1},\ell_r) \nonumber\right.
\\ \hspace*{-15mm}
&&\left. \times\sum_{\ell^\prime_1\ldots \ell^\prime_{r^\prime}}
G(\ell^\prime_0,\ell^\prime_1)\ldots
G(\ell^\prime_{r^\prime-1},\ell^\prime_{r^\prime})~p^{r+r^\prime}
 \Big\bra\!\Big\bra~
 \bra\phi_{\ell_r}
 \phi_{\ell^\prime_{r^\prime}}\ket_{\{\phi|A,Z\}}
 \overline{W}[\ell_0,\ell_0^\prime;\{A,Z\}]
 \nonumber \right.
\\ \hspace*{-15mm} &&
\left. \times~ \Big[\prod_{i=1}^r
\overline{W}[\ell_{i-1},\ell_i;\{A,Z\}] \Big]
\Big[\prod_{j=1}^{r^\prime}
\overline{W}[\ell_{j-1}^\prime,\ell^\prime_j;\{A,Z\}]
 \Big]~
 \Big\ket\!
\Big\ket_{\!\{A,Z\}}\right\}\Big|_{\ell_0=\frac{t}{\del},\ell^\prime_0=\frac{t^\prime}{\del}}
\nonumber
\\ \hspace*{-15mm}
&=&
 \lim_{\del\to 0} \left\{ \sum_{r,r^\prime\geq
0}(-\del)^{r+r^\prime} \sum_{\ell_1\ldots
\ell_r}G(\ell_0,\ell_1)\ldots G(\ell_{r-1},\ell_r)
\nonumber\right.
\\ \hspace*{-15mm}
&&\left.\hspace*{20mm} \times\sum_{\ell^\prime_1\ldots
\ell^\prime_{r^\prime}} G(\ell^\prime_0,\ell^\prime_1)\ldots
G(\ell^\prime_{r^\prime-1},\ell^\prime_{r^\prime})~\big[1+C(\ell_r,\ell^\prime_{r^\prime})\big]
 \nonumber \right.
\\ \hspace*{-15mm} &&
\left.\hspace*{20mm} \times~
\tilde{\Delta}_{r+r^\prime+2}(\ell_0,\ldots,\ell_r;\ell^\prime_0,\ldots,\ell^\prime_{r^{\prime}})
\room\right\}\Big|_{\ell_0=\frac{t}{\del},\ell^\prime_0=\frac{t^\prime}{\del}}
 \label{eq:workoutSigma}
\end{eqnarray}
The limits $\del\to 0$ in
(\ref{eq:workoutR},\ref{eq:workoutSigma}) are well-defined. Each
time summation combines with a factor $\del$ to generate an
integral, whereas pairwise identical times in
(\ref{eq:workoutSigma}) leave a `bare' factor $\del$ but can also
be anticipated to cause $\tilde{\Delta}_{r+r^\prime+2}(\ldots)$
gaining a factor $p=\rate/2\del$ in compensation.

Equations (\ref{eq:workoutR},\ref{eq:workoutSigma}) show, upon
inserting the respective definitions (\ref{eq:W_LU},\ref{eq:W_IP})
of the function $\overline{W}[.;.]$, that replacing lookup table
strategy definitions by inner product ones
 does have implications.
For instance, the following relation holds for lookup table MGs
but appears to be no longer valid for inner product ones:
\begin{equation}
\tilde{\Delta}_{r+r^\prime+2}(\ell_0,\ldots,\ell_{r};\ell^\prime_0,\ldots,\ell^\prime_{r^\prime})=
\Delta_{r+r^\prime+2}(\ell_0,\ldots,\ell_{r},\ell^\prime_0,\ldots,\ell^\prime_{r^\prime})
\end{equation}

\section{Inner product MGs with inconsistent fake market history}

\noindent The simplest instance of our inner product MG model is the case
$\zeta=1$, i.e.  fake market history, of the inconsistent type.
Here the function
$\overline{W}[\ell,\ell^\prime;\{A,Z\}]=p^{-1}\sum_{\lambda}f[Z(\ell,\lambda)]f[Z(\ell^\prime,\lambda)]$
is no longer dependent on the real market bids $A$, and $\bra
Z(\ell,\lambda)Z(\ell^\prime,\lambda^\prime)\ket=S^2
\delta_{\ell\ell^\prime}\delta_{\lambda\lambda^\prime}$. Even for
this case no exact macroscopic solution has so far been published.
We define the short-hands
$\tilde{Z}_{\ell,\lambda}=f[Z(\ell,\lambda)]$ and
$\kappa_n=\int\!Dz ~f^n[Sz]$ with the Gaussian measure
$Dz=(2\pi)^{-\frac{1}{2}}\rme^{-\frac{1}{2}z^2}dz$ (since $f[A]$ is anti-symmetric, $\kappa_n=0$ for all odd $n$).

\subsection{The retarded self-interaction kernel $R$}

\noindent To proceed we have to calculate the two functions
(\ref{eq:general_Deltas},\ref{eq:general_Deltahat}). They
occur only in expressions (\ref{eq:workoutR}) and
(\ref{eq:workoutSigma}), where causality of the
response function $G$ enforces a helpful ordering of the time
arguments. The kernel $\Delta_{r+1}(\ell_0,\ldots,\ell_{r})$ as
occurring in  (\ref{eq:workoutR}) is easy to evaluate, since here
we may use $\ell_0>\ell_1>\ldots>\ell_r$. This property, in
combination with $\kappa_1=\bra \tilde{Z}_{\ell,\lambda}\ket=0$
for all $(\ell,\lambda)$, implies that the only non-zero
contributions are those where the Gaussian variables
$\{\tilde{Z}\}$ have pair-wise identical indices:
\begin{eqnarray}
\hspace*{-5mm}
 \Delta_{r+1}(\ell_0,\ldots,\ell_{r})&=& \frac{1}{p}\sum_{\lambda_0
 \ldots\lambda_r=1}^p
\Big\bra \tilde{Z}_{\ell_0,\lambda_0} \tilde{Z}_{\ell_r,\lambda_0}
\prod_{i=1}^{r}
\tilde{Z}_{\ell_{i-1},\lambda_i}\tilde{Z}_{\ell_i,\lambda_i}
 \Big\ket_{\!\{Z\}}
\nonumber
\\ \hspace*{-5mm}&=& \frac{1}{p}\sum_{\lambda_0
 \ldots\lambda_r=1}^p
\bra \tilde{Z}_{\ell_0,\lambda_0}
 \tilde{Z}_{\ell_{0},\lambda_1}\ket
\Big( \prod_{i=1}^{r-1}
 \bra \tilde{Z}_{\ell_i,\lambda_i}
\tilde{Z}_{\ell_i,\lambda_{i+1}} \ket\Big) \bra
\tilde{Z}_{\ell_r,\lambda_r} \tilde{Z}_{\ell_r,\lambda_0}\ket
\nonumber
\\ \hspace*{-5mm}&=& \frac{1}{p}~\kappa_2^{r+1}\sum_{\lambda_0
 \ldots\lambda_r=1}^p
\delta_{\lambda_0,\lambda_1} \Big( \prod_{i=1}^{r-1}
 \delta_{\lambda_i,\lambda_{i+1}}\Big)
 \delta_{\lambda_r,\lambda_0}=\kappa_2^{r+1}
\end{eqnarray}
Insertion into (\ref{eq:workoutR}) then gives us the fully
explicit form
\begin{eqnarray}
\hspace*{-5mm}
 R(t,t^\prime)
 &=&\kappa_2\Big\{
\delta(t-t^\prime)+\lim_{\del\to 0}
\frac{1}{\del}\sum_{r>0}(-\del)^{r} \kappa_2^r\sum_{\ell_1\ldots
\ell_{r-1}} G(\ell_0,\ell_1)\ldots G(\ell_{r-1},\ell_r)\Big\}
\nonumber
\\ \hspace*{-5mm}
&=& \kappa_2\Big\{ \delta(t-t^\prime)+\sum_{r>0}(-\kappa_2)^r
G^r(t,t^\prime)\Big\}\nonumber
\\ \hspace*{-5mm} &=&\kappa_2(\one+\kappa_2G)^{-1}(t,t^\prime)
 \label{eq:finishedR}
\end{eqnarray}
This result for inner product MGs depends on the
function $f[.]$ and the variance $S$ only via a single parameter
$\kappa_2=\int\!Dz ~f^2[Sz]$, and reduces to the
corresponding expression found earlier \cite{CoolenHeimel2001} for
look-up table MGs with fake histories when $\kappa_2=1$.

\subsection{The effective noise covariance kernel $\Sigma$}

\noindent We now turn to the function (\ref{eq:general_Deltahat}).
This is needed in (\ref{eq:workoutSigma}) but only for time
combinations with $\ell_0>\ell_1>\ldots >\ell_r$ and
$\ell^\prime_0>\ell^\prime_1>\ldots >\ell^\prime_{r^\prime}$.
Again we first group together the various terms that have
identical time labels, and  we also introduce two auxiliary
summation indices $\lambda$ that allow us to separate the terms
with times of the type $\{\ell_i\}$ from those of the type
$\{\ell^\prime_j\}$ in a clean way:
\begin{eqnarray}
\hspace*{-20mm} && \hspace*{-10mm}
 \tilde{\Delta}_{r+r^\prime+2}(\ell_0,\ldots,\ell_{r};\ell^\prime_0,\ldots,\ell^\prime_{r^\prime})\nonumber
 \\
 \hspace*{-20mm}
&=&\frac{1}{p}\sum_{\lambda_0\ldots\lambda_r=1}^p
\sum_{\lambda^\prime_1\ldots\lambda^\prime_{r^\prime+1}=1}^p
 \Big\bra
\tilde{Z}_{\ell_0,\lambda_0}\tilde{Z}_{\ell^\prime_0,\lambda_0}
\tilde{Z}_{\ell_r,\lambda^\prime_{r^\prime+1}}\tilde{Z}_{\ell^\prime_{r^\prime},\lambda^\prime_{r^\prime+1}}
\nonumber \\
 \hspace*{-20mm} && \hspace*{40mm} \times
\Big[\prod_{i=1}^{r}
\tilde{Z}_{\ell_{i-1},\lambda_i}\tilde{Z}_{\ell_i,\lambda_i} \Big]
\Big[\prod_{j=1}^{r^\prime}
\tilde{Z}_{\ell^\prime_{j-1},\lambda^\prime_j}
\tilde{Z}_{\ell^\prime_{j},\lambda^\prime_j} \Big]
 \Big\ket_{\!\{Z\}}\nonumber
 \\
 \hspace*{-20mm}
 &=&
\frac{1}{p}\sum_{\lambda_0\ldots\lambda_{r+1}=1}^p
\sum_{\lambda^\prime_0\ldots\lambda^\prime_{r^\prime+1}=1}^p
\delta_{\lambda_{r+1},\lambda^\prime_{r^\prime+1}}\delta_{\lambda^\prime_0,\lambda_0}
 \Big\bra
\tilde{Z}_{\ell_0,\lambda_0}\tilde{Z}_{\ell^\prime_0,\lambda^\prime_0}
\tilde{Z}_{\ell_r,\lambda_{r+1}}\tilde{Z}_{\ell^\prime_{r^\prime},\lambda^\prime_{r^\prime+1}}
\nonumber
\\
\hspace*{-20mm} &&\hspace*{40mm} \times \Big[\prod_{i=1}^{r}
\tilde{Z}_{\ell_{i-1},\lambda_i}\tilde{Z}_{\ell_i,\lambda_i} \Big]
\Big[\prod_{j=1}^{r^\prime}
\tilde{Z}_{\ell^\prime_{j-1},\lambda^\prime_j}
\tilde{Z}_{\ell^\prime_{j},\lambda^\prime_j} \Big]
 \Big\ket_{\!\{Z\}}~~~~~~~~
 \nonumber
 \\ \hspace*{-20mm}
 &=&
 \frac{1}{p}\sum_{\lambda_0\ldots\lambda_{r+1}=1}^p
\sum_{\lambda^\prime_0\ldots\lambda^\prime_{r^\prime+1}=1}^p
\delta_{\lambda_{r+1},\lambda^\prime_{r^\prime+1}}\delta_{\lambda^\prime_0,\lambda_0}
 \Big\bra
\Big[\prod_{i=0}^r \tilde{Z}_{\ell_i,\lambda_i}
 \tilde{Z}_{\ell_{i},\lambda_{i+1}}\Big]
 \Big[\prod_{j=0}^{r^\prime} \tilde{Z}_{\ell^\prime_j,\lambda^\prime_j}
 \tilde{Z}_{\ell^\prime_{j},\lambda^\prime_{j+1}}\Big]
 \Big\ket_{\!\{Z\}}
 \label{eq:intermediate_Sdelta}
\end{eqnarray}
Due to the time ordering relations of the problem, there can never
be time coincidences amongst members of the set
$\{\ell_0,\ldots,\ell_r\}$, nor amongst members of the set
$\{\ell^\prime_0,\ldots,\ell^\prime_{r^\prime}\}$. Thus the
various factors within the first pair of round brackets
$\big(\prod_{i=0}^r\ldots\big)$ in (\ref{eq:intermediate_Sdelta})
are all independent, and so are those within the second pair of
round brackets $\big(\prod_{j=0}^{r^\prime}\ldots\big)$. The only
possible time coincidences are {\em pairings} between un-primed
times $\{\ell_i\}$ and primed times $\{\ell^\prime_j\}$. Thus we
have to consider the following cases:
\begin{itemize}
\item
If there are no time coincidences, i.e. $\ell_i\neq \ell^\prime_j$
for all $(i,j)$,  the evaluation of (\ref{eq:intermediate_Sdelta})
proceeds similarly to that of the previous functions
$\Delta_{r+1}(\ldots)$:
\begin{eqnarray}\hspace*{-5mm}
&&\hspace*{-15mm}
 \tilde{\Delta}_{r+r^\prime+2}(\ell_0,\ldots,\ell_{r};\ell^\prime_0,\ldots,\ell^\prime_{r^\prime})\nonumber
 \\
 \hspace*{-5mm}
&=&
 \frac{1}{p}~\kappa_2^{r+r^\prime+2}\sum_{\lambda_0\ldots\lambda_{r+1}=1}^p
\sum_{\lambda^\prime_0\ldots\lambda^\prime_{r^\prime+1}=1}^p
\delta_{\lambda_{r+1},\lambda^\prime_{r^\prime+1}}\delta_{\lambda^\prime_0,\lambda_0}
\Big[\prod_{i=0}^r \delta_{\lambda_i,\lambda_{i+1}}\Big]
 \Big[\prod_{j=0}^{r^\prime}
 \delta_{\lambda^\prime_j,\lambda^\prime_{j+1}}\Big]
 \nonumber
  \\
  \hspace*{-5mm}
&=&
 \frac{1}{p}~\kappa_2^{r+r^\prime+2}\sum_{\lambda_0=1}^p
\sum_{\lambda^\prime_0=1}^p
\delta_{\lambda^\prime_0,\lambda_0}=\kappa_2^{r+r^\prime+2}
\end{eqnarray}
\item
Now consider the effect of a time pairing, where
$\ell_i=\ell_j^\prime$ (note: their can be multiple pairings, but
the number of coinciding times is two at most, due to the
built-in time ordering). The contribution of the Gaussian averages
with times $(\ell_i,\ell^\prime_j)$  to the sum
$\sum_{\lambda_0\ldots\lambda_{r+1}}\sum_{\lambda^\prime_0\ldots\lambda^\prime_{r^\prime+1}}$
in (\ref{eq:intermediate_Sdelta}), which in the absence of pairing
equaled simply
\begin{eqnarray*}
\hspace*{-5mm} {\rm no~pairing:}~~~\bra
\tilde{Z}_{\ell_i,\lambda_i}
 \tilde{Z}_{\ell_{i},\lambda_{i+1}}
\tilde{Z}_{\ell^\prime_j,\lambda^\prime_j}
 \tilde{Z}_{\ell^\prime_{j},\lambda^\prime_{j+1}}\ket&=&\bra \tilde{Z}_{\ell_i,\lambda_i}
 \tilde{Z}_{\ell_{i},\lambda_{i+1}}\ket\bra
\tilde{Z}_{\ell^\prime_j,\lambda^\prime_j}
 \tilde{Z}_{\ell^\prime_{j},\lambda^\prime_{j+1}}\ket\\
 \hspace*{-5mm}&=&
\kappa_2^2\delta_{\lambda_i,\lambda_{i+1}}\delta_{\lambda^\prime_j,\lambda^\prime_{j+1}}
\end{eqnarray*}
now becomes:
\begin{eqnarray*}
\hspace*{-5mm} \ell_i=\ell^\prime_j:~~&&\bra
\tilde{Z}_{\ell_i,\lambda_i}
 \tilde{Z}_{\ell_{i},\lambda_{i+1}}
\tilde{Z}_{\ell^\prime_j,\lambda^\prime_j}
 \tilde{Z}_{\ell^\prime_{j},\lambda^\prime_{j+1}}\ket=
(\kappa_4-3\kappa_2^2)~
\delta_{\lambda_i,\lambda_{i+1}}\delta_{\lambda^\prime_{j},\lambda^\prime_{j+1}}\delta_{\lambda_i,\lambda^\prime_j}\\
\hspace*{-5mm} &&\hspace*{10mm}
 +\kappa_2^2~[
\delta_{\lambda_i,\lambda_{i+1}}\delta_{\lambda^\prime_j,\lambda^\prime_{j+1}}
+\delta_{\lambda_i,\lambda^\prime_j}\delta_{\lambda_{i+1},\lambda^\prime_{j+1}}+\delta_{\lambda_i,\lambda^\prime_{j+1}}
\delta_{\lambda_{i+1},\lambda^\prime_{j}}]
\end{eqnarray*}
\end{itemize}
 \begin{figure}[t]{\tiny
\hspace*{80mm}\setlength{\unitlength}{0.52mm}
\begin{picture}(130,200)
\put(10,30){\vertex}\put(30,30){\vertex}\put(70,30){\vertex}\put(90,30){\vertex}\put(130,30){\vertex}\put(150,30){\vertex}
\put(10,10){\vertex}\put(30,10){\vertex}\put(70,10){\vertex}\put(90,10){\vertex}\put(130,10){\vertex}\put(150,10){\vertex}
\put(10,35){\here{$0$}}\put(30,35){\here{$1$}}\put(70,35){\here{$i$}}\put(90,35){\here{$i+1$}}\put(130,35){\here{$r$}}\put(150,35){\here{$r+1$}}
\put(10,5){\here{$0$}}\put(30,5){\here{$1$}}\put(70,5){\here{$j$}}\put(90,5){\here{$j+1$}}\put(130,5){\here{$r^\prime$}}\put(150,5){\here{$r^\prime+1$}}

\put(9,10){\line(0,1){20}}\put(149,10){\line(0,1){20}}
\put(10,10){\line(1,0){60}}\put(150,10){\line(-1,0){60}}\put(10,30){\line(1,0){60}}\put(150,30){\line(-1,0){60}}
\put(69,10){\line(1,1){20}}\put(89,10){\line(-1,1){20}}
\put(69,10){\line(0,1){20}}

\put(10,80){\vertex}\put(30,80){\vertex}\put(70,80){\vertex}\put(90,80){\vertex}\put(130,80){\vertex}\put(150,80){\vertex}
\put(10,60){\vertex}\put(30,60){\vertex}\put(70,60){\vertex}\put(90,60){\vertex}\put(130,60){\vertex}\put(150,60){\vertex}
\put(10,85){\here{$0$}}\put(30,85){\here{$1$}}\put(70,85){\here{$i$}}\put(90,85){\here{$i+1$}}\put(130,85){\here{$r$}}\put(150,85){\here{$r+1$}}
\put(10,55){\here{$0$}}\put(30,55){\here{$1$}}\put(70,55){\here{$j$}}\put(90,55){\here{$j+1$}}\put(130,55){\here{$r^\prime$}}\put(150,55){\here{$r^\prime+1$}}

\put(9,60){\line(0,1){20}}\put(149,60){\line(0,1){20}}
\put(10,60){\line(1,0){60}}\put(150,60){\line(-1,0){60}}\put(10,80){\line(1,0){60}}\put(150,80){\line(-1,0){60}}
\put(69,60){\line(0,1){20}}\put(89,60){\line(0,1){20}}

\put(10,130){\vertex}\put(30,130){\vertex}\put(70,130){\vertex}\put(90,130){\vertex}\put(130,130){\vertex}\put(150,130){\vertex}
\put(10,110){\vertex}\put(30,110){\vertex}\put(70,110){\vertex}\put(90,110){\vertex}\put(130,110){\vertex}\put(150,110){\vertex}
\put(10,135){\here{$0$}}\put(30,135){\here{$1$}}\put(70,135){\here{$i$}}\put(90,135){\here{$i+1$}}\put(130,135){\here{$r$}}\put(150,135){\here{$r+1$}}
\put(10,105){\here{$0$}}\put(30,105){\here{$1$}}\put(70,105){\here{$j$}}\put(90,105){\here{$j+1$}}\put(130,105){\here{$r^\prime$}}\put(150,105){\here{$r^\prime+1$}}

\put(9,110){\line(0,1){20}}\put(149,110){\line(0,1){20}}
\put(10,110){\line(1,0){60}}\put(150,110){\line(-1,0){60}}\put(10,130){\line(1,0){60}}\put(150,130){\line(-1,0){60}}
\put(69,110){\line(1,1){20}}\put(89,110){\line(-1,1){20}}

\put(10,180){\vertex}\put(30,180){\vertex}\put(70,180){\vertex}\put(90,180){\vertex}\put(130,180){\vertex}\put(150,180){\vertex}
\put(10,160){\vertex}\put(30,160){\vertex}\put(70,160){\vertex}\put(90,160){\vertex}\put(130,160){\vertex}\put(150,160){\vertex}
\put(10,185){\here{$0$}}\put(30,185){\here{$1$}}\put(70,185){\here{$i$}}\put(90,185){\here{$i+1$}}\put(130,185){\here{$r$}}\put(150,185){\here{$r+1$}}
\put(10,155){\here{$0$}}\put(30,155){\here{$1$}}\put(70,155){\here{$j$}}\put(90,155){\here{$j+1$}}\put(130,155){\here{$r^\prime$}}\put(150,155){\here{$r^\prime+1$}}

\put(9,160){\line(0,1){20}}\put(149,160){\line(0,1){20}}\put(10,180){\line(1,0){140}}\put(10,160){\line(1,0){140}}

\put(-150,170){\small\sl no pairings:
$~~\delta_{\lambda_i,\lambda_{i+1}}\delta_{\lambda^\prime_j,\lambda^\prime_{j+1}}$}

\put(-150,120){\small\sl $\ell_i=\ell^\prime_j$:
$~~\delta_{\lambda_i,\lambda_{i+1}}\delta_{\lambda^\prime_j,\lambda^\prime_{j+1}}\!\to
\delta_{\lambda_i,\lambda^\prime_{j+1}}\delta_{\lambda_{i+1},\lambda^\prime_j}$}

\put(-150,70){\small\sl $\ell_i=\ell^\prime_j$:
$~~\delta_{\lambda_i,\lambda_{i+1}}\delta_{\lambda^\prime_j,\lambda^\prime_{j+1}}\!\to
\delta_{\lambda_i,\lambda^\prime_{j}}\delta_{\lambda_{i+1},\lambda^\prime_{j+1}}$}

\put(-150,20){\small\sl $\ell_i=\ell^\prime_j$:
$~~\delta_{\lambda_i,\lambda_{i+1}}\delta_{\lambda^\prime_j,\lambda^\prime_{j+1}}\!\to
\delta_{\lambda_i,\lambda^\prime_j}\delta_{\lambda_i,\lambda_{i+1}}\delta_{\lambda^\prime_j,\lambda^\prime_{j+1}}$}
\end{picture}}
\vspace*{3mm} \caption{Diagrammatical representation of the
different contributions of the Gaussian averages to the function
$\tilde{\Delta}_{r+r^\prime+2}$.  Each label
$i\in\{0,\ldots,r+1\}$ and each label
$j\in\{0,\ldots,r^\prime+1\}$ is drawn as a distinct vertex of a
graph. Each factor $\delta_{\lambda_i,\lambda^\prime_j}$ is drawn
as a line segment connecting the vertices $i$ and $j$. Top graph:
the case where there are no time coincidences. Bottom three
graphs: the three different new contributions that are generated
by the occurrence of a time pairing where $\ell_i=\ell^\prime_j$.
} \label{fig:diagrams}
\end{figure}
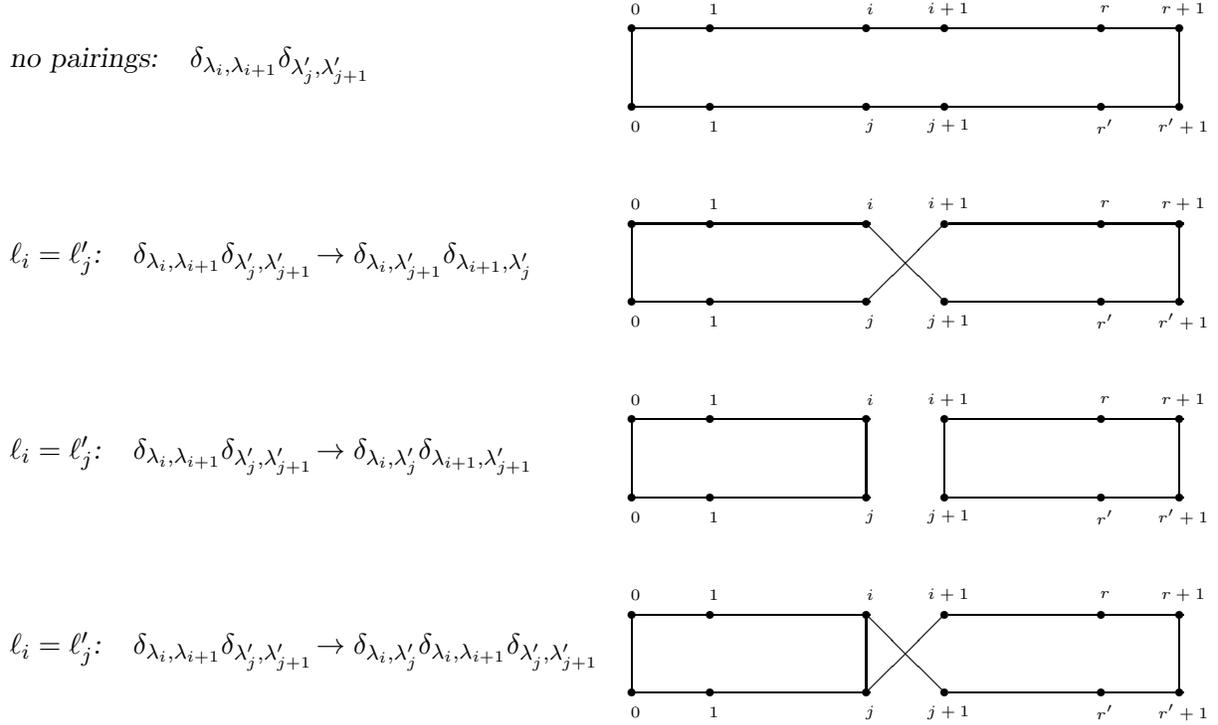
The effect of time pairings can be represented
diagrammatically; see figure \ref{fig:diagrams}.

We now make the crucial observation that in expression
(\ref{eq:workoutSigma}) any time pairing would inevitably generate
a `bare' factor $\del$ that would no longer be absorbed into an
integration via $\sum_{\ell_k}\del\to \int\!dt_k$. Hence the only
contributions to $\tilde{\Delta}_{r+r^\prime+2}$ to survive the
limit $\del\to 0$ in (\ref{eq:workoutSigma}) are those where each
time pairing also generates an extra $\order(p)$ factor to
compensate for the emerging $\del$. Let us inspect the various
diagrams, see figure \ref{fig:diagrams} (and the higher order
versions generated due to multiple time coincidences) and their
contributions to $\tilde{\Delta}_{r+r^\prime+2}$. Each {\em
connected} diagram implies that ultimately in the summation
$\sum_{\lambda_0\ldots\lambda_r}\sum_{\lambda^\prime_0\ldots
\lambda^\prime_{r^\prime}}$ we are restricted to
$\lambda_0=\lambda_1=\ldots
\lambda_r=\lambda^\prime_0=\ldots=\lambda^\prime_{r^\prime}$. This
leaves a sum over $\lambda_0$ only, and a final $\order(1)$
contribution to $\tilde{\Delta}_{r+r^\prime+2}$. On the other
hand, in the case of a {\em disconnected} diagram, each internally
connected sub-diagram will give an $\order(p)$ factor. Here  the
order of the final contribution to $\tilde{\Delta}_{r+r^\prime+2}$
is $p^L$ where $L$ denotes the total number of time pairings that
gave rise to the diagram cuts. Hence the only relevant diagrams in
figure \ref{fig:diagrams} are the top one (when there are no
pairings) and the third from the top (when there are pairings, but
where these also generate compensating $\order(p)$ factors),
including the higher order diagrams with multiple vertical
connections (in the case of multiple time pairings).

 It follows, in combination with the contribution coming from
 the unpaired terms and from
 each possible time pairing as calculated earlier,  and taking into account the crucial time orderings in (\ref{eq:workoutSigma}), that
\begin{eqnarray}
\hspace*{-15mm}
 \tilde{\Delta}_{r+r^\prime+2}(\ell_0,\ldots,\ell_{r};\ell^\prime_0,\ldots,\ell^\prime_{r^\prime})
&=&
 \kappa_2^{r+r^\prime+2}\prod_{i=1}^{r}\prod_{j=1}^{r^\prime}\left[1+\delta_{\ell_i,\ell^\prime_j}(p+\order(p^0)\right]\nonumber
 \\ \hspace*{-15mm}
 &=& \kappa_2^{r+r^\prime+2}\prod_{i=1}^{r}\prod_{j=1}^{r^\prime}\left[1+\frac{\rate}{2\del}\delta_{\ell_i,\ell^\prime_j}(1+\order(\del)\right]
\end{eqnarray}
Insertion into (\ref{eq:workoutSigma}) gives
\begin{eqnarray}
\hspace*{-15mm} \Sigma(t_0,t_0^\prime) &=& \kappa_2^2
  \sum_{r,r^\prime\geq
0}(-\kappa_2)^{r+r^\prime} \int_0^\infty\!dt_1\ldots dt_r
dt^\prime_1 \ldots
dt^\prime_{r^\prime}\prod_{i=1}^{r}\prod_{j=1}^{r^\prime}\left[1+\frac{1}{2}\rate\delta(t_i-t^\prime_j)\right]\nonumber
\\
\hspace*{-15mm} &&\times \big[1+C(t_r,t^\prime_{r^\prime})\big]
G(t_0,t_1)\ldots G(t_{r-1},t_r)G(t_0^\prime,t^\prime_1)\ldots
G(t^\prime_{r^\prime-1},t^\prime_{r^\prime})
 \label{eq:finishedSigma}
\end{eqnarray}
We see again, as with the retarded self-interaction kernel $R$,
that also this result for inner product MGs reduces to the
corresponding expression found earlier \cite{CoolenHeimel2001} for
look-up table MGs with fake histories when $\kappa_2=1$.

\subsection{Summary, TTI stationary state and phase diagram}

\noindent The full and closed dynamical equations for the inner
product MGs with inconsistent fake market information are thus
found to be given by
\begin{equation}
C(t,t^\prime)=\bra \sigma[q(t)]\sigma[q(t^\prime)]\ket_\star
~~~~~~~~G(t,t^\prime)=\frac{\delta}{\delta\theta(t^\prime)}\bra
\sigma[q(t)]\ket_\star \label{eq:finalCGagain}
\end{equation}
Averages are defined with respect to the effective single agent
process with a retarded self-interaction and an effective
zero-average Gaussian noise $\eta(t)$:
\begin{eqnarray}
\hspace*{-15mm} \frac{d}{dt}q(t)&=&
\theta(t)-\alpha\kappa_2\int_0^t\!dt^\prime~(\one+\kappa_2
G)^{-1}(t,t^\prime)~\sigma[q(t^\prime)] +\sqrt{\alpha}\eta(t)
\label{eq:single_trader_again}
\\
\hspace*{-15mm} \bra \eta(t_0)\eta(t_0^\prime)\ket&=&\kappa_2^2
  \sum_{r,r^\prime\geq
0}(-\kappa_2)^{r+r^\prime} \int_0^\infty\!dt_1\ldots dt_r
dt^\prime_1 \ldots
dt^\prime_{r^\prime}\prod_{i=1}^{r}\prod_{j=1}^{r^\prime}\left[1+\frac{1}{2}\rate\delta(t_i-t^\prime_j)\right]\nonumber
\\
\hspace*{-15mm} &&\times \big[1+C(t_r,t^\prime_{r^\prime})\big]
G(t_0,t_1)\ldots G(t_{r-1},t_r)G(t_0^\prime,t^\prime_1)\ldots
G(t^\prime_{r^\prime-1},t^\prime_{r^\prime})
 \label{eq:Sigma_again}
\end{eqnarray}
We notice that the role of the remaining parameter $\kappa_2$ is
to define a characteristic time scale for the process, and that
apart from this our equations reduce in the limit of infinitesimal
perturbation fields exactly to the theory of the fake history
look-up table MGs as derived in \cite{CoolenHeimel2001}. To see
this we re-define
$G(t,t^\prime)=\kappa_2^{-1}\hat{G}(t,t^\prime)$,
$\theta(t)=\kappa_2\hat{\theta(t)}$ and
$\eta(t)=\kappa_2\hat{\eta(t)}$, and find
\begin{eqnarray}
\hspace*{-15mm} C(t,t^\prime)&=&\bra
\sigma[q(t)]\sigma[q(t^\prime)]\ket_\star
~~~~~~~~\hat{G}(t,t^\prime)=\frac{\delta}{\hat{\theta}(t^\prime)}\bra
\sigma[q(t)]\ket_\star \label{eq:finalCGagain2}
\\
\hspace*{-15mm} \frac{1}{\kappa_2}\frac{d}{dt}q(t)&=&
\hat{\theta}(t)-\alpha\int_0^t\!dt^\prime~(\one+\hat{G})^{-1}(t,t^\prime)~\sigma[q(t^\prime)]
+\sqrt{\alpha}\hat{\eta}(t) \label{eq:single_trader_again2}
\\
\hspace*{-15mm} \bra \hat{\eta}(t_0)\hat{\eta}(t_0^\prime)\ket&=&
  \sum_{r,r^\prime\geq
0}(-1)^{r+r^\prime} \int_0^\infty\!dt_1\ldots dt_r dt^\prime_1
\ldots
dt^\prime_{r^\prime}\prod_{i=1}^{r}\prod_{j=1}^{r^\prime}\left[1+\frac{1}{2}\rate\delta(t_i-t^\prime_j)\right]\nonumber
\\
\hspace*{-15mm} &&\times \big[1+C(t_r,t^\prime_{r^\prime})\big]
\hat{G}(t_0,t_1)\ldots
\hat{G}(t_{r-1},t_r)\hat{G}(t_0^\prime,t^\prime_1)\ldots
\hat{G}(t^\prime_{r^\prime-1},t^\prime_{r^\prime})
 \label{eq:Sigma_again2}
\end{eqnarray}
It will no longer come as a surprise that upon using
(\ref{eq:finishedR},\ref{eq:finishedSigma}) to calculate  in
time-translation invariant stationary states the persistent order
parameters $\chiR=\int\!dt~R(t)$ and $S_0^2=\Sigma(\infty)$, we
find that the only difference between the stationary state order
parameter equations of lookup table MGs and the present inner
product MGs is a re-scaling of the static susceptibility $\chi$:
\begin{eqnarray}
u&=& \sigma[\infty]\sqrt{\alpha/2(1+c)} \label{eq:u_again}
\\
 \phi&=& 1- {\rm Erf}[u]
\label{eq:phi_eqn_again}
\\
c &=&\sigma^2[\infty]\Big\{ 1-{\rm Erf}[u]+\frac{1}{2u^2}{\rm
Erf}[u]-\frac{1}{u\sqrt{\pi}}\rme^{-u^2}\Big\} \label{eq:c_eqn_again}
\\
\chi&=& {\rm Erf}[u]/\kappa_2(\alpha-{\rm Erf}[u])
 \label{eq:chi_eqn_again}
\end{eqnarray}
Since phase transitions in the present type of MGs are defined by
a divergence of $\chi$, we conclude that not only the values
of the static observables $\{\phi,c\}$ but also the phase diagrams
of the two fake history MG model families are identical.

\subsection{The volatility}

\noindent Finally we calculate the volatility.
 Since the average
overall bid is zero, the volatility  is here defined by
$\sigma^2=\lim_{t\to\infty}\lim_{\del\to 0}\bra\bra
A^2(\ell)\ket\ket_{\{A,Z\}}|_{\ell=t/\del}$. Inserting
 (\ref{eq:Aexplicit})  into the definition of $\sigma$, with $A_e\to
 0$, gives
\begin{eqnarray}
\sigma^2 &=& \lim_{t\to\infty} \lim_{\del\to 0}\left\{
\sum_{r,r^\prime\geq 0}(-\del)^{r+r^\prime} \sum_{\ell_0\ldots
\ell_r}G(\ell_0,\ell_1)\ldots G(\ell_{r-1},\ell_r)
\nonumber\right.
\\
&&\left. \times~\sum_{\ell^\prime_0\ldots \ell^\prime_{r^\prime}}
G(\ell^\prime_0,\ell^\prime_1)\ldots
G(\ell^\prime_{r^\prime-1},\ell^\prime_{r^\prime})~\delta_{\ell,\ell_0}\delta_{\ell,\ell^\prime_0}
 p^{r+r^\prime}
 \Big\bra\!\Big\bra~
 \bra\phi_{\ell_r}
 \phi_{\ell^\prime_{r^\prime}}\ket_{\{\phi|A,Z\}}
 \nonumber\right.
\\ &&
\left. \times~ \Big[\prod_{i=1}^r
\overline{W}[\ell_{i-1},\ell_i;\{A,Z\}] \Big]
\Big[\prod_{j=1}^{r^\prime}
\overline{W}[\ell_{j-1}^\prime,\ell^\prime_j;\{A,Z\}]
 \Big]~
 \Big\ket\!
\Big\ket_{\!\{A,Z\}}\right\}\Big|_{\ell=\frac{t}{\del}} \nonumber
\\
&=& \lim_{t\to\infty} \lim_{\del\to 0} \left\{
\frac{1}{2}\sum_{r,r^\prime\geq 0}(-\del)^{r+r^\prime}
\sum_{\ell_0\ldots \ell_r}G(\ell_0,\ell_1)\ldots
G(\ell_{r-1},\ell_r) \nonumber\right.
\\
&&\left.\hspace*{20mm} \times~\sum_{\ell_0^\prime\ldots
\ell^\prime_{r^\prime}} G(\ell^\prime_0,\ell^\prime_1)\ldots
G(\ell^\prime_{r^\prime-1},\ell^\prime_{r^\prime})~\big[1+C(\ell_r,\ell^\prime_{r^\prime})\big]\delta_{\ell,\ell_0}\delta_{\ell,\ell^\prime_0}
 \nonumber \right.
\\ &&
\left.\hspace*{20mm} \times~
\tilde{\tilde{\Delta}}_{r+r^\prime+1}(\ell_0,\ldots,\ell_r;\ell^\prime_0,\ldots,\ell^\prime_{r^{\prime}})
\room\right\}\Big|_{\ell=t/\del}
 \label{eq:workout_vola}
\end{eqnarray}
with
\begin{eqnarray}
 \hspace*{-15mm}
  &&\hspace*{-8mm}
\tilde{\tilde{\Delta}}_{r+r^\prime+1}(\ell_0,\ldots,\ell_r;\ell^\prime_0,\ldots,\ell^\prime_{r^{\prime}})
\label{eq:general_Deltahathat}
\\
\hspace*{-15mm} &=&
 p^{r+r^\prime}
 \Big\bra\!\Big\bra\overline{W}[\ell_r,\ell^\prime_{r^\prime};\{A,Z\}]
 \Big[\prod_{i=1}^{r}
\overline{W}[\ell_{i-1},\ell_i;\{A,Z\}]\Big]\Big[\prod_{j=1}^{r^\prime}
\overline{W}[\ell^\prime_{j-1},\ell^\prime_j;\{A,Z\}]\Big]
 \Big\ket\!\Big\ket_{\!\{A,Z\}}
\nonumber
\end{eqnarray}
 We need this function
(\ref{eq:general_Deltahathat}) only for times with
$\ell_0>\ell_1>\ldots >\ell_r$ and
$\ell^\prime_0>\ell^\prime_1>\ldots >\ell^\prime_{r^\prime}$. For
the case of inner product MG, where
$\overline{W}[\ell,\ell^\prime;\{A,Z\}]=p^{-1}\sum_{\lambda=1}^p
\tilde{Z}_{\ell,\lambda}\tilde{Z}_{\ell^\prime,\lambda}$, its
evaluation is very similar to that of (\ref{eq:general_Deltahat}):
\begin{eqnarray}
\hspace*{-15mm}
 \tilde{\tilde{\Delta}}_{r+r^\prime+1}(\ell_0,\ldots,\ell_{r};\ell^\prime_0,\ldots,\ell^\prime_{r^\prime})
 &=&\frac{1}{p}\sum_{\lambda_1\ldots\lambda_{r+1}=1}^p
\sum_{\lambda^\prime_1\ldots\lambda^\prime_{r^\prime+1}=1}^p
\delta_{\lambda_{r+1},\lambda^\prime_{r^\prime+1}}
 \nonumber
 \\
\hspace*{-15mm}
 &&\hspace*{-20mm}\times
 \Big\bra \tilde{Z}_{\ell_0,\lambda_1}\tilde{Z}_{\ell^\prime_0,\lambda_1^\prime}\Big[\prod_{i=1}^r \tilde{Z}_{\ell_i,\lambda_i}
 \tilde{Z}_{\ell_{i},\lambda_{i+1}}\Big]
 \Big[\prod_{j=1}^{r^\prime} \tilde{Z}_{\ell^\prime_j,\lambda^\prime_j}
 \tilde{Z}_{\ell^\prime_{j},\lambda^\prime_{j+1}}\Big]
 \Big\ket_{\!\{Z\}}~~~~~
\end{eqnarray}
Due to the time ordering, the factor
$\tilde{Z}_{\ell_0,\lambda_1}\tilde{Z}_{\ell^\prime_0,\lambda_1^\prime}$
is statistically independent of all others, so that
\begin{eqnarray}
\hspace*{-15mm}
 \tilde{\tilde{\Delta}}_{r+r^\prime+1}(\ell_0,\ldots,\ell_{r};\ell^\prime_0,\ldots,\ell^\prime_{r^\prime})
 &=&\frac{\kappa_2}{p}\sum_{\lambda_1\ldots\lambda_{r+1}=1}^p
\sum_{\lambda^\prime_1\ldots\lambda^\prime_{r^\prime+1}=1}^p
\delta_{\lambda_{r+1},\lambda^\prime_{r^\prime+1}}
 \nonumber
\hspace*{-15mm} \\
 && \times
 \Big\bra \Big[\prod_{i=1}^r \tilde{Z}_{\ell_i,\lambda_i}
 \tilde{Z}_{\ell_{i},\lambda_{i+1}}\Big]
 \Big[\prod_{j=1}^{r^\prime} \tilde{Z}_{\ell^\prime_j,\lambda^\prime_j}
 \tilde{Z}_{\ell^\prime_{j},\lambda^\prime_{j+1}}\Big]
 \Big\ket_{\!\{Z\}}~~~~~
 \label{eq:intermediate_sigmadelta}
\end{eqnarray}
Again we inspect the effect of time pairings:
\begin{itemize}
\item
If there are no time coincidences, i.e. $\ell_i\neq \ell^\prime_j$
for all $(i,j)$,  we simply find
\begin{eqnarray}&&\hspace*{-15mm}
 \tilde{\tilde{\Delta}}_{r+r^\prime+1}(\ell_0,\ldots,\ell_{r};\ell^\prime_0,\ldots,\ell^\prime_{r^\prime})\nonumber
 \\
&=&
 \frac{1}{p}~\kappa_2^{r+r^\prime+1}\sum_{\lambda_1\ldots\lambda_{r+1}=1}^p
\sum_{\lambda^\prime_1\ldots\lambda^\prime_{r^\prime+1}=1}^p
\delta_{\lambda_{r+1},\lambda^\prime_{r^\prime+1}}
\Big[\prod_{i=1}^r \delta_{\lambda_i,\lambda_{i+1}}\Big]
 \Big[\prod_{j=1}^{r^\prime}
 \delta_{\lambda^\prime_j,\lambda^\prime_{j+1}}\Big]
 \nonumber
  \\
&=&
 \frac{1}{p}~\kappa_2^{r+r^\prime+1}\sum_{\lambda_1=1}^p
\sum_{\lambda^\prime_1=1}^p
\delta_{\lambda^\prime_1,\lambda_1}=\kappa_2^{r+r^\prime+1}
\end{eqnarray}
\item
The effect of a time pairing $\ell_i=\ell_j^\prime$ on the
contribution of the Gaussian averages with times
$(\ell_i,\ell^\prime_j)$ is exactly as before:
\begin{eqnarray*}
\hspace*{-15mm} {\rm no~pairing:~~}\bra
\tilde{Z}_{\ell_i,\lambda_i}
 \tilde{Z}_{\ell_{i},\lambda_{i+1}}
\tilde{Z}_{\ell^\prime_j,\lambda^\prime_j}
 \tilde{Z}_{\ell^\prime_{j},\lambda^\prime_{j+1}}\ket&=&\bra \tilde{Z}_{\ell_i,\lambda_i}
 \tilde{Z}_{\ell_{i},\lambda_{i+1}}\ket\bra
\tilde{Z}_{\ell^\prime_j,\lambda^\prime_j}
 \tilde{Z}_{\ell^\prime_{j},\lambda^\prime_{j+1}}\ket\\
\hspace*{-15mm} &=&
\kappa_2^2\delta_{\lambda_i,\lambda_{i+1}}\delta_{\lambda^\prime_j,\lambda^\prime_{j+1}}
\end{eqnarray*}
\begin{eqnarray*}
\hspace*{-15mm} \ell_i=\ell^\prime_j:~~&&\bra
\tilde{Z}_{\ell_i,\lambda_i}
 \tilde{Z}_{\ell_{i},\lambda_{i+1}}
\tilde{Z}_{\ell^\prime_j,\lambda^\prime_j}
 \tilde{Z}_{\ell^\prime_{j},\lambda^\prime_{j+1}}\ket=
(\kappa_4-3\kappa_2^2)~
\delta_{\lambda_i,\lambda_{i+1}}\delta_{\lambda^\prime_{j},\lambda^\prime_{j+1}}\delta_{\lambda_i,\lambda^\prime_j}\\
\hspace*{-15mm} &&\hspace*{20mm}
 +~\kappa_2^2\left(
\delta_{\lambda_i,\lambda_{i+1}}\delta_{\lambda^\prime_j,\lambda^\prime_{j+1}}
+\delta_{\lambda_i,\lambda^\prime_j}\delta_{\lambda_{i+1},\lambda^\prime_{j+1}}+\delta_{\lambda_i,\lambda^\prime_{j+1}}\delta_{\lambda_{i+1},\lambda^\prime_{j}}
\right)
\end{eqnarray*}
\end{itemize}
 \begin{figure}[t]{\tiny
\hspace*{80mm}\setlength{\unitlength}{0.52mm}
\begin{picture}(130,200)
\put(10,30){\vertex}\put(30,30){\vertex}\put(70,30){\vertex}\put(90,30){\vertex}\put(130,30){\vertex}\put(150,30){\vertex}
\put(10,10){\vertex}\put(30,10){\vertex}\put(70,10){\vertex}\put(90,10){\vertex}\put(130,10){\vertex}\put(150,10){\vertex}
\put(10,35){\here{$1$}}\put(30,35){\here{$2$}}\put(70,35){\here{$i$}}\put(90,35){\here{$i+1$}}\put(130,35){\here{$r$}}\put(150,35){\here{$r+1$}}
\put(10,5){\here{$1$}}\put(30,5){\here{$2$}}\put(70,5){\here{$j$}}\put(90,5){\here{$j+1$}}\put(130,5){\here{$r^\prime$}}\put(150,5){\here{$r^\prime+1$}}

\put(149,10){\line(0,1){20}}
\put(10,10){\line(1,0){60}}\put(150,10){\line(-1,0){60}}\put(10,30){\line(1,0){60}}\put(150,30){\line(-1,0){60}}
\put(69,10){\line(1,1){20}}\put(89,10){\line(-1,1){20}}
\put(69,10){\line(0,1){20}}

\put(10,80){\vertex}\put(30,80){\vertex}\put(70,80){\vertex}\put(90,80){\vertex}\put(130,80){\vertex}\put(150,80){\vertex}
\put(10,60){\vertex}\put(30,60){\vertex}\put(70,60){\vertex}\put(90,60){\vertex}\put(130,60){\vertex}\put(150,60){\vertex}
\put(10,85){\here{$1$}}\put(30,85){\here{$2$}}\put(70,85){\here{$i$}}\put(90,85){\here{$i+1$}}\put(130,85){\here{$r$}}\put(150,85){\here{$r+1$}}
\put(10,55){\here{$1$}}\put(30,55){\here{$2$}}\put(70,55){\here{$j$}}\put(90,55){\here{$j+1$}}\put(130,55){\here{$r^\prime$}}\put(150,55){\here{$r^\prime+1$}}

\put(149,60){\line(0,1){20}}
\put(10,60){\line(1,0){60}}\put(150,60){\line(-1,0){60}}\put(10,80){\line(1,0){60}}\put(150,80){\line(-1,0){60}}
\put(69,60){\line(0,1){20}}\put(89,60){\line(0,1){20}}

\put(10,130){\vertex}\put(30,130){\vertex}\put(70,130){\vertex}\put(90,130){\vertex}\put(130,130){\vertex}\put(150,130){\vertex}
\put(10,110){\vertex}\put(30,110){\vertex}\put(70,110){\vertex}\put(90,110){\vertex}\put(130,110){\vertex}\put(150,110){\vertex}
\put(10,135){\here{$1$}}\put(30,135){\here{$2$}}\put(70,135){\here{$i$}}\put(90,135){\here{$i+1$}}\put(130,135){\here{$r$}}\put(150,135){\here{$r+1$}}
\put(10,105){\here{$1$}}\put(30,105){\here{$2$}}\put(70,105){\here{$j$}}\put(90,105){\here{$j+1$}}\put(130,105){\here{$r^\prime$}}\put(150,105){\here{$r^\prime+1$}}

\put(149,110){\line(0,1){20}}
\put(10,110){\line(1,0){60}}\put(150,110){\line(-1,0){60}}\put(10,130){\line(1,0){60}}\put(150,130){\line(-1,0){60}}
\put(69,110){\line(1,1){20}}\put(89,110){\line(-1,1){20}}

\put(10,180){\vertex}\put(30,180){\vertex}\put(70,180){\vertex}\put(90,180){\vertex}\put(130,180){\vertex}\put(150,180){\vertex}
\put(10,160){\vertex}\put(30,160){\vertex}\put(70,160){\vertex}\put(90,160){\vertex}\put(130,160){\vertex}\put(150,160){\vertex}
\put(10,185){\here{$1$}}\put(30,185){\here{$2$}}\put(70,185){\here{$i$}}\put(90,185){\here{$i+1$}}\put(130,185){\here{$r$}}\put(150,185){\here{$r+1$}}
\put(10,155){\here{$1$}}\put(30,155){\here{$2$}}\put(70,155){\here{$j$}}\put(90,155){\here{$j+1$}}\put(130,155){\here{$r^\prime$}}\put(150,155){\here{$r^\prime+1$}}

\put(149,160){\line(0,1){20}}\put(10,180){\line(1,0){140}}\put(10,160){\line(1,0){140}}

\put(-150,170){\small\sl no pairings:
$~~\delta_{\lambda_i,\lambda_{i+1}}\delta_{\lambda^\prime_j,\lambda^\prime_{j+1}}$}

\put(-150,120){\small\sl $\ell_i=\ell^\prime_j$:
$~~\delta_{\lambda_i,\lambda_{i+1}}\delta_{\lambda^\prime_j,\lambda^\prime_{j+1}}\!\to
\delta_{\lambda_i,\lambda^\prime_{j+1}}\delta_{\lambda_{i+1},\lambda^\prime_j}$}

\put(-150,70){\small\sl $\ell_i=\ell^\prime_j$:
$~~\delta_{\lambda_i,\lambda_{i+1}}\delta_{\lambda^\prime_j,\lambda^\prime_{j+1}}\!\to
\delta_{\lambda_i,\lambda^\prime_{j}}\delta_{\lambda_{i+1},\lambda^\prime_{j+1}}$}

\put(-150,20){\small\sl $\ell_i=\ell^\prime_j$:
$~~\delta_{\lambda_i,\lambda_{i+1}}\delta_{\lambda^\prime_j,\lambda^\prime_{j+1}}\!\to
\delta_{\lambda_i,\lambda^\prime_j}\delta_{\lambda_i,\lambda_{i+1}}\delta_{\lambda^\prime_j,\lambda^\prime_{j+1}}$}
\end{picture}}
\vspace*{3mm} \caption{Diagrammatical representation of the
different contributions of the Gaussian averages to the function
$\tilde{\tilde{\Delta}}_{r+r^\prime+1}$.  Each label
$i\in\{1,\ldots,r+1\}$ and each label
$j\in\{1,\ldots,r^\prime+1\}$ is drawn as a distinct vertex of a
graph. Each factor $\delta_{\lambda_i,\lambda^\prime_j}$ is drawn
as a line segment connecting the vertices $i$ and $j$. Top graph:
the case where there are no time coincidences. Bottom three
graphs: the three different new contributions that are generated
by the occurrence of a time pairing where $\ell_i=\ell^\prime_j$.
} \label{fig:diagrams2}
\end{figure}
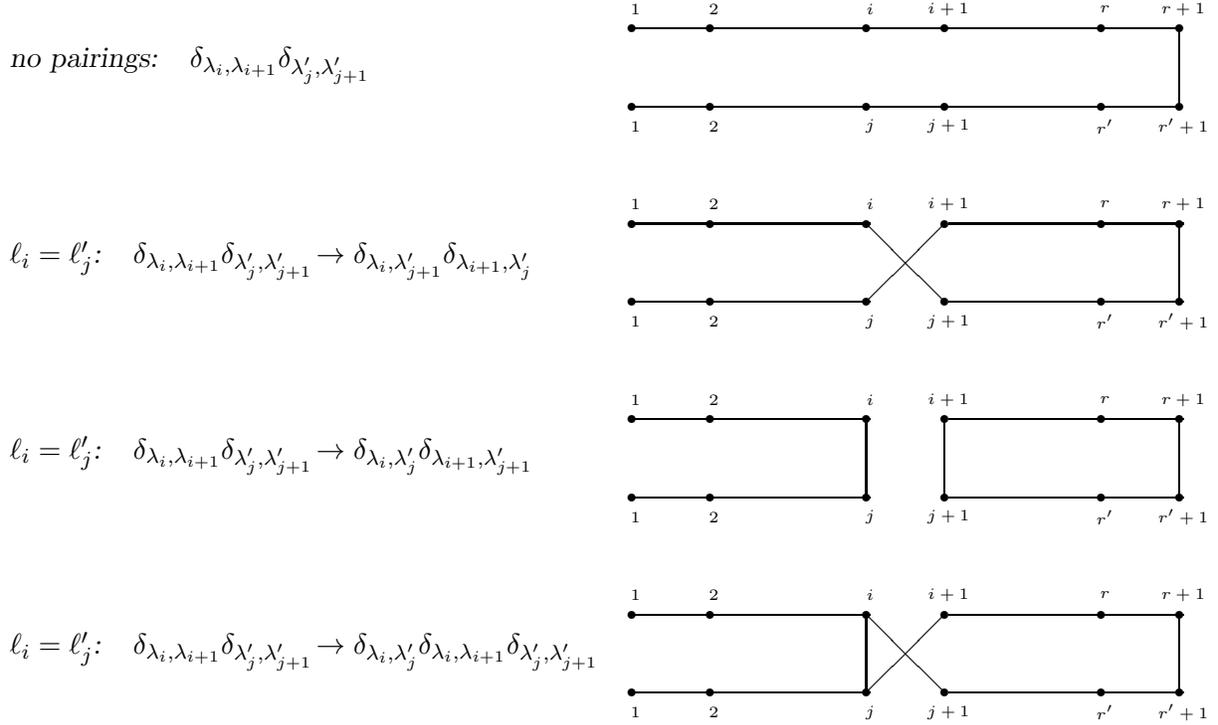
The corresponding diagrammatic representation is shown in figure
\ref{fig:diagrams2}.  Since each time pairing will have to
be compensated by an $\order(p)$ factor to retain relevance in the
limit $\del\to 0$, the only relevant diagrams in figure
\ref{fig:diagrams2} are again the top one (no pairings) and the
third from the top (a time pairing with vertical connections that
cause a diagram cut), including the higher order diagrams with
multiple vertical connections (in the case of multiple time
pairings). We conclude that
\begin{eqnarray}
\hspace*{-15mm}
 \tilde{\tilde{\Delta}}_{r+r^\prime+1}(\ell_0,\ldots,\ell_{r};\ell^\prime_0,\ldots,\ell^\prime_{r^\prime})
&=&
 \kappa_2^{r+r^\prime+1}\prod_{i=1}^{r}\prod_{j=1}^{r^\prime}\left[1+\delta_{\ell_i,\ell^\prime_j}(p+\order(p^0)\right]\nonumber
 \\ \hspace*{-15mm}
 &=& \kappa_2^{r+r^\prime+1}\prod_{i=1}^{r}\prod_{j=1}^{r^\prime}\left[1+\frac{\rate}{2\del}\delta_{\ell_i,\ell^\prime_j}(1+\order(\del)\right]
\end{eqnarray}
and hence
\begin{eqnarray}
\hspace*{-15mm} \sigma^2 &=& \frac{1}{2}\kappa_2\lim_{t\to\infty}
  \sum_{r,r^\prime\geq
0}(-\kappa_2)^{r+r^\prime} \int_0^\infty\!dt_1\ldots dt_r
dt^\prime_1 \ldots
dt^\prime_{r^\prime}\prod_{i=1}^{r}\prod_{j=1}^{r^\prime}\left[1+\frac{1}{2}\rate\delta(t_i-t^\prime_j)\right]\nonumber
\\
\hspace*{-15mm} &&\times \big[1+C(t_r,t^\prime_{r^\prime})\big]
G(t,t_1)G(t_1,t_2)\ldots
G(t_{r-1},t_r)G(t,t^\prime_1)G(t^\prime_1,t^\prime_2)\ldots
G(t^\prime_{r^\prime-1},t^\prime_{r^\prime})\nonumber
\\
\hspace*{-15mm} &=& \lim_{\tau\to\infty}\frac{\kappa_2}{2\tau}
\int_0^\tau\!dt~ \sum_{r,r^\prime\geq 0}(-\kappa_2)^{r+r^\prime}
\int_0^\infty\!dt_1\ldots dt_r dt^\prime_1 \ldots
dt^\prime_{r^\prime}\prod_{i=1}^{r}\prod_{j=1}^{r^\prime}\left[1+\frac{1}{2}\rate\delta(t_i-t^\prime_j)\right]\nonumber
\\
\hspace*{-15mm} &&\times  G(t,t_1)G(t_1,t_2)\ldots
G(t_{r-1},t_r)G(t,t^\prime_1)G(t^\prime_1,t^\prime_2)\ldots
G(t^\prime_{r^\prime-1},t^\prime_{r^\prime}) \nonumber
\\
\hspace*{-15mm} &&\times \big[1+C(t_r,t^\prime_{r^\prime})\big]
 ~~~~\label{eq:finishedvola}
\end{eqnarray}
Comparison with the corresponding expression in
\cite{CoolenHeimel2001} for look-up table MGs shows that the only
difference is in the various occurrences of $\kappa_2$. If we
implement the transformation that maps the effective single agent
exactly onto that of the lookup-table models, viz.
$G(t,t^\prime)=\kappa_2^{-1}\hat{G}(t,t^\prime)$, we find that the
inner-product volatility differs from that of the lookup table
models by a factor $\kappa_2$:
\begin{eqnarray}
\hspace*{-15mm} \sigma^2 &=&
\lim_{\tau\to\infty}\frac{\kappa_2}{2\tau} \int_0^\tau\!dt~
\sum_{r,r^\prime\geq 0}(-1)^{r+r^\prime}
\int_0^\infty\!dt_1\ldots dt_r dt^\prime_1 \ldots
dt^\prime_{r^\prime}\prod_{i=1}^{r}\prod_{j=1}^{r^\prime}\left[1+\frac{1}{2}\rate\delta(t_i-t^\prime_j)\right]\nonumber
\\
\hspace*{-15mm} &&\times \hat{G}(t,t_1)\hat{G}(t_1,t_2)\ldots
\hat{G}(t_{r-1},t_r)\hat{G}(t,t^\prime_1)\hat{G}(t^\prime_1,t^\prime_2)\ldots
\hat{G}(t^\prime_{r^\prime-1},t^\prime_{r^\prime}) \nonumber
\\
\hspace*{-15mm} &&\times \big[1+C(t_r,t^\prime_{r^\prime})\big]
 \label{eq:finishedvola2}
\end{eqnarray}

\section{Inner product MGs with real market history}

\noindent The previous cases could be solved exactly and in full,
due to the absence of real history.  We now turn to the more
demanding situation where $\zeta<1$ in definition (\ref{eq:W_IP}),
so that the information vector truly depends on the past market
history. Here it will turn out advantageous to define the
stochastic and time dependent symmetric $p\times p$ matrices
$\bB(\ell)$ with entries
\begin{equation}
B_{\lambda\lambda^\prime}(\ell)=\F_\lambda[\ell,A,Z]\F_{\lambda^\prime}[\ell,A,Z]
\label{eq:matricesB}
\end{equation}
with $\F_\lambda[\ldots]$ as defined in (\ref{eq:Flambda_IP}).
This allows us to write the relevant functions
$\Delta_{r+1}(\ldots)$,  $\tilde{\Delta}_{r+r^\prime+2}(\ldots)$
and $\tilde{\tilde{\Delta}}_{r+r^\prime+1}(\ldots)$ (that occur in
the kernels $R$ and $\Sigma$ and in the volatility) as averages
over a trace:
\begin{eqnarray}
\hspace*{-20mm}&&
 \Delta_{r+1}(\ell_0,\ldots,\ell_{r})=
 \frac{1}{p}{\rm Tr}
\Big\bra\!\Big\bra \bB(\ell_0) \bB(\ell_1) \ldots \bB(\ell_r)
 \Big\ket\!\Big\ket_{\!\{A,Z\}}
 \label{eq:Del1}
\\
\hspace*{-25mm}&&
 \tilde{\Delta}_{r+r^\prime+2}(\ell_0,\ldots,\ell_{r};\ell^\prime_0,\ldots,\ell^\prime_{r^\prime}\!)=
 \frac{1}{p}{\rm Tr}
 \Big\bra\!\Big\bra
 \bB(\ell_0) \bB(\ell_1) \ldots \bB(\ell_r)
\bB(\ell^\prime_{r^\prime})\ldots \bB(\ell^\prime_1)
\bB(\ell^\prime_{0})
 \Big\ket\!\Big\ket_{\!\{A,Z\}}
 \nonumber
 \\
 \hspace*{-20mm}&&
 \label{eq:Del2}
\\
\hspace*{-20mm}&&
\tilde{\tilde{\Delta}}_{r+r^\prime+1}(\ell_0,\ell_1\ldots,\ell_r;\ell_0,\ell^\prime_1,\ldots,\ell^\prime_{r^{\prime}}\!)
=
 \frac{1}{p}{\rm Tr} \Big\bra\!\Big\bra
\bB(\ell_{0}) \bB(\ell_{1}) \ldots
\bB(\ell_r)\bB(\ell^\prime_{r^\prime}) \ldots \bB(\ell^\prime_1)
 \Big\ket\!\Big\ket_{\!\{A,Z\}}
 \nonumber
 \\
 \hspace*{-20mm}&&
 \label{eq:Del3}
\end{eqnarray}
These expressions are of course still exact, but not easy to
evaluate.

\subsection{Short history correlation times in TTI stationary states}

\noindent
 We now make an approximation for the TTI stationary state solution in the spirit of the short
 history correlation times ansatz made in \cite{Coolen2005} for lookup table
 MGs; in fact we will find below that for lookup table MGs the two are identical. In working out our expressions for $\chi_R=\int\!dt~R(t)$, $\Sigma(\infty)=\lim_{\tau\to\infty}\Sigma(t+\tau,t)$ and
 the volatility $\sigma^2$ we will replace the kernels
 (\ref{eq:Del1},\ref{eq:Del2},\ref{eq:Del3}) by the values
 they will take for times which are sufficiently widely separated to
 de-correlate the random matrices $\bB(\ell)$, i.e. we replace
 \begin{equation}
 \prod_{i=1}^r\bB(\ell_i)~\to~\bB^r(A,Z)~~~~~~~~\bB(A,Z)=\lim_{L\to\infty}\frac{1}{L}\sum_{\ell\leq
 L}\bB(\ell)
 \label{eq:short_corr_times}
  \end{equation}
 For TTI states  this allows us to
 express everything in terms of the average eigenvalue distribution
 $\varrho(\mu)$ of the now time independent random $p\times p$ matrix $\bB(A,Z)$:
\begin{eqnarray}
 \Delta_{r+1}(\ell_0,\ldots,\ell_{r})&=&
 \frac{1}{p}~{\rm Tr}~\bra\bra \bB^{r+1}(A,Z)\ket\ket_{\{A,Z\}}
 \nonumber
 \\ &=& \int_0^\infty\!d\mu~\varrho(\mu)~ \mu^{r+1}
 \label{eq:Del1approx}
\\
 \tilde{\Delta}_{r+r^\prime+2}(\ell_0,\ldots,\ell_{r};\ell^\prime_0,\ldots,\ell^\prime_{r^\prime})&=&
 \frac{1}{p}~{\rm Tr}~\bra\bra
 \bB^{r+r^\prime+2}(A,Z)\ket\ket_{\{A,Z\}}\nonumber \\ &=& \int_0^\infty\!d\mu~\varrho(\mu)~\mu^{r+r^\prime+2}
 \label{eq:Del2approx}
\\
\tilde{\tilde{\Delta}}_{r+r^\prime+1}(\ell_0,\ell_1\ldots,\ell_r;\ell_0,\ell^\prime_1,\ldots,\ell^\prime_{r^{\prime}})
&=&
 \frac{1}{p}~{\rm
 Tr}~\bra\bra\bB^{r+r^\prime+1}(A,Z)\ket\ket_{\{A,Z\}}\nonumber \\
 &=& \int_0^\infty\!d\mu~\varrho(\mu)~\mu^{r+r^\prime+1}~~~~~~
 \label{eq:Del3approx}
\end{eqnarray}
Since $\bB(A,Z)$ is non-negative definite one always has $\mu\geq
0$.
Insertion of
(\ref{eq:Del1approx},\ref{eq:Del2approx},\ref{eq:Del3approx}) into
expressions (\ref{eq:workoutR},\ref{eq:workoutSigma}) for the
kernels $R$ and $\Sigma$ followed by appropriate integration and
re-summation of the series leads us to the following
 expressions for $\chi_R$ and
$S_0^2=\Sigma(\infty)$:
\begin{equation}
\chi_R=\int_0^\infty\!d\mu~\varrho(\mu)\frac{\mu}{1+\mu\chi}~~~~~~~~
S_0^2 = (1+c)\int_0^\infty\!d\mu~\varrho(\mu) \frac{\mu^2}{(1+\mu
\chi)^2} \label{eq:ChiRandS0}
\end{equation}
The effects of having real memory are concentrated
fully in the eigenvalue distribution $\varrho(\mu)$. Once this distribution has
been calculated in terms of $\{c,\phi,\chi\}$, the persistent
order parameters of the MG will be given by a closed
set of equations. Upon introducing a convenient parameter $\omega\in[0,1]$ via
\begin{eqnarray}
\omega&=& \frac{\int\!d\mu~\varrho(\mu)\mu\chi/(1+\mu\chi)}
{\sqrt{\int\!d\mu~\varrho(\mu)[\mu\chi/(1+\mu\chi)]^2}}
\label{eq:persistent0}
\end{eqnarray}
our closed order parameter equations take the form
\begin{eqnarray}
u&=&\frac{\omega\sigma[\infty]\sqrt{\alpha}}{\sqrt{2(1+c)}},~~~~~~\frac{1-\phi}{\alpha}=\int_0^\infty\!\frac{d\mu~\varrho(\mu)}{1+(\mu\chi)^{-1}},~~~~~~
\phi= 1- {\rm Erf}[u]
\label{eq:persistent1}
\\
c &=&\sigma^2[\infty]\Big\{ 1-{\rm Erf}[u]+\frac{1}{2u^2}{\rm
Erf}[u]-\frac{1}{u\sqrt{\pi}}\rme^{-u^2}\Big\}
\label{eq:persistent2}
\end{eqnarray}
From this we can already extract several results, and explain observations made in the past for MG types other than look-up table ones on the basis of numerical simulations:
\begin{itemize}
\item
Irrespective of the extent to which the histories are real, and irrespective of the nature of the function $f[A]$, the ergodicity-breaking phase transition point marked by a divergence of $\chi$ occurs {\em always} at the value $\alpha_c(T)$ that was found originally
for fake history look-up table MG models \cite{Replicas1,Replicas2,HeimelCoolen2001,CoolenHeimelSherr2001}, as solved from the closed set
\begin{eqnarray}
u&=&\sigma[\infty]\sqrt{\alpha/2(1+c)},~~~~~~{\rm Erf}[u]=\alpha,~~~~~~
\\
c &=&\sigma^2[\infty]\Big\{ 1-{\rm Erf}[u]+\frac{1}{2u^2}{\rm
Erf}[u]-\frac{1}{u\sqrt{\pi}}\rme^{-u^2}\Big\}
\end{eqnarray}
\item
Equations (\ref{eq:persistent0},\ref{eq:persistent1},\ref{eq:persistent2}) return to those derived earlier for inconsistent random histories when $\omega=1$, whereas real histories affect the persistent observables as soon as $\omega<1$. For finite $\chi$ we thus arrive at the conclusion, via (\ref{eq:persistent0}), that our observables are
affected by having real histories if and only if $\varrho(\mu)$ is {\em not} a $\delta$-distribution.
A nonzero width of $\varrho(\mu)$ is the fingerprint of real histories influencing the behaviour of the MG.
\end{itemize}
We note that the combined equations
(\ref{eq:ChiRandS0},\ref{eq:persistent0},\ref{eq:persistent1},\ref{eq:persistent2}) are
identical to those found for the lookup table MGs
with true market histories \cite{Coolen2005}, with the eigenvalue
distribution $\varrho(\mu)$ taking over the role of the relative
history frequency distribution in \cite{Coolen2005}. This can be
understood. Our choice of notation allows us to apply the present
derivation also to the lookup table models, simply by making in
(\ref{eq:matricesB}) the substitutions
$\lambda\in\{1,\ldots,p\}\to\blambda\in\{-1,1\}^M$ and
$\F_\lambda[\ell,A,Z]\to
\sqrt{p}~\delta_{\blambda,\blambda(\ell,A,Z)}$, where
$\blambda(\ell,A,Z)$ is the history string (potentially partly
fake) as observed by the agents at step $\ell$. In lookup table
MGs one would therefore have $B_{\lambda\lambda^\prime}(\ell)=p
\delta_{\blambda,\blambda(\ell,A,Z)}\delta_{\blambda^\prime,\blambda(\ell,A,Z)}$
and hence
 \begin{equation}
  B_{\lambda\lambda^\prime}(A,Z)=
p\delta_{\blambda\blambda^\prime}\lim_{L\to\infty}\frac{1}{L}\sum_{\ell\leq
L}  \delta_{\blambda,\blambda(\ell,A,Z)}
=p\delta_{\blambda\blambda^\prime}\pi_{\blambda}(A,Z)
\end{equation}
Here $\pi_\blambda(A,Z)$ denotes the stationary state frequency of
occurrence of history string $\blambda$. Since this matrix
$\bB(A,Z)$ is of a diagonal form, its eigenvalues are simply the
diagonal entries $p\pi_{\blambda}(A,Z)$, and the asymptotic
eigenvalue distribution becomes
\begin{equation}
\varrho(\mu)=\lim_{p\to\infty} \frac{1}{p}\sum_{\blambda}\bra\bra
\delta\left[\mu-p \pi_{\blambda}(A,Z)\right]\ket\ket_{\{A,Z\}}
\end{equation}
Thus, for lookup table MGs the eigenvalue distribution
$\varrho(\mu)$ is indeed identical to the relative history
frequency distribution of \cite{Coolen2005}.

The question of whether and to what extent having real as opposed to fake market histories affects
the MG thus boils down to assessing the shape of the eigenvalue distribution $\varrho(\mu)$. For look-up table MGs this distribution
is known to become nontrivial above the critical point \cite{ChalMars00,Coolen2005}. Numerical simulations show that the same is true for
inner product MGs, so also here models with real history will behave differently from those with fake histories.

\subsection{Calculation of eigenvalue spectrum $\varrho(\mu)$ for weakly correlated bids}

\begin{figure}[t]
\vspace*{-0mm} \hspace*{23mm} \setlength{\unitlength}{0.45mm}
\begin{picture}(300,200)

   \put(0,100){\includegraphics[height=95\unitlength,width=130\unitlength]{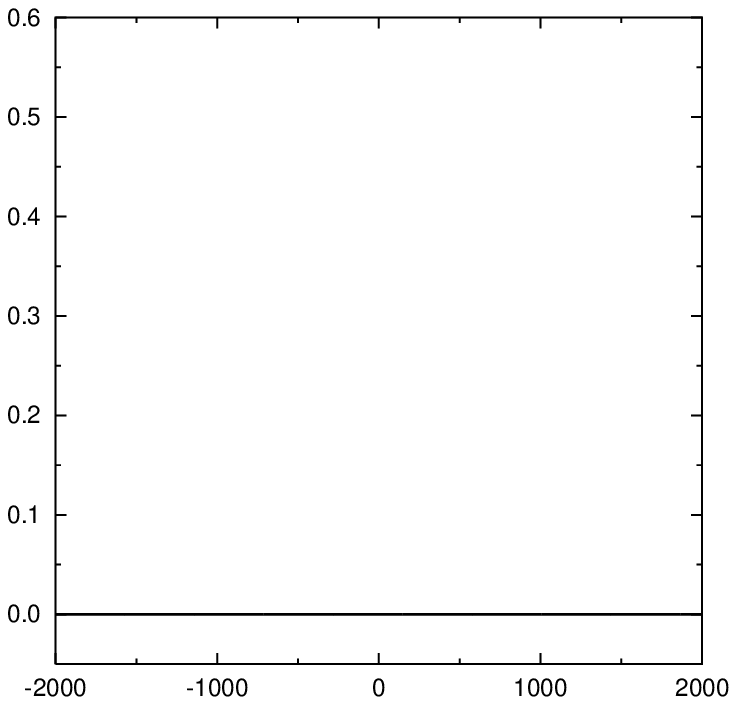}}
  \put(79,181){\small\here{$\zeta\!=\!0$}}

 \put(100,100){\includegraphics[height=95\unitlength,width=130\unitlength]{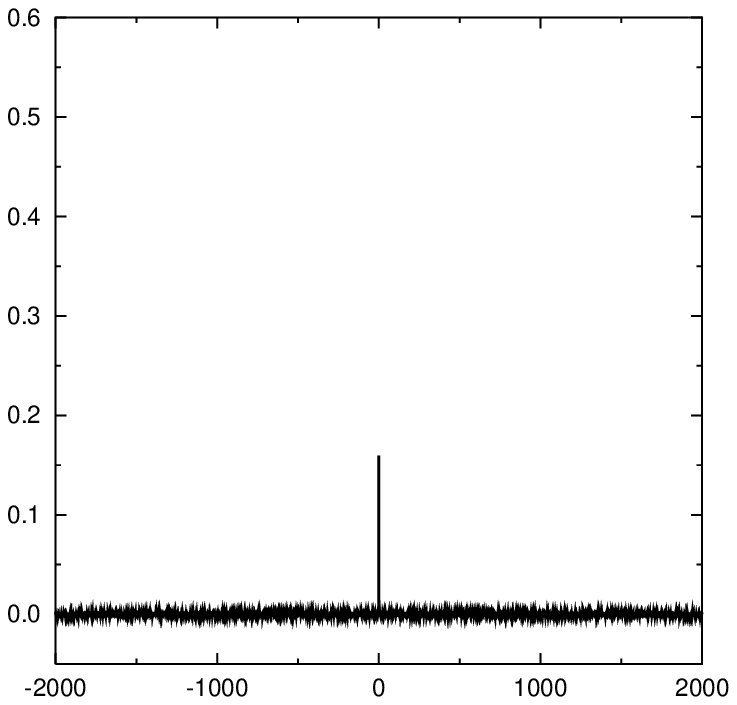}}
  \put(179,181){\small\here{$\zeta\!=\!\frac{1}{2}$}}

 \put(200,100){\includegraphics[height=95\unitlength,width=130\unitlength]{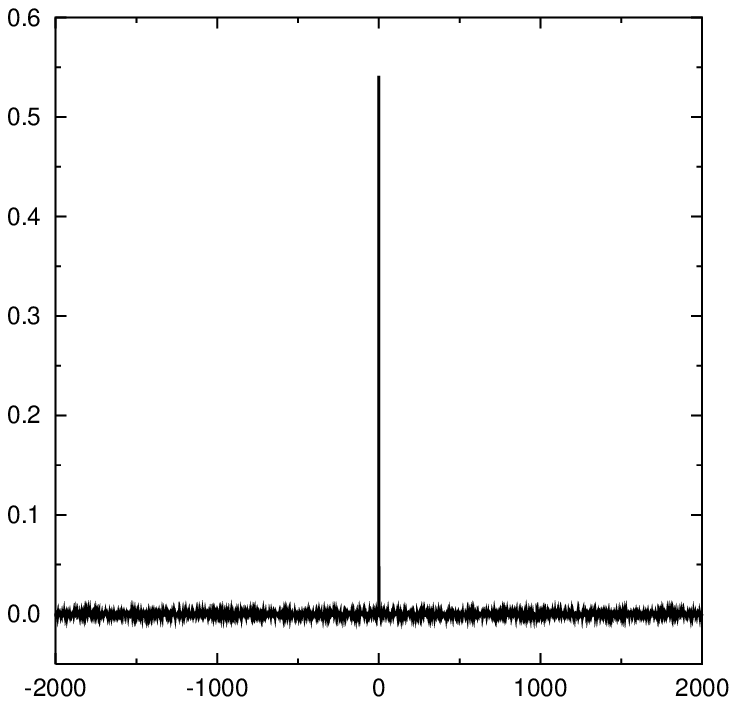}}
   \put(279,181){\small\here{$\zeta\!=\!1$}}

   \put(0,0){\includegraphics[height=95\unitlength,width=130\unitlength]{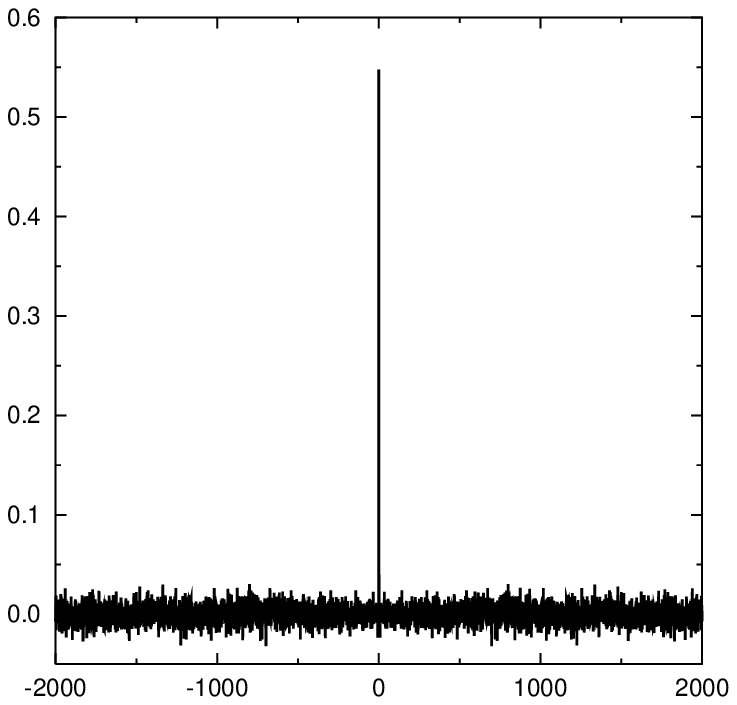}}
     \put(79,81){\small\here{$\zeta\!=\!0$}}

 \put(100,0){\includegraphics[height=95\unitlength,width=130\unitlength]{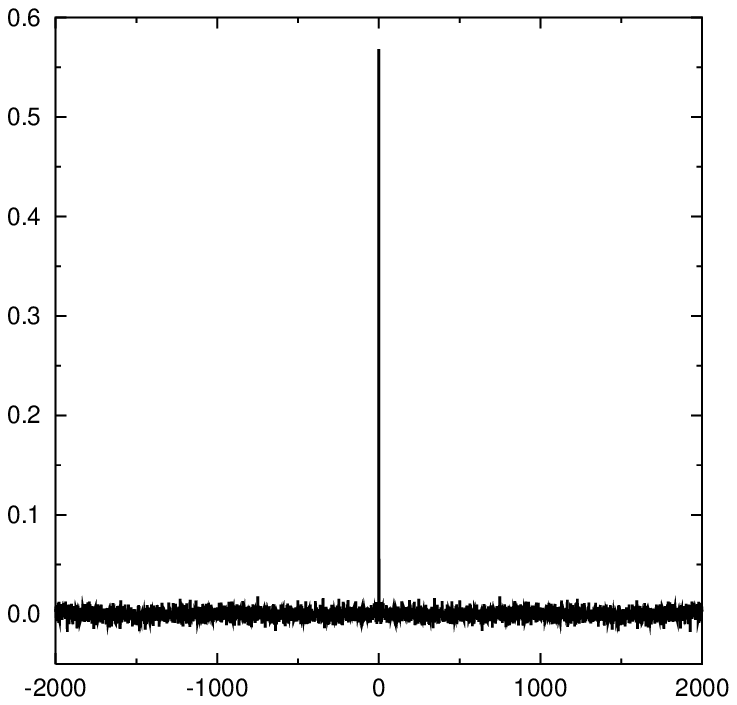}}
 \put(179,81){\small\here{$\zeta\!=\!\frac{1}{2}$}}

 \put(200,0){\includegraphics[height=95\unitlength,width=130\unitlength]{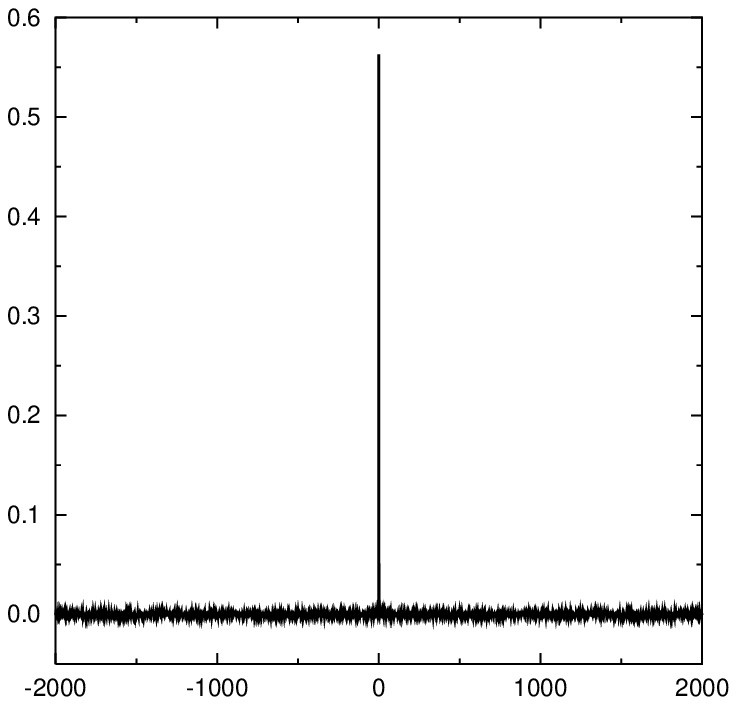}}
   \put(279,81){\small\here{$\zeta\!=\!1$}}

  \put(49,-14){\small $\ell$}    \put(149,-14){\small $\ell$}     \put(249,-14){\small $\ell$}
  \put(-65,50){\small $f[A]=\sgn(A):$}    \put(-65,150){\small $f[A]=A:$}

\end{picture}
 \vspace*{2mm}
\caption{Time-translation invariant bid covariances $\Xi(\ell)=\bra A(t+\ell)A(t)\ket$  of inner product MGs (\ref{eq:inp_B}) measured in numerical simulations, with $N=1025$ and $\alpha=2$. Covariances are calculated over $200.N$ time steps, following equilibration. Top row: $f[A]=A$, with $\zeta\in\{0,\frac{1}{2},1\}$  Bottom row: $f[A]=\sgn(A)$, with $\zeta\in\{0,\frac{1}{2},1\}$. The left pictures refer to strictly real histories,
the right pictures to strictly fake histories, with the middle ones representing a mixing of the two. These data are representative
of the typical behaviour throughout the ergodic regime. The bid correlations are {\em always} seen to be weak, which justifies our approximate calculation of the spectrum $\varrho(\mu)$. We also observe qualitative differences between $f[A]=A$ and $f[A]=\sgn(A)$, e.g. vanishing volatility as $\zeta\to 0$ (strictly real histories) for $f[A]=A$, which will be addressed later.
}\label{fig:covariances}
\end{figure}

In the remainder of this paper we will deal exclusively with the
inner product version of the MG with real history and inconsistent bid noise,   i.e. from now on we will have
\begin{eqnarray}
\hspace*{-15mm}&&
B_{\lambda\lambda^\prime}(A,Z)=\frac{1}{L}\sum_{\ell\leq
 L}f[(1\!-\!\zeta) A(\ell\!-\!\lambda) \!+\! \zeta
 Z(\ell,\lambda)]
 f[(1\!-\!\zeta) A(\ell\!-\!\lambda^\prime) \!+\! \zeta Z(\ell,\lambda^\prime)]~~~~
\label{eq:inp_B}
 \end{eqnarray}
 where $\bra Z(\ell,\lambda)Z(\ell^\prime,\lambda^\prime)\ket=S^2 \delta_{\ell\ell^\prime}\delta_{\lambda\lambda^\prime}$, and where
 we focus on the regime where  $1\ll p\ll L$.
We try to understand the origin of the nontrivial eigenvalue distribution $\varrho(\mu)$ in inner product MGs and its impact on the
 order parameter equations.
  Numerical simulations show that, even when the histories are real, the bid correlations are still very weak, see e.g. figure \ref{fig:covariances}; this  we will exploit in our analysis.
 The
 remaining complications arise from the fact that, although
 the matrix $\bB(A,Z)$ is an average over a diverging number of at most
 weakly correlated variables, the number of entries of $\bB(A,Z)$ also diverges.
To prepare the stage we first write the time-translation invariant bid covariances
 as $\Xi(\lambda)=\bra A(\ell)A(\ell+\lambda)\ket_{\{A,Z\}}$, and introduce the short-hands
\begin{eqnarray}
Q_0&=& \int\!Dz~f^2[z\sqrt{(1-\zeta)^2\sigma^2+\zeta^2 S^2}]
\label{eq:Q0}
\\
Q_1&=& \int\!Dy~\Big[\int\!Dz~f[(1\!-\!\zeta)
\sigma y \!+\! \zeta Sz]\Big]^2
\label{eq:Q1}
 \end{eqnarray}
 with $\sigma^2=\lim_{\ell\to\infty}\bra\bra A^2(\ell)\ket\ket_{\{A,Z\}}=\Xi(0)$, and $0\leq Q_1\leq Q_0$.
 As for the look-up table MGs, one expect also for inner product MGs the distribution $\varrho(\mu)$ to be well-defined as $p\to\infty$, since
 $\int\!d\mu~\varrho(\mu)\mu=p^{-1}\sum_{\lambda\leq p}\bra\bra B_{\lambda\lambda}(A,Z)\ket\ket_{A,Z}=Q_0$.

 In the absence of bid correlations, as for fully fake histories, $\bB(A,Z)$ will be self-averaging for $L\to\infty$
and fixed $p$; this implies (due to the time-translation invariance of the bid correlations) that $\bB$ will be of the Toeplitz form.
One expects time translation invariance to hold also if the correlations between bids at different times are sufficiently small, and also $\bB(A,Z)$
to remain self-averaging at least with respect to the inconsistent noise variables, although possibly not with respect to the bids.
 Here we explore the consequences for the spectrum $\varrho(\mu)$
of assuming the bid correlations $\Xi(\lambda)$ for $\lambda\neq 0$ to be small,
and of taking $\bB(A,Z)$ to be Toeplitz and self-averaging with respect to the $\{Z(\ell,\lambda)\}$, i.e. $B_{\lambda\lambda^\prime}(A,Z)=B(\lambda-\lambda^\prime|A)$.
If $\Xi(\lambda)$ decays to zero or to small random values sufficiently fast, we may continue $\bB(A)$
$L$-periodically, so that it becomes a circular Toeplitz matrix. Its $L$ eigenvalues $\mu_r$ $(r=0,\ldots,L-1)$ will now be given by
\begin{eqnarray}
\mu_r&=& \sum_{\ell^\prime=0}^{L-1}B(\lambda|A)\rme^{-2\pi \rmi\ell^\prime r/L} =
Q_0 +S_r(A)
\end{eqnarray}
with $Q_0$ as given in (\ref{eq:Q0}) and, upon replacing $\sum_{\ell\leq L}$ by $\sum_{\ell<L}$ in (\ref{eq:inp_B}) (which is allowed
since $L\to\infty$):
\begin{eqnarray}
\hspace*{-15mm}
S_r(A)&=& \frac{1}{L}\sum_{\ell\ell^\prime=1}^{L-1}\rme^{-2\pi \rmi\ell^\prime r/L}
 \Big[\int\!Dz~f[(1\!-\!\zeta)
A(\ell\!-\!\ell^\prime) \!+\! \zeta Sz]\Big]\Big[ \int\!Dz~f[(1\!-\!\zeta)
A(\ell) \!+\! \zeta S z]\Big]
\nonumber
\\
\hspace*{-15mm}
&=&
 \frac{1}{\sqrt{L}}\sum_{\ell=1}^{L-1}\int\!Dz~f[(1\!-\!\zeta)
A(\ell) \!+\! \zeta Sz]\rme^{-2\pi \rmi\ell r/L}
\nonumber
\\
\hspace*{-15mm}
&&\hspace*{10mm}\times
\frac{1}{\sqrt{L}} \sum_{\ell^\prime=\ell-L+1}^{\ell-1}\int\!Dz~f[(1\!-\!\zeta)
A(\ell^\prime) \!+\! \zeta S z]\rme^{2\pi \rmi \ell^\prime r/L}
\label{eq:Sr_intermediate}
 \end{eqnarray}
The first and second line are both sums over a large number of nearly independent random variables. If the correlations between the bids at different times
are sufficiently small, we may take the second sum to be independent of $\ell$, which allows us to put $\ell=L$ in this term, provided we correct for the inappropriate generation in doing so of terms where the bids involve identical times (i.e. where $\ell=\ell^\prime$, as such terms were absent initially):
\begin{eqnarray}
S_r(A)&=&  |z_r(A)|^2-Q_1
\label{eq:Sr}
\\
z_r(A)&=& \frac{1}{\sqrt{L}}\sum_{\ell=1}^{L-1}\int\!Dz~f[(1\!-\!\zeta)
A(\ell) \!+\! \zeta Sz]\rme^{2\pi \rmi\ell r/L}
\label{eq:zr}
 \end{eqnarray}
 Clearly $z_{L-r}=\overline{z}_r$.
Given our assumption of weakly correlated bids and given that $f[A]$ is by definition anti-symmetric,
the $L$ complex random variables $z_r$ (where $r=0\ldots L-1$) will for $L\to\infty$ acquire zero-average
Gaussian statistics (central limit theorem), with
\begin{eqnarray}
\hspace*{-10mm}
\bra z_r(A) z_{r^\prime}(A)\ket&=& \frac{Q_1}{L\!-\!1}\sum_{\ell=1}^{L-1} \rme^{2\pi \rmi\ell (r+r^\prime)/L}
=\frac{Q_1L}{L\!-\!1}\delta_{r+r^\prime,{\rm mod}L}-\frac{Q_1}{L\!-\!1}
\\
\hspace*{-10mm}
\bra \overline{z}_r(A) \overline{z}_{r^\prime}(A)\ket&=& \frac{Q_1}{L\!-\!1}\sum_{\ell=1}^{L-1} \rme^{-2\pi \rmi\ell (r+r^\prime)/L}
=\frac{Q_1L}{L\!-\!1}\delta_{r+r^\prime,{\rm mod}L}-\frac{Q_1}{L\!-\!1}
\\
\hspace*{-10mm}
\bra z_r(A) \overline{z}_{r^\prime}(A)\ket&=& \frac{Q_1}{L\!-\!1}\sum_{\ell=1}^{L-1} \rme^{2\pi \rmi\ell (r-r^\prime)/L}
=\Big(Q_1+\frac{Q_1}{L\!-\!1}\Big)\delta_{r-r^\prime,{\rm mod}L}-\frac{Q_1}{L\!-\!1}
\end{eqnarray}
 If the spectrum is well-defined as $L\to\infty$, we may take $L$ to be odd and write the spectrum as an average over
$\{z_1,\ldots,z_{L/2-1/2}\}$, using $z_{L-r}=\overline{z}_r$ to deal with the remaining  $\{z_{L/2-1/2},\ldots,z_{L-1}\}$.
We may disregard $r=0$, since it gives only a vanishing contribution to the spectrum. The remaining $\frac{L-1}{2}$ complex Gaussian variables
are written in terms of their real and imaginary parts as $z_r=x_r+\rmi y_r$, for which one finds
\begin{eqnarray}
\hspace*{-10mm}
\bra x_r x_{r^\prime}\ket&=& \frac{Q_1 L}{2(L\!-\!1)}\delta_{rr^\prime}-\frac{Q_1}{L\!-\!1},~~~~~
\bra y_r y_{r^\prime}\ket= \frac{Q_1 L}{2(L\!-\!1)}\delta_{rr^\prime},~~~~~
\bra x_r y_{r^\prime}\ket= 0
\end{eqnarray}
If for $L\to\infty$ we are allowed to neglect the $\order(1/L)$ correlations between the Gaussian $x_r$,
 we may use the law of large numbers and write the spectrum $\varrho(\mu)$ as an average over their distribution, giving
 (with $|z_r|^2=x_r^2+y_r^2$):
\begin{eqnarray}
\varrho(\mu)&=& \int\!DuDv~\delta[\mu
-Q_0+Q_1 - \frac{1}{2}Q_1(u^2+v^2)]\nonumber
\\
&=& Q_1^{-1}\theta[\mu-Q_0+Q_1]\rme^{-(\mu-Q_0+Q_1)/Q_1}
\label{eq:spectrum_calculated}
\end{eqnarray}
It is encouraging to see that
this simple expression passes the two tests at our disposal.
Firstly, it obeys the exact equation $\int\!d\mu~\mu\varrho(\mu)=Q_0$.
Second, since $\lim_{\zeta\to 1}Q_1=0$ one always has $\lim_{\zeta\to 1}\varrho(\mu)=\delta[\mu-Q_0]$, so in the fake history limit
it leads us correctly to the standard stationary state equations of the fake history MGs.
The remaining deviations very likely to be finite size effects. Having a finite value of $L$ in measurements
removes the self-averaging of the matrix $B_{\lambda\lambda^\prime}(A,Z)$ with respect to the noise variables $Z(\ell,\lambda)$, which our $L\to\infty$ calculation assumed (and thereby deforms the matrix in the
direction of the random matrices as studied in e.g. \cite{Hertz,Stanley,Bouchaud}), and furthermore we have in these simulation experiments only been able to sample in equilibrium states matrices of limited dimension $p\in\{1025,2049\}$. Both these limitations result in a smoothening of the spectrum as measured in practice, compared to the asymptotic shape.
The effect of retaining the $\order(L^{-1})$ correlations between the $x_r$ in the above calculation turns out to be an order $L^{-1}$
reduction of the width of $\varrho(\mu)$; furthermore, when including $\order(L^{-1})$ terms finds that the condition  $\int\!d\mu~\mu\varrho(\mu)=Q_0$ is violated at order $L^{-1}$, which implies (as expected) that in that order we can no longer treat the $z_r$ as Gaussian variables.

\begin{figure}[t]
\vspace*{-0mm} \hspace*{6mm} \setlength{\unitlength}{0.51mm}
\begin{picture}(300,200)

   \put(0,100){\includegraphics[height=100\unitlength,width=140\unitlength]{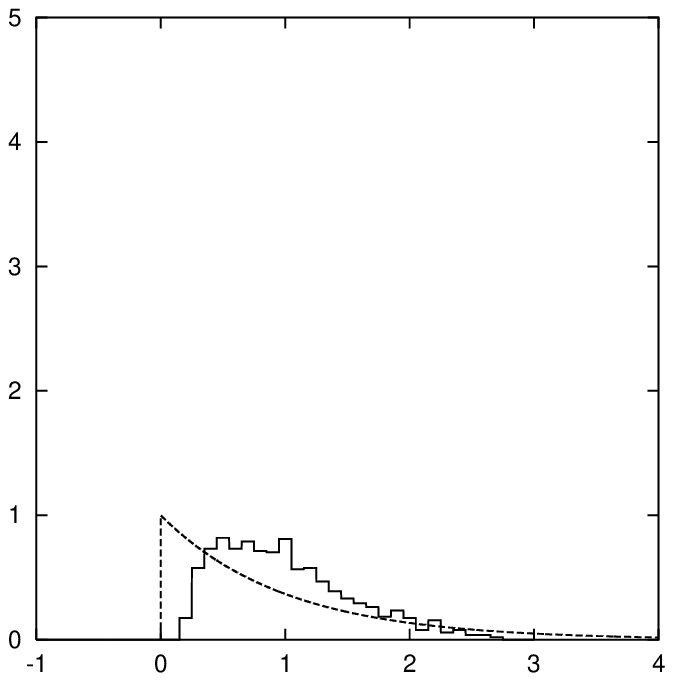}}
  \put(70,185){\small\here{$\alpha\!=\!1,~\zeta\!=\!0$}}

 \put(100,100){\includegraphics[height=100\unitlength,width=140\unitlength]{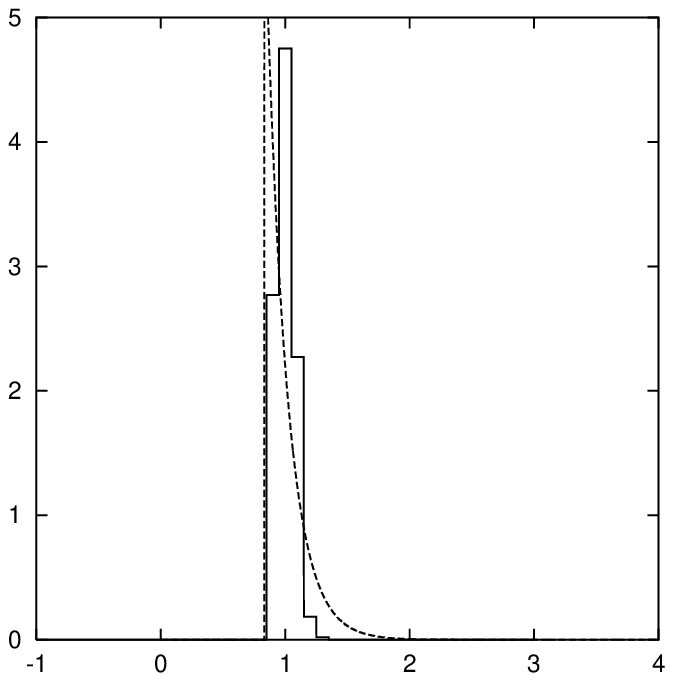}}
  \put(170,185){\small\here{$\alpha\!=\!1,~\zeta\!=\!\frac{1}{2}$}}

 \put(200,100){\includegraphics[height=100\unitlength,width=140\unitlength]{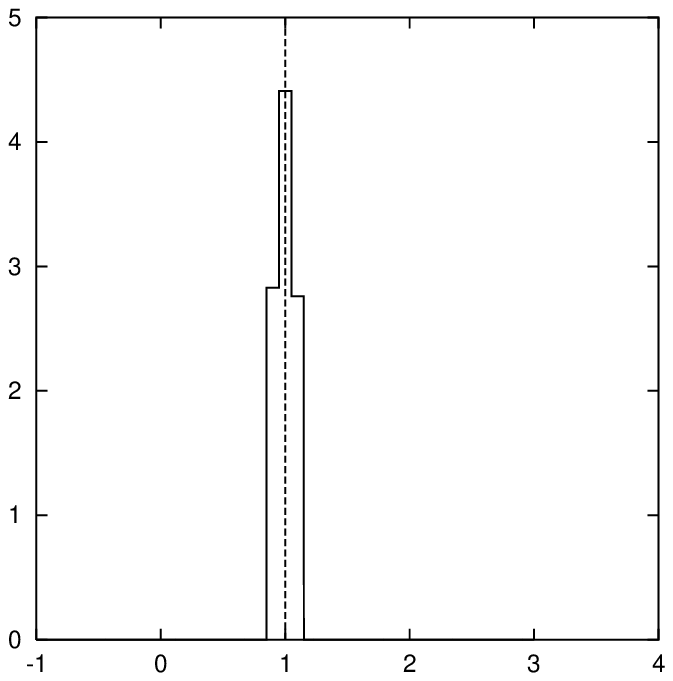}}
   \put(270,185){\small\here{$\alpha\!=\!1,~\zeta\!=\!1$}}

   \put(0,0){\includegraphics[height=100\unitlength,width=140\unitlength]{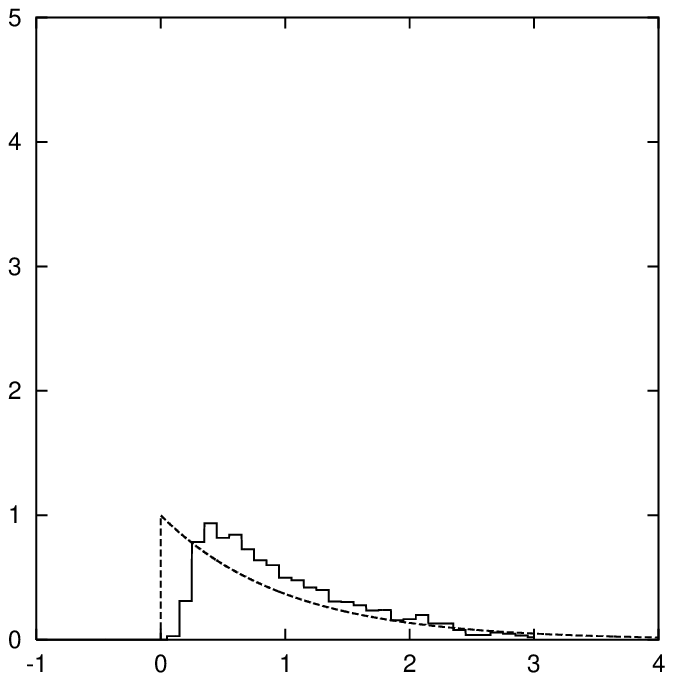}}
     \put(70,85){\small\here{$\alpha\!=\!2,~\zeta\!=\!0$}}

 \put(100,0){\includegraphics[height=100\unitlength,width=140\unitlength]{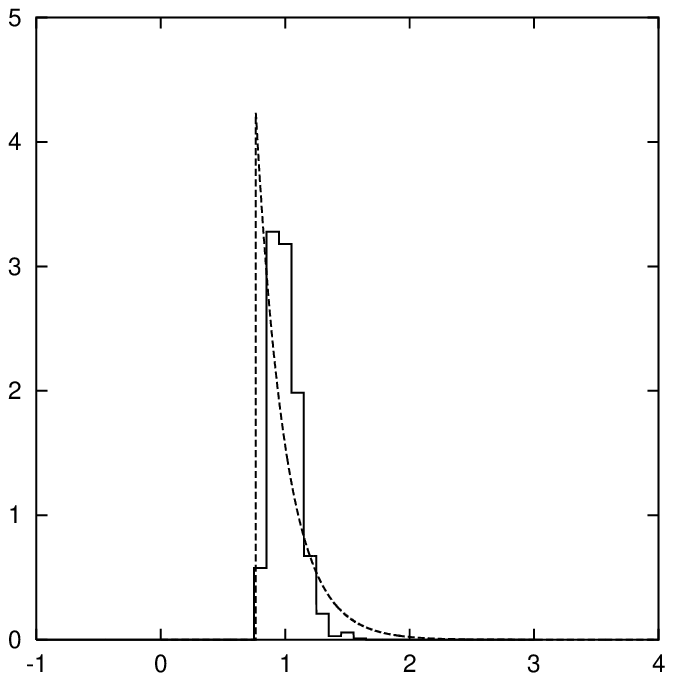}}
 \put(170,85){\small\here{$\alpha\!=\!2,~\zeta\!=\!\frac{1}{2}$}}

 \put(200,0){\includegraphics[height=100\unitlength,width=140\unitlength]{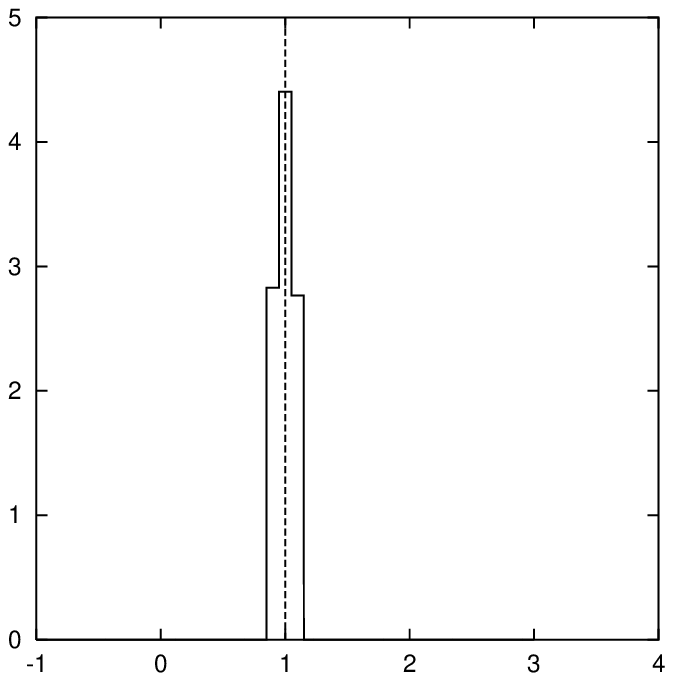}}
   \put(270,85){\small\here{$\alpha\!=\!2,~\zeta\!=\!1$}}

  \put(47,-6){\small $\mu$}    \put(147,-6){\small $\mu$}     \put(247,-6){\small $\mu$}
  \put(-18,50){\small $\varrho(\mu)$}    \put(-18,150){\small $\varrho(\mu)$}

\end{picture}
 \vspace*{2mm}
\caption{Comparison between eigenvalue spectra $\varrho(\mu)$ of the stochastic matrix $\bB(A,Z)$ of inner product MGs (\ref{eq:inp_B}) as measured in numerical simulations (histograms, for $N=1025$) versus the approximate prediction (\ref{eq:spectrum_sgnA}) based on assuming weak bid correlations, for $f[A]=\sgn(A)$. Top row: $\alpha=1$, with $\zeta\in\{0,\frac{1}{2},1\}$. Bottom row: $\alpha=2$, with $\zeta\in\{0,\frac{1}{2},1\}$. All matrix entries in simulations were calculated over $L=400.N$ time steps, following
equilibration. For $\zeta=\frac{1}{2}$ we used in (\ref{eq:spectrum_sgnA}) the value for $\sigma$ as measured in the simulation; for $\zeta\in\{0,1\}$ the volatility drops out of the formula. Note that there are at least two obvious sources of deviation between prediction and experiment (apart from the
crude approximation used in the theory): finite size effects ($N=1025$ rather than $N\to\infty$) and finite  observation time effects ($L\leq 400.N$ rather than $L\to\infty$). Given these limitations, the qualitative agreement between prediction and experiment is surprisingly acceptable.
Spectra of random matrices are indeed known to have finite size effects which decay much slower than $N^{-1/2}$, however, we will find later that at the level
of the MG's conventional order parameters the finite size deviations between theory and simulations are of the expected order $N^{-1/2}$.
} \label{fig:spectra_sgnA} \vspace*{-2mm}
\end{figure}

We note that the above reasoning would also apply for $\Xi(\lambda)=\sigma^2\delta_{\lambda 0}$, a situation indistinguishable from replacing the actual bids
in the history signal by consistent fake information, so we must conclude that the {\em origin} of nontrivial
eigenvalue distributions $\varrho(\mu)$ is the {\em consistency} of the histories (although of course the precise shape of $\varrho(\mu)$ will depend on whether or not the histories are real). We note also that allowing for $\bB(A,Z)$ not being self-averaging
with respect to the bids has been
crucial in our argument. Had we set the bid correlations to zero and averaged $\bB(A,Z)$ over the bid process,
we would have found the trivial spectrum $\varrho(\mu)=\delta[\mu-Q_0]$.
\vsp

Let inspect what our general results imply for the two canonical choices for $f[A]$. For $f[A]=A$ one obtains
 $Q_0= (1-\zeta)^2\sigma^2+\zeta^2 S^2$ and $Q_1=(1-\zeta)^2\sigma^2$, so that
 \begin{eqnarray}
\varrho(\mu)=\frac{\theta[\mu-\zeta^2 S^2]}{(1-\zeta)^2\sigma^2}~\rme^{-(\mu-\zeta^2 S^2)/(1-\zeta)^2\sigma^2}
\label{eq:spectrum_A}
 \end{eqnarray}
 Fully real history would give $\lim_{\zeta\to 0}\varrho(\mu)=\theta[\mu]\sigma^{-2}\rme^{-\mu/\sigma^2}$.
 For $f[A]=\sgn(A)$, in contrast, one has $Q_0=1$ and
\begin{eqnarray}
&& Q_1=\int\!Dy~{\rm Erf}^2\Big[\frac{(1\!-\!\zeta)
\sigma y}{\zeta S\sqrt{2}}\Big]=\frac{4}{\pi}{\rm arctan}[\sqrt{1+2\frac{(1\!-\!\zeta)^2
\sigma^2}{\zeta^2 S^2}}]-1
\end{eqnarray}
and so
 \begin{eqnarray}
 \hspace*{-25mm}
\varrho(\mu)=\frac{\theta\Big[\mu\!-\!2\!+\!\frac{4}{\pi}{\rm arctan}[\sqrt{1\!+\!2\frac{(1\!-\!\zeta)^2
\sigma^2}{\zeta^2 S^2}}]\Big]}{ \frac{4}{\pi}{\rm arctan}[\sqrt{1\!+\!2\frac{(1\!-\!\zeta)^2
\sigma^2}{\zeta^2 S^2}}]\!-\!1}~ \rme^{-\Big[\mu-2+\frac{4}{\pi}{\rm arctan}[\sqrt{1+2\frac{(1\!-\!\zeta)^2
\sigma^2}{\zeta^2 S^2}}]\Big]/\Big[\frac{4}{\pi}{\rm arctan}[\sqrt{1+2\frac{(1\!-\!\zeta)^2
\sigma^2}{\zeta^2 S^2}}]-1\Big]}\hspace*{-5mm}
\nonumber \\[-4mm]
\label{eq:spectrum_sgnA}
\end{eqnarray}
Here the limit of fully real history gives $\lim_{\zeta\to 0}Q_1=1$ and $\lim_{\zeta\to 0}\varrho(\mu)=\theta[\mu]e^{-\mu}$.
The persistent order parameter equations (\ref{eq:persistent0},\ref{eq:persistent1},\ref{eq:persistent2}) do not involve the volatility,
but it turns out that we generally need to know the volatility in order to use (\ref{eq:spectrum_sgnA}) and (\ref{eq:spectrum_A}) for closure, unless $\zeta\in\{0,1\}$, i.e. unless the histories are strictly real or strictly fake.
In figure \ref{fig:spectra_sgnA} we compare the above predictions with the eigenvalue spectra as measured in numerical simulations of the
inner product MG with $f[A]=\sgn(A)$, where (for $\zeta\not\in \{0,1\}$) we inserted in our formulae the value of $\sigma$ as measured.
Given the severe finite size and finite time effects in such simulations, the agreement is surprisingly acceptable.

 \subsection{The volatility}

\begin{figure}[t]
\vspace*{-0mm} \hspace*{23mm} \setlength{\unitlength}{0.45mm}
\begin{picture}(300,200)

   \put(0,100){\includegraphics[height=95\unitlength,width=130\unitlength]{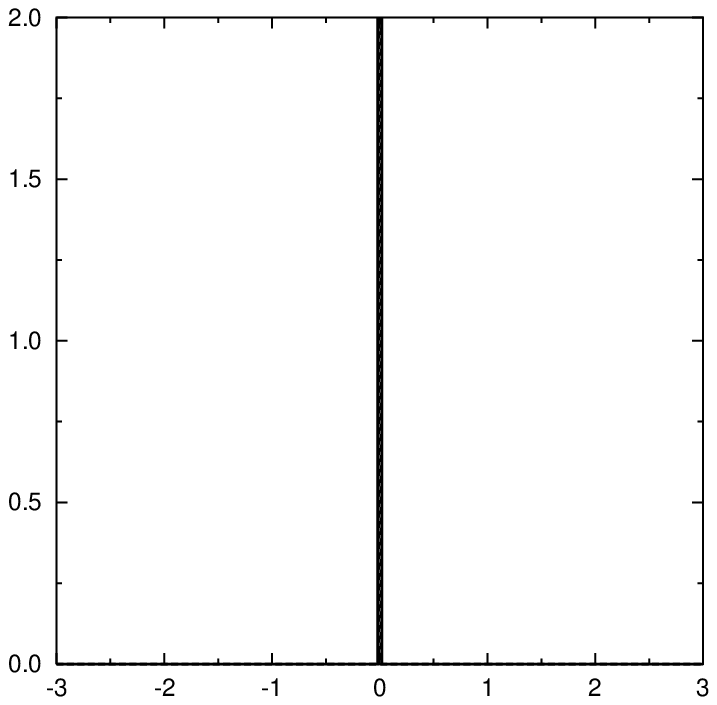}}
  \put(79,181){\small\here{$\zeta\!=\!0$}}

 \put(100,100){\includegraphics[height=95\unitlength,width=130\unitlength]{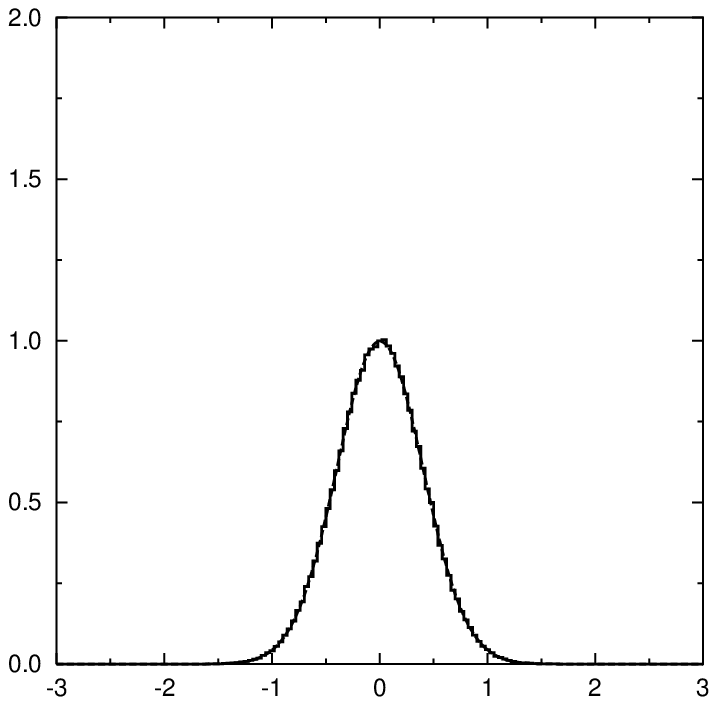}}
  \put(179,181){\small\here{$\zeta\!=\!\frac{1}{2}$}}

 \put(200,100){\includegraphics[height=95\unitlength,width=130\unitlength]{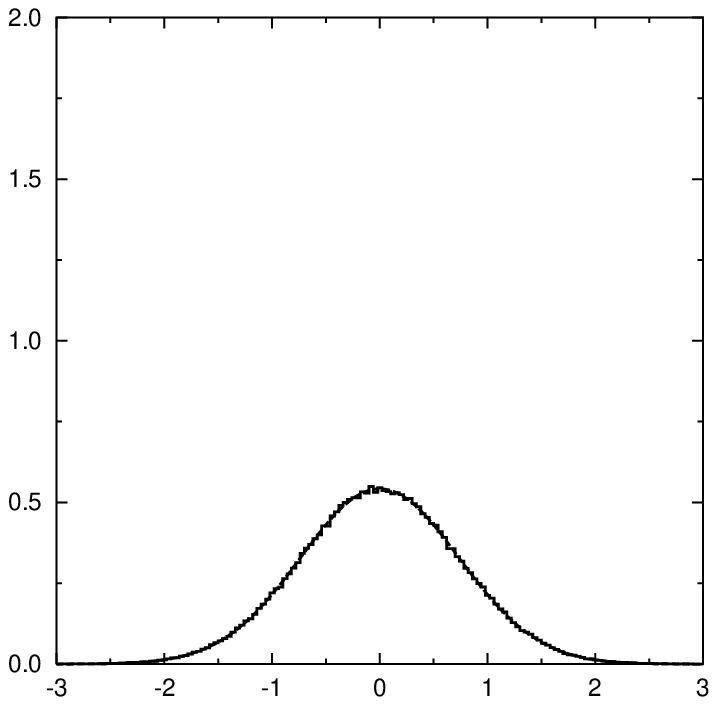}}
   \put(279,181){\small\here{$\zeta\!=\!1$}}

   \put(0,0){\includegraphics[height=95\unitlength,width=130\unitlength]{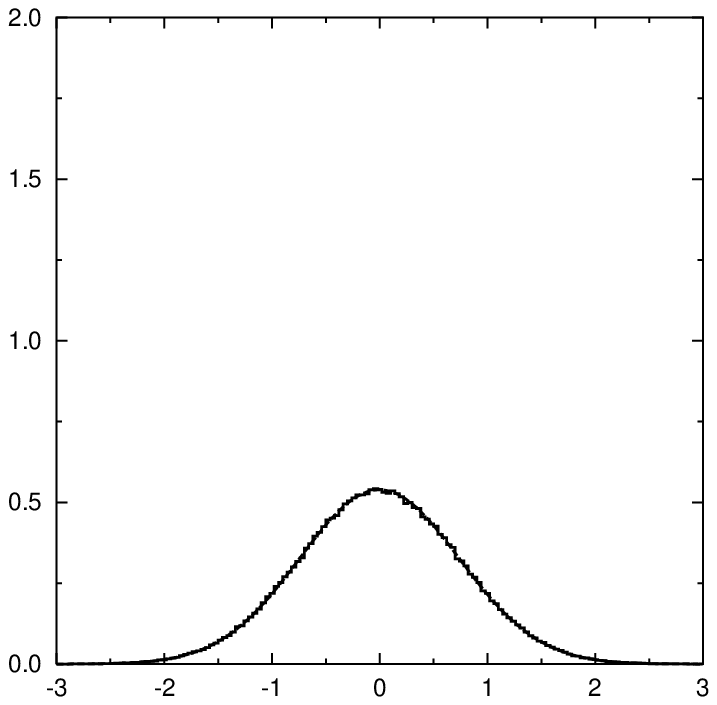}}
     \put(79,81){\small\here{$\zeta\!=\!0$}}

 \put(100,0){\includegraphics[height=95\unitlength,width=130\unitlength]{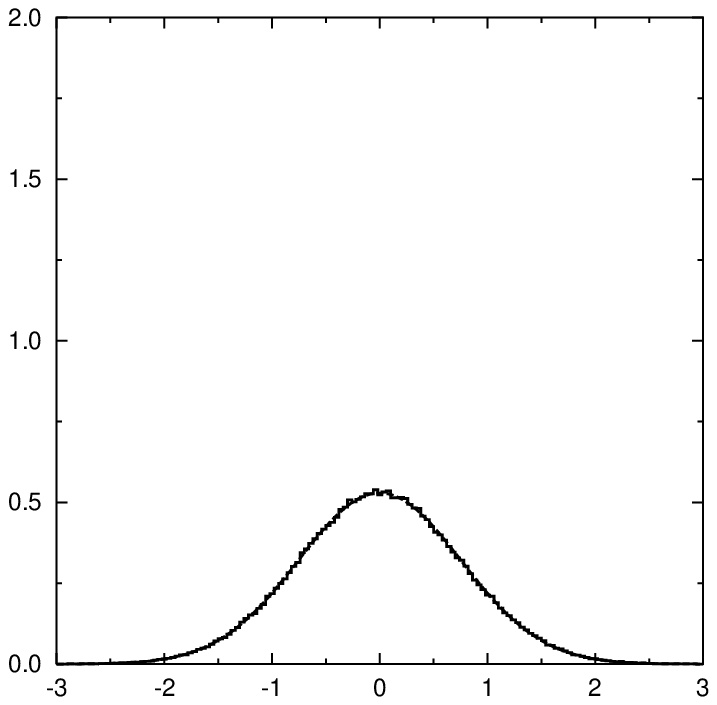}}
 \put(179,81){\small\here{$\zeta\!=\!\frac{1}{2}$}}

 \put(200,0){\includegraphics[height=95\unitlength,width=130\unitlength]{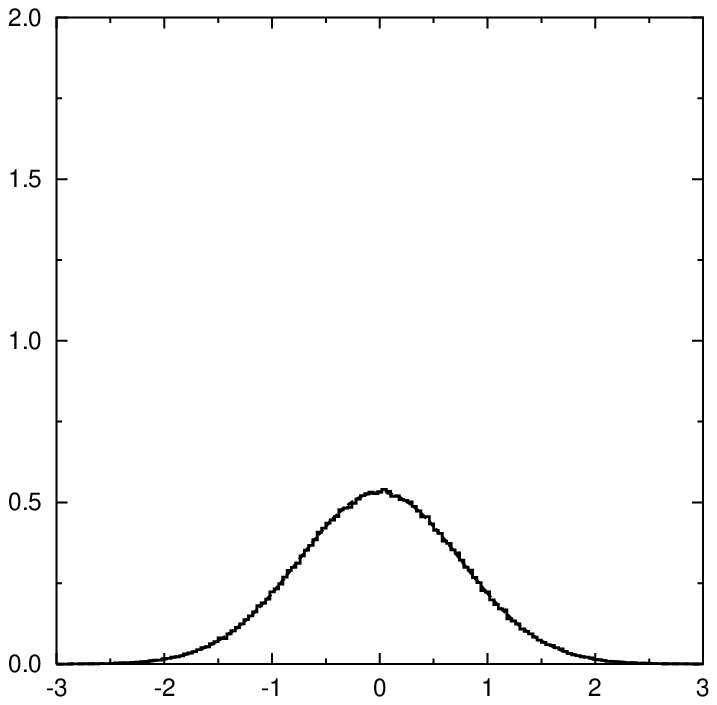}}
   \put(279,81){\small\here{$\zeta\!=\!1$}}

  \put(49,-14){\small $A$}    \put(149,-14){\small $A$}     \put(249,-14){\small $A$}
  \put(-65,50){\small $f[A]=\sgn(A):$}    \put(-65,150){\small $f[A]=A:$}

\end{picture}
 \vspace*{2mm}
\caption{The bid distribution $P(A)=L^{-1}\sum_{\ell\leq L}\bra \delta[A-A(\ell)]\ket$  of inner product MGs (\ref{eq:inp_B}), as measured in numerical simulations (histograms), with $N=1025$ and $\alpha=2$. Statistics are calculated over $200.N$ time steps, following equilibration.
The solid lines (barely visible, since they are virtually coinciding with the histograms) are Gaussian distributions with average and variance identical to those observed, for comparison.
Top row: $f[A]=A$, with $\zeta\in\{0,\frac{1}{2},1\}$  Bottom row: $f[A]=\sgn(A)$, with $\zeta\in\{0,\frac{1}{2},1\}$. The left pictures refer to strictly real histories,
the right pictures to strictly fake histories, with the middle ones representing a mixing of the two. These data (corresponding to the same experiments as those of figure \ref{fig:covariances}) are representative
of the typical behaviour throughout the ergodic regime. The bids are {\em always} indistinguishable from Gaussian variables.
}\label{fig:gaussianbids}
\end{figure}

 We turn to the calculation of the volatility, which measures the fluctuations in the bids $A(\ell)$.
It turns out that these bids are always Gaussian random variables, see figure \ref{fig:gaussianbids}, which is not obvious since the
bids obey a generally nonlinear stochastic equation.
 The first step is to substitute (\ref{eq:Del3}) (obtained upon assuming short history correlation times in ergodic stationary states)  into the as yet exact formula  (\ref{eq:workout_vola}) for the volatility, giving
\begin{eqnarray}
\hspace*{-15mm}
\sigma^2
&=& \lim_{t\to\infty} \lim_{\del\to 0} \Big\{
\frac{1}{2}\sum_{r,r^\prime\geq 0}(-\del)^{r+r^\prime}
\int_0^\infty\!d\mu~\varrho(\mu)~\mu^{r+r^\prime+1}
\sum_{\ell_0\ldots \ell_r}G(\ell_0,\ell_1)\ldots
G(\ell_{r-1},\ell_r) \nonumber
\\
\hspace*{-15mm}
&&
\hspace*{10mm} \times\sum_{\ell_0^\prime\ldots
\ell^\prime_{r^\prime}} G(\ell^\prime_0,\ell^\prime_1)\ldots
G(\ell^\prime_{r^\prime-1},\ell^\prime_{r^\prime})~\big[1+C(\ell_r,\ell^\prime_{r^\prime})\big]\delta_{\ell,\ell_0}\delta_{\ell,\ell^\prime_0}
 \Big\}\Big|_{\ell=t/\del}
 \label{eq:getting_sigma1}
\end{eqnarray}
As always we next have to approximate non-persistent dynamical order parameters by expressions involving only persistent ones.
Following procedures similar to those used in the past \cite{BookAC} we assume the agent correlation function $C(\ell)$ to decay very fast, so that
we may replace $C(\ell)\to c +\delta_{\ell 0}(1-c)$. Substitution into (\ref{eq:getting_sigma1}), followed by carrying out various sums analytically (using $\del\sum_\ell G(\ell)=\chi$) leads us to the approximation
\begin{eqnarray}
\hspace*{-25mm}
\sigma^2
&=&
\frac{1}{2}(1+c)\int_0^\infty\!d\mu~\varrho(\mu)\frac{\mu}{(1+\mu\chi)^2}
\nonumber
\\
\hspace*{-25mm}
&&+\frac{1}{2}(1\!-\!c)
 \lim_{t\to\infty} \lim_{\del\to 0} \Big\{
\sum_{r,r^\prime\geq 0}(-\del)^{r+r^\prime}
\int_0^\infty\!d\mu~\varrho(\mu)~\mu^{r+r^\prime+1}
\sum_{\ell_0\ldots \ell_r}G(\ell_0,\ell_1)\ldots
G(\ell_{r-1},\ell_r) \nonumber
\\
\hspace*{-25mm}
&&
\hspace*{40mm} \times\sum_{\ell_0^\prime\ldots
\ell^\prime_{r^\prime}} G(\ell^\prime_0,\ell^\prime_1)\ldots
G(\ell^\prime_{r^\prime-1},\ell^\prime_{r^\prime})~\delta_{\ell_r,\ell^\prime_{r^\prime}}\delta_{\ell,\ell_0}\delta_{\ell,\ell^\prime_0}
 \Big\}\Big|_{\ell=t/\del}
 \label{eq:getting_sigma2}
\end{eqnarray}
In the second term we see that, unless $r=r^\prime=0$ (where $\delta_{\ell_r,\ell^\prime_{r^\prime}}$ reduces directly to
$\delta_{\ell_0,\ell^\prime_{0}}=\delta_{\ell\ell}=1$), the factor $\delta_{\ell_r,\ell^\prime_{r^\prime}}$ will always leave us with one residual
factor $\del$ that is no longer compensated by a time summation, resulting in the removal of the corresponding term for $\del\to 0$. Hence one
retains only
\begin{eqnarray}
\sigma^2
&=&
\frac{1}{2}(1+c)\int_0^\infty\!d\mu~\varrho(\mu)\frac{\mu}{(1+\mu\chi)^2}+
\frac{1}{2}(1-c)\int_0^\infty\!d\mu~\varrho(\mu)\mu
\nonumber
\\
&=& \frac{1}{2}(1+c)\int_0^\infty\!d\mu~\varrho(\mu)\frac{\mu}{(1+\mu\chi)^2}+
\frac{1}{2}(1-c)Q_0
\label{eq:sigma_done}
\end{eqnarray}
For $\zeta \to 1$ (strictly inconsistently fake histories), where $\varrho(\mu)\to \delta[\mu-Q_0]$ and $Q_0\to \kappa_2$, this expression
 reduces to the familiar approximation formula known from inner product MGs (apart from various factors $\kappa_2$, which appear exactly
 as predicted by   (\ref{eq:chi_eqn_again}) and (\ref{eq:finishedvola2}):
\begin{eqnarray}
\lim_{\zeta\to 1}\sigma^2
&=&\kappa_2\Big\{
\frac{1}{2}\frac{1+c}{(1+\kappa_2\chi)^2}+
\frac{1}{2}(1-c)\Big\}
\end{eqnarray}

\subsection{Compactification of the theory and universality for $\zeta\to 0$}

The final equations which describe the static limit of our theory, describing time-translation invariant stationary states,
can be compactified further. The simple exponential shape of the spectrum (\ref{eq:spectrum_calculated}) allows
us to express  the two relevant averages in (\ref{eq:persistent0},\ref{eq:persistent1})
in terms of the exponential integral $E_1(z)=\int_z^\infty\!dt~t^{-1}\rme^{-t}$ \cite{AbramStegun}, resulting in the following set of coupled  equations
for the static order parameters $\{c,\phi,\chi\}$:
\begin{eqnarray}
\hspace*{-15mm}
\omega
&=& \frac{\chi Q_1-\rme^{[1+\chi(Q_0-Q_1)]/\chi Q_1}~E_1(\frac{1+\chi(Q_0-Q_1)}{\chi Q_1})}
{\sqrt{(\chi Q_1)^2\!-\rme^{[1+\chi(Q_0-Q_1)]/\chi Q_1}~E_1(\frac{1+\chi(Q_0-Q_1)}{\chi Q_1})\big(2\chi Q_1\!+\!1\big)
+\frac{\chi Q_1}{1+\chi(Q_0-Q_1)}}}~~~
\label{eq:final_omega}
\\
\hspace*{-15mm}
u&=&\omega\sigma[\infty]\sqrt{\alpha}/\sqrt{2(1+c)}
\label{eq:final_u}
\\
\hspace*{-15mm}
\phi&=&1- {\rm Erf}[u]
\label{eq:final_phi}
\\
\hspace*{-15mm}
{\rm Erf}[u]&=& \alpha\Big\{1-\frac{1}{\chi Q_1}~\rme^{[1+\chi(Q_0-Q_1)]/\chi Q_1}~E_1\Big(\frac{1\!+\!\chi(Q_0\!-\!Q_1)}{\chi Q_1}\Big)\Big\}
\label{eq:final_chi}
\\
\hspace*{-15mm}
c &=&\sigma^2[\infty]\Big\{ 1-{\rm Erf}[u]+\frac{1}{2u^2}{\rm
Erf}[u]-\frac{1}{u\sqrt{\pi}}\rme^{-u^2}\Big\}
\label{eq:final_c}
\end{eqnarray}
Equations (\ref{eq:final_omega},\ref{eq:final_u}) simply define $\omega$ and $u$ as short-hands for complicated functions of $\{c,\phi,\chi\}$.
The volatility and the factors $Q_{0,1}$ are calculated from
\begin{eqnarray}
\hspace*{-15mm}
\sigma^2
&=& \frac{1\!+\!c}{2Q_1\chi^2}\left\{\frac{1\!+\!Q_1\chi}{Q_1\chi}
\rme^{[1+\chi(Q_0-Q_1)]/\chi Q_1}~E_1\Big(\frac{1\!+\!\chi(Q_0\!-\!Q_1)}{\chi Q_1}\Big)-\frac{1}{1\!+\!\chi(Q_0\!-\!Q_1)}\right\}
\nonumber
\\
\hspace*{-15mm}&&
+
\frac{1}{2}(1\!-\!c)Q_0
\nonumber
\\
\hspace*{-15mm}
&=&
\frac{1\!+\!c}{2Q_1\chi^2}\left\{
(1\!+\!Q_1\chi)\frac{\alpha\!-\!{\rm Erf}[u]}{\alpha}-\frac{1}{1\!+\!\chi(Q_0\!-\!Q_1)}\right\}
+
\frac{1}{2}(1-c)Q_0
\label{eq:final_sigma}
\\
\hspace*{-15mm}
Q_0&=& \int\!Dz~f^2[z\sqrt{(1-\zeta)^2\sigma^2+\zeta^2 S^2}]
\label{eq:final_Q0}
\\
\hspace*{-15mm}
Q_1&=& \int\!Dy~\Big[\int\!Dz~f[(1\!-\!\zeta)
\sigma y \!+\! \zeta Sz]\Big]^2
\label{eq:final_Q1}
 \end{eqnarray}
All the details regarding the history definitions (viz. the sampling function $f[A]$ and the degree of `fakeness' $\zeta$)
are fully contained within the factors $Q_{0,1}$. Our stationary state order parameter equations are now fully closed, and can be solved numerically.
\vsp

Of special interest is the limit $\zeta\to 0$, where the histories are strictly real.
In this limit  one can show that
the persistent order parameters $\{\phi,c\}$ become functions of $\alpha$ that no longer depend on the choice made for $f[A]$.
The reason for this universality is that according to (\ref{eq:Q0},\ref{eq:Q1}) for $\zeta\to 0$ one always has $\lim_{\zeta\to 0}Q_1=\lim_{\zeta\to 0}Q_0=
\int\!Dz~f^2[\sigma z]$. We may now define $\tilde{\chi}=Q_0\chi$ and find for $\zeta\to 0$ our order parameter equations for $\{c,\phi,\tilde{\chi}\}$ reducing to
\begin{eqnarray}
\omega
&=& \frac{\tilde{\chi}-\rme^{1/\tilde{\chi}}~E_1(1/\tilde{\chi})}
{\sqrt{\tilde{\chi}^2-\rme^{1/\tilde{\chi}}~E_1(1/\tilde{\chi})\big(2\tilde{\chi}\!+\!1\big)
+\tilde{\chi}}}~~~
\label{eq:universal_omega}
\\
u&=&\omega\sigma[\infty]\sqrt{\alpha}/\sqrt{2(1+c)}
\label{eq:universal_u}
\\
\phi&=&1- {\rm Erf}[u]
\label{eq:universal_phi}
\\
{\rm Erf}[u]&=& \alpha\Big\{1-\tilde{\chi}^{-1}\rme^{1/\tilde{\chi}}~E_1(1/\tilde{\chi})\Big\}
\label{eq:universal_chi}
\\
c &=&\sigma^2[\infty]\Big\{ 1-{\rm Erf}[u]+\frac{1}{2u^2}{\rm
Erf}[u]-\frac{1}{u\sqrt{\pi}}\rme^{-u^2}\Big\}
\label{eq:universal_c}
\end{eqnarray}
The values of $\{c,\phi,\tilde{\chi}\}$ as solved from the above closed set can only depend on the remaining control
parameter $\alpha$.
Since $\lim_{z\to 0} z \rme^zE_1(z)=0$, the ergodicity-breaking transition point where $\tilde{\chi}\to\infty$, corresponds as always to  $\phi=1-\alpha$.
We would only need to know the choice made for $f[A]$ if we wished to disentangle $\chi$ and $Q_0$ from $\tilde{\chi}$, or to calculate the volatility
which would be the solution of
\begin{eqnarray}
\sigma^2&=& \Big\{\frac{1\!+\!c}{2\tilde{\chi}^2}\Big[(1\!+\!\tilde{\chi})\frac{\alpha\!-\!1\!+\!\phi}{\alpha}-1\Big]+\frac{1}{2}(1\!-\!c)\Big\}
\int\!Dz~f^2[\sigma z]
\label{eq:vola_zeta0}
\end{eqnarray}
For $f[A]=A$, however, we have a degenerate situation at $\zeta=0$. Here we find (\ref{eq:vola_zeta0}) converting into
\begin{eqnarray}
\sigma^2\Big\{1-
\frac{1}{\tilde{\chi}^2}\Big[
(1\!+\!\tilde{\chi})\frac{\alpha\!-\!1\!+\!\phi}{\alpha}-1\Big]\Big\}=0
\label{eq:zero_sigma}
\end{eqnarray}
We find that (possibly apart from one degenerate case) that the only stationary solution has $\sigma=0$,
so also $Q_0=\sigma^2=0$. This, in turn, tells us that $\chi=\infty$ for all $\alpha$, so the $f[A]=A$ system is always critical and the ergodicity assumption
is always inconsistent.

\section{Theory versus numerical simulations of inner product MGs}

Below we test the predictions of our theory further against numerical simulations of inner product MGs without
agents' decision noise, so $T=0$ and $\sigma[\infty]=1$. Our equations allow for such noise,
but it represents a simple generalization
that is well understood and delays considerably the equilibration in simulations.
In contrast to look-up table MGs, where the non-Markovian nature of the microscopic process involves delayed forces going back over
a time interval $\Delta t=\order(N^{-1}\log N$), in inner product MGs this time interval scales as $\Delta t=\order(N^0$).
This already limits severely the scope of numerical simulations, and, given the multitude of control parameters to be varied,
system sizes will have to remain modest, here mostly $N=\order(10^3)$. Even then, finite size and finite time effects remain a serious problem,
especially for large $\alpha$.

\subsection{Inner product MG with $f[A]=\sgn(A)$}

\begin{figure}[t]
\vspace*{-0mm} \hspace*{1mm} \setlength{\unitlength}{0.45mm}
\begin{picture}(300,100)

   \put(0,0){\includegraphics[height=100\unitlength,width=140\unitlength]{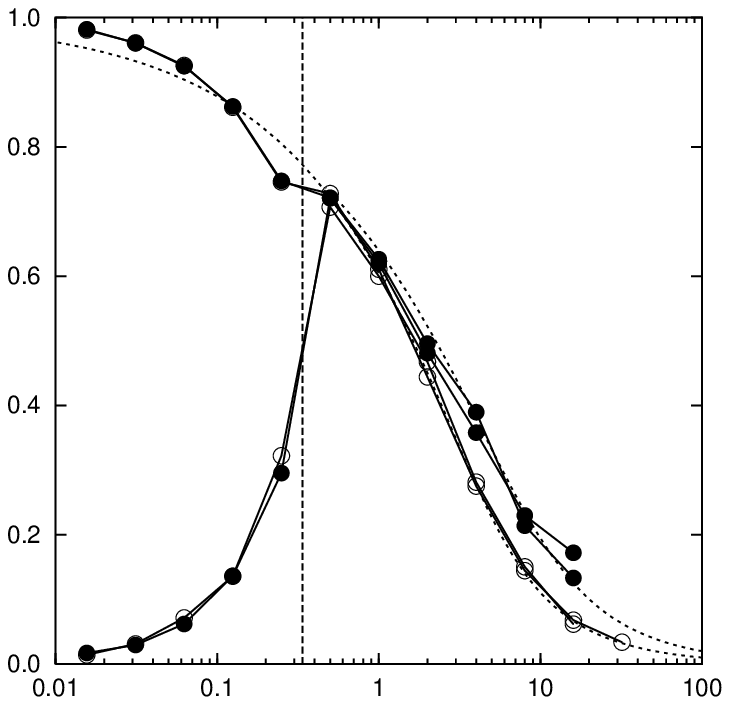}}
     \put(-3,50){\small\here{$c$}}

 \put(120,0){\includegraphics[height=100\unitlength,width=140\unitlength]{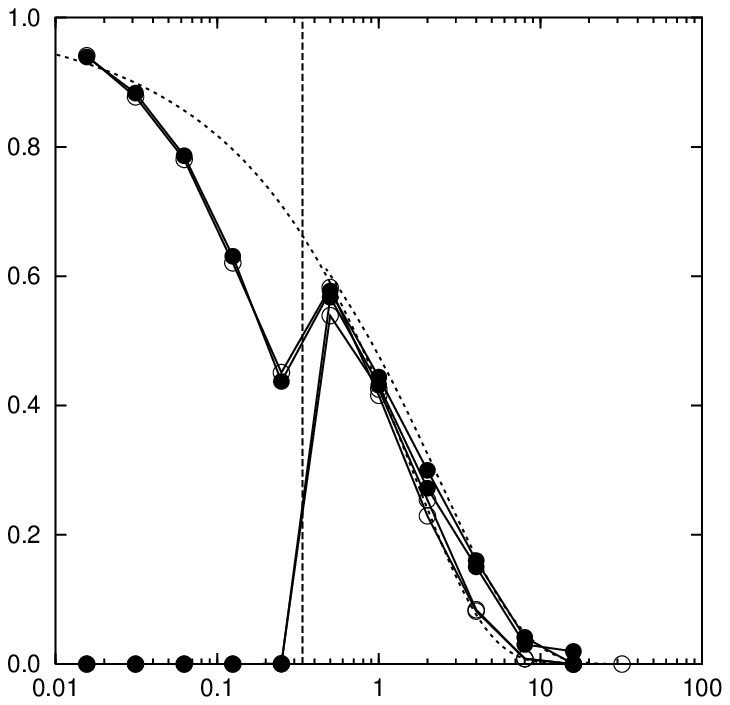}}
 \put(117,50){\small\here{$\phi$}}

 \put(240,0){\includegraphics[height=100\unitlength,width=140\unitlength]{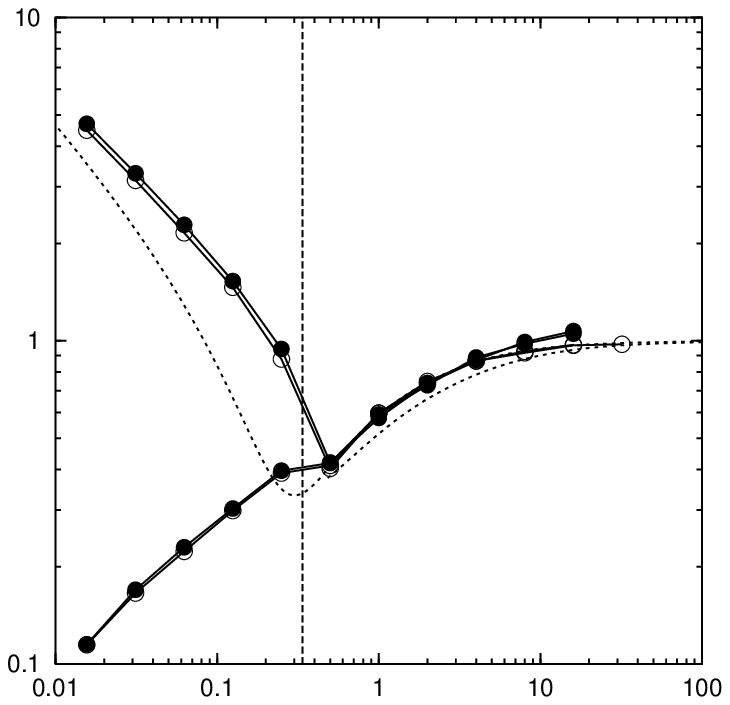}}
   \put(237,50){\small\here{$\sigma$}}

  \put(49,-12){\small $\alpha$}    \put(169,-12){\small $\alpha$}     \put(289,-12){\small $\alpha$}
\end{picture}
 \vspace*{7mm}
\caption{Observables $c$, $\phi$ and $\sigma$ as functions of $\alpha=p/N$, measured in numerical simulations for $f[A]=\sgn(A)$ and $S=1$.
  Measurements are taken over $200.p$ time steps, after an equilibration of $200.p$ time steps; system sizes: $pN^2=2^{30}$ (except for $\alpha=16$ and $\alpha=32$, where $N=1025$ and $N=513$, respectively). Each panel
  shows the result of four simulation runs: $\zeta=0$ (strictly real histories, connected full circles) with biased and unbiased initial conditions, and $\zeta=1$
  (strictly fake histories, connected open circles) with biased and unbiased initial conditions. Biased initial conditions: $q_i(0)=\order(1)$ (zero average random); unbiased initial conditions: $q_i(0)=\order(10^{-5})$ (zero average random). Dotted curves give the theoretical predictions for $\zeta=0$ and $\zeta=1$, upon assuming ergodic stationary states, with the vertical dashed line marking the transition value $\alpha_c$ below which the theory no longer applies.
}\label{fig:observables_v_alpha_sgnA}
\end{figure}

\begin{figure}[t]
\vspace*{-0mm} \hspace*{1mm} \setlength{\unitlength}{0.45mm}
\begin{picture}(300,100)

   \put(0,0){\includegraphics[height=100\unitlength,width=140\unitlength]{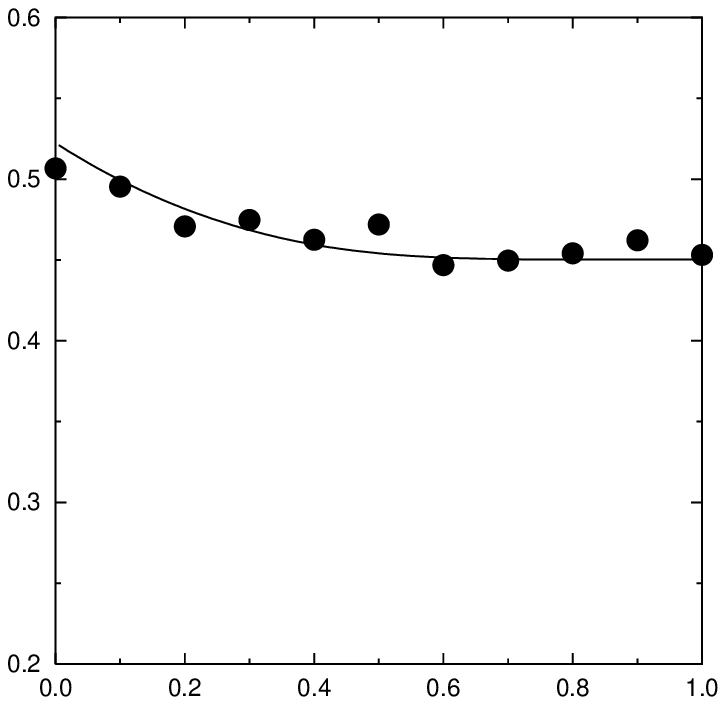}}
     \put(-3,50){\small\here{$c$}}

 \put(120,0){\includegraphics[height=100\unitlength,width=140\unitlength]{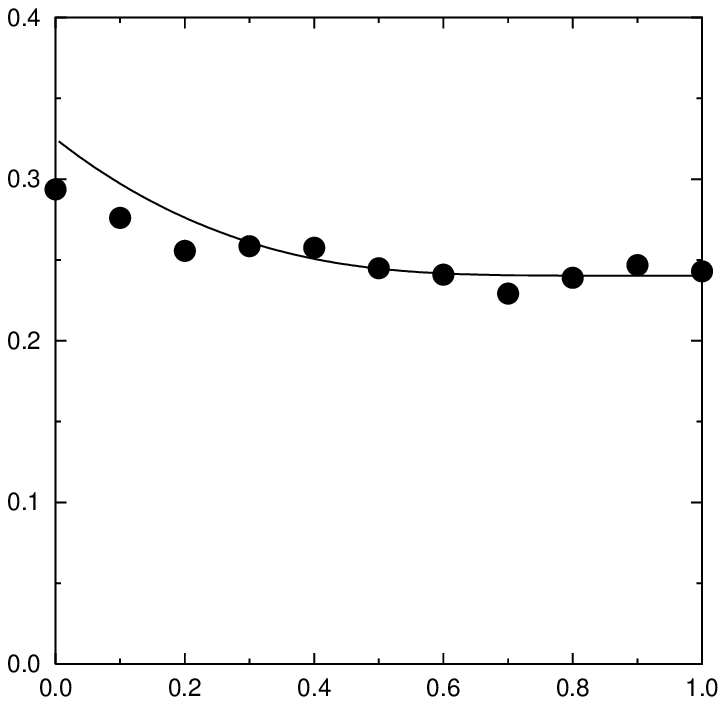}}
 \put(117,50){\small\here{$\phi$}}

 \put(240,0){\includegraphics[height=100\unitlength,width=140\unitlength]{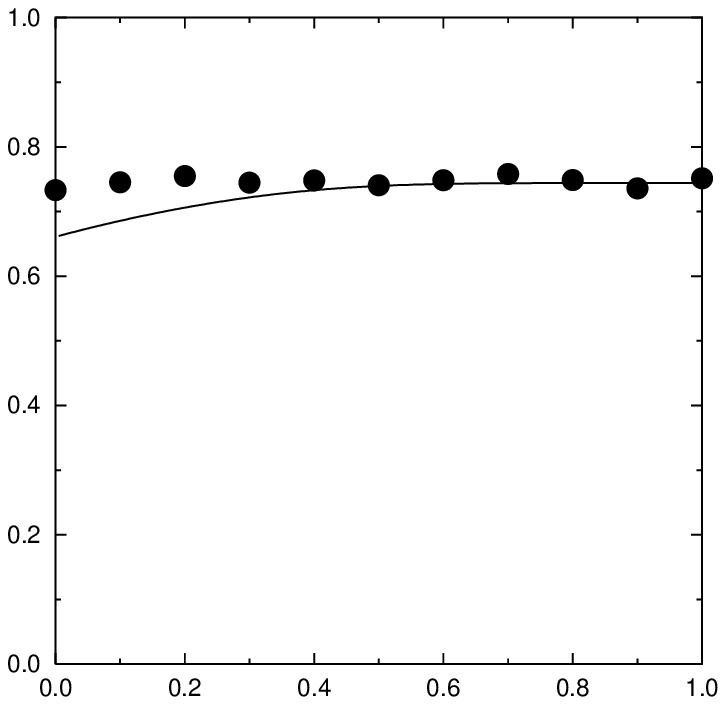}}
   \put(237,50){\small\here{$\sigma$}}

  \put(49,-12){\small $\zeta$}    \put(169,-12){\small $\zeta$}     \put(289,-12){\small $\zeta$}
\end{picture}
 \vspace*{7mm}
\caption{Observables $c$, $\phi$ and $\sigma$ as functions of $\zeta$, measured in numerical simulations for $f[A]=\sgn(A)$, $\alpha=2$, and $S=1$.
  Measurements are taken over $400.N$ time steps, after an equilibration of $400.N$ time steps; system sizes: $N=1025$. Circles give
   the simulation results. Curves give the theoretical predictions, upon assuming ergodic stationary states. Strictly real histories and stricly fake histories correspond to $\zeta=0$ and $\zeta=1$, respectively. The random deviations between theory and simulations are compatible with the expected finite
   size effects of order $N^{-1/2}\approx 0.03$, except for the volatility where fluctuations are small and
   the weak predicted trend is not confirmed.
}\label{fig:observables_v_zeta_sgnA}
\end{figure}

In an earlier section we have already compared the observed spectra of the history covariance matrix
to our theoretical predictions, see figure
\ref{fig:spectra_sgnA}, showing quite reasonable agreement. In figure \ref{fig:observables_v_alpha_sgnA}
we compare the predicted values of the observables $\{c,\phi,\sigma\}$  (obtained by solving our order closed equations
(\ref{eq:final_omega}-\ref{eq:final_Q1}) numerically)
to numerical simulation data, for $\zeta\in\{0,1\}$, $S=1$, and different values of $\alpha$.
We see our prediction that the transition point is independent of $\zeta$, i.e. of whether  histories are real or fake, is confirmed quite convincingly. In the ergodic regime $\alpha>\alpha_c$, where our theory applies, we also observe a good agreement in terms of the values of $c$ and
$\phi$ (except for large values of $\alpha$, where finite size and finite time effects become severe). In terms of the volatility $\sigma$, for which we had to make further approximations (to express non-persistent terms in persistent ones) theory and experiment exhibit deviations: the experimentally
observed value for $\sigma$ appears roughly independent of $\zeta$ in the regime where the data are still reliable (if not a higher value for $\zeta=0$ compared to $\zeta=1$), whereas the theory predicts a slight
volatility {\em reduction} due to having real histories compared to the fake ones. This suggests that, at least for $f[A]=\sgn(A)$, the usual crude step to replace $C(\ell)\to
c+(1-c)\delta_{\ell\ell^\prime}$ in the true formula (\ref{eq:getting_sigma1}) for the volatility, resulting in the approximation
 (\ref{eq:sigma_done}), is less appropriate for real than for fake histories. A careful analysis of the short-time behaviour of the kernels $C$ and $G$ will be required to understand why this is so.
 Fortunately, the volatility drops out of our order parameter equations for $\zeta\in\{0,1\}$.
 Upon measuring order parameters as a function of $\zeta$, in order to probe those cases $0<\zeta<1$ where the volatility does play a vital role in closing our equations, one still observes excellent agreement in those regimes where simulation data are reliable.

\subsection{Inner product MG with $f[A]=A$}

\begin{figure}[t]
\vspace*{-0mm} \hspace*{1mm} \setlength{\unitlength}{0.45mm}
\begin{picture}(300,100)

   \put(0,0){\includegraphics[height=100\unitlength,width=140\unitlength]{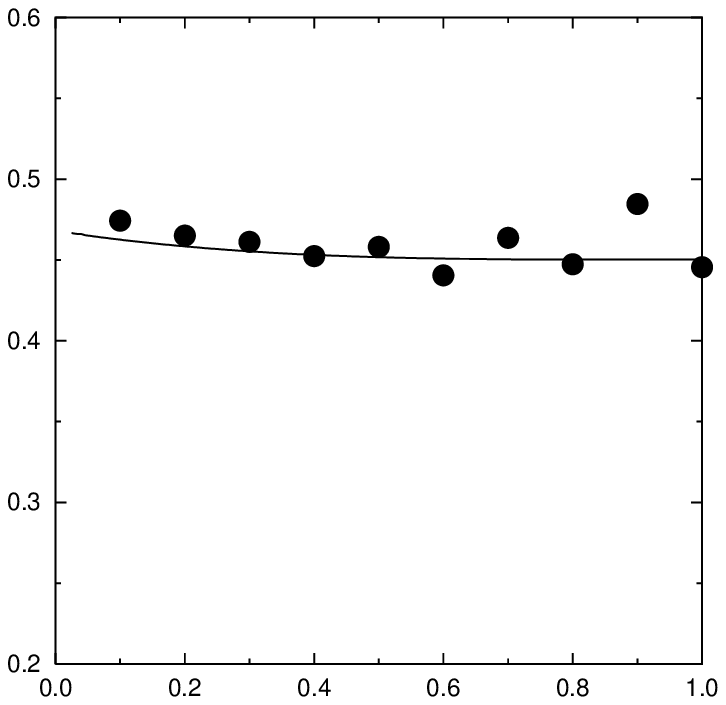}}
     \put(-3,50){\small\here{$c$}}

 \put(120,0){\includegraphics[height=100\unitlength,width=140\unitlength]{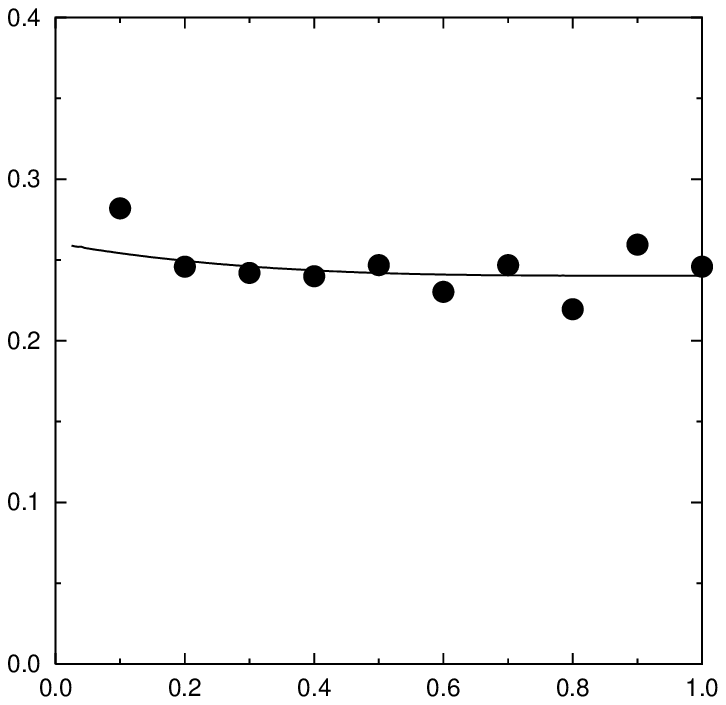}}
 \put(117,50){\small\here{$\phi$}}

 \put(240,0){\includegraphics[height=100\unitlength,width=140\unitlength]{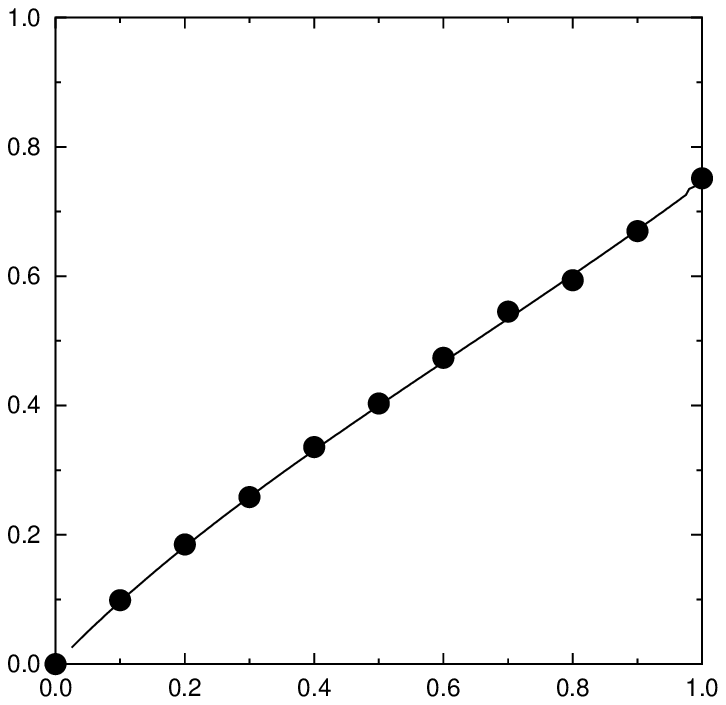}}
   \put(237,50){\small\here{$\sigma$}}

  \put(49,-12){\small $\zeta$}    \put(169,-12){\small $\zeta$}     \put(289,-12){\small $\zeta$}
\end{picture}
 \vspace*{7mm}
\caption{Observables $c$, $\phi$ and $\sigma$ as functions of $\zeta$, measured in numerical simulations for $f[A]=A$, $\alpha=2$, and $S=1$.
  Measurements are taken over $400.N$ time steps, after an equilibration of $400.N$ time steps; system sizes: $N=1025$. Circles give
   the simulation results. Curves give the theoretical predictions, upon assuming ergodic stationary states. Strictly real histories and stricly fake histories correspond to $\zeta=0$ and $\zeta=1$, respectively. The deviations between theory and simulations are compatible with the expected finite
   size effects of order $N^{-1/2}\approx 0.03$.
}\label{fig:observables_v_zeta_A}
\end{figure}

\begin{figure}[t]
\vspace*{-0mm} \hspace*{1mm} \setlength{\unitlength}{0.27mm}
\begin{picture}(500,340)



     \put(0,170){\includegraphics[height=160\unitlength,width=160\unitlength]{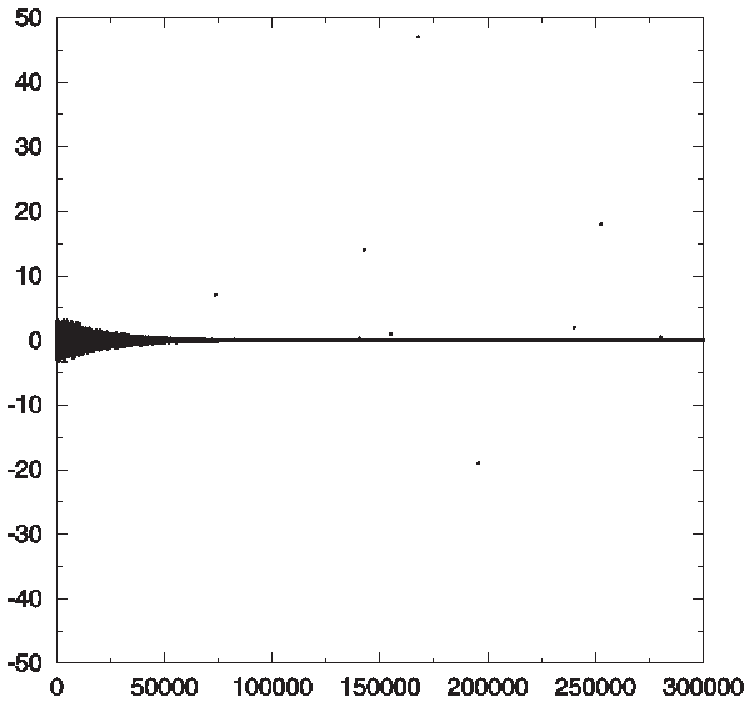}}
   \put(120,170){\includegraphics[height=160\unitlength,width=160\unitlength]{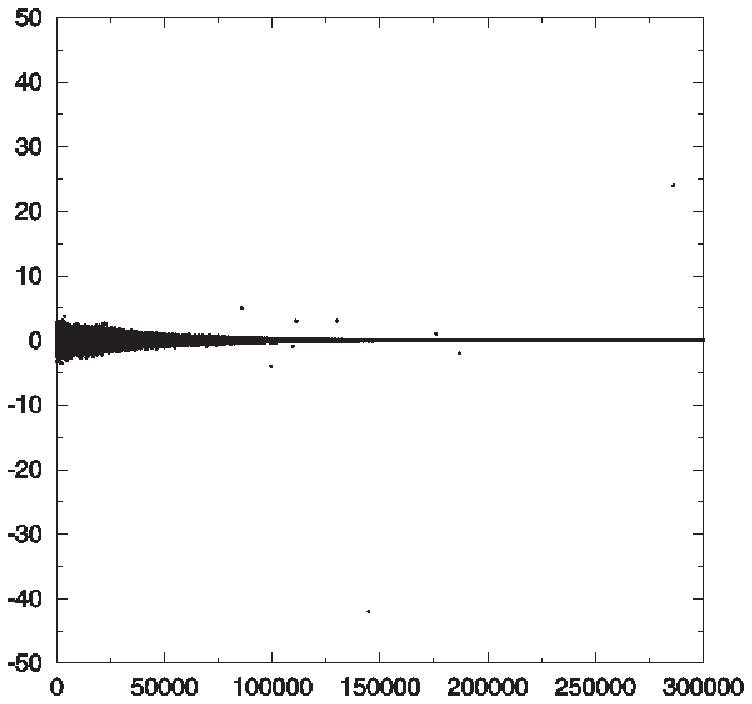}}
   \put(240,170){\includegraphics[height=160\unitlength,width=160\unitlength]{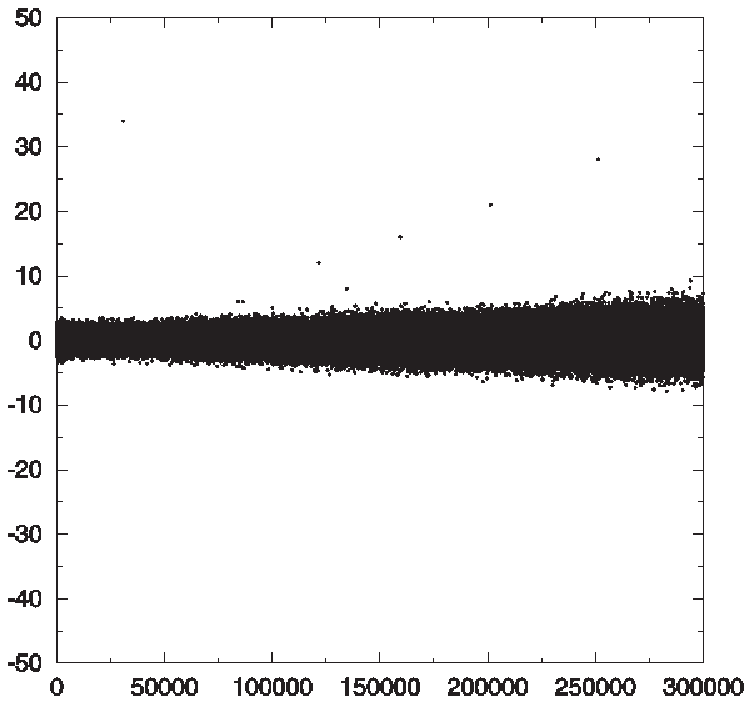}}
   \put(360,170){\includegraphics[height=160\unitlength,width=160\unitlength]{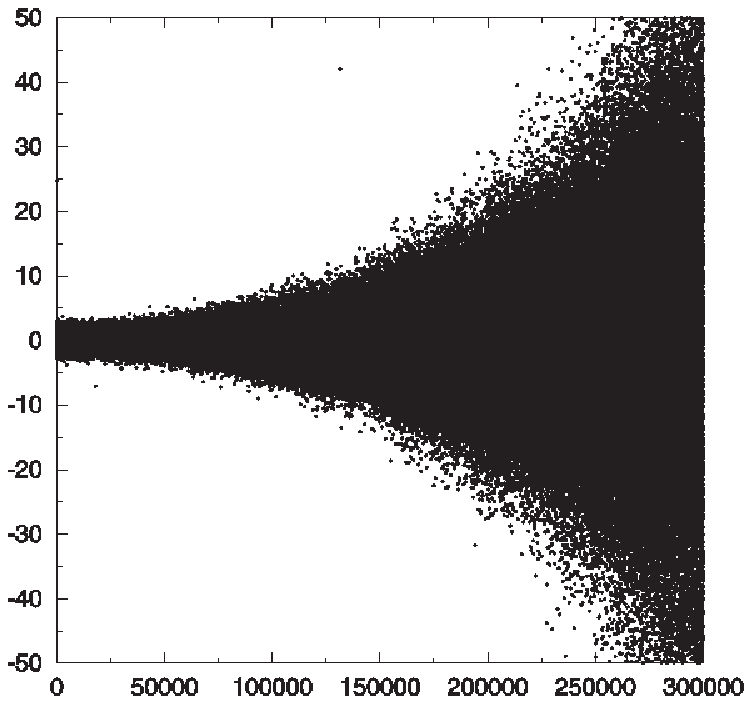}}
   \put(480,170){\includegraphics[height=160\unitlength,width=160\unitlength]{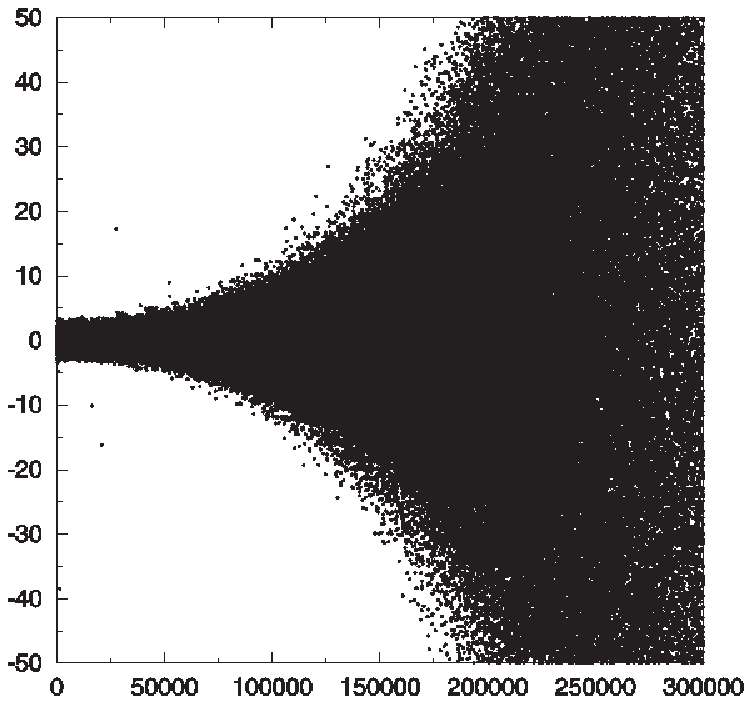}}

     \put(0,0){\includegraphics[height=160\unitlength,width=160\unitlength]{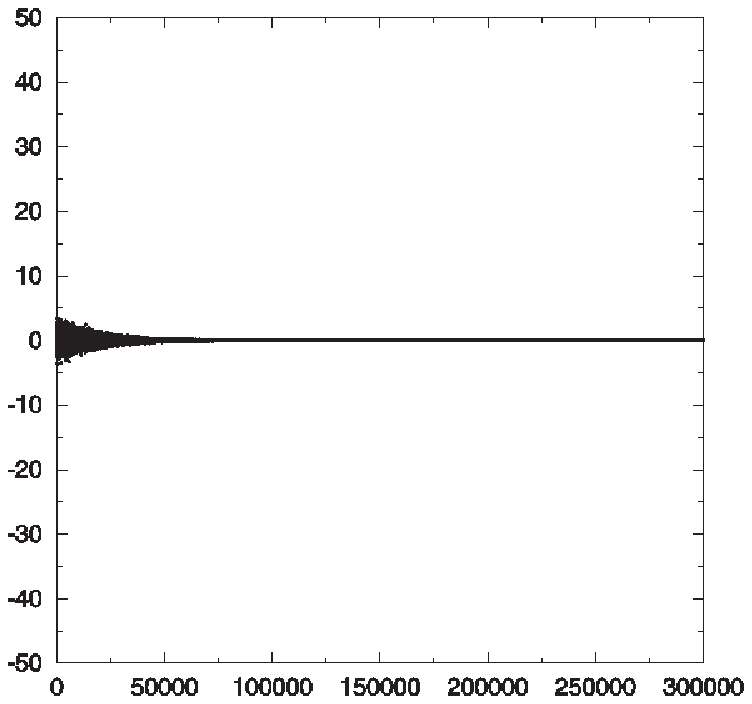}}
   \put(120,0){\includegraphics[height=160\unitlength,width=160\unitlength]{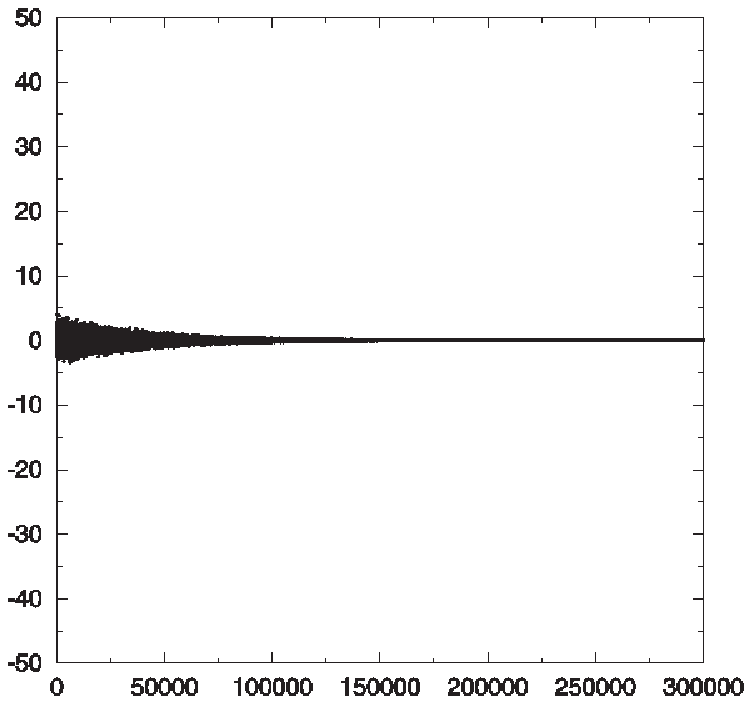}}
   \put(240,0){\includegraphics[height=160\unitlength,width=160\unitlength]{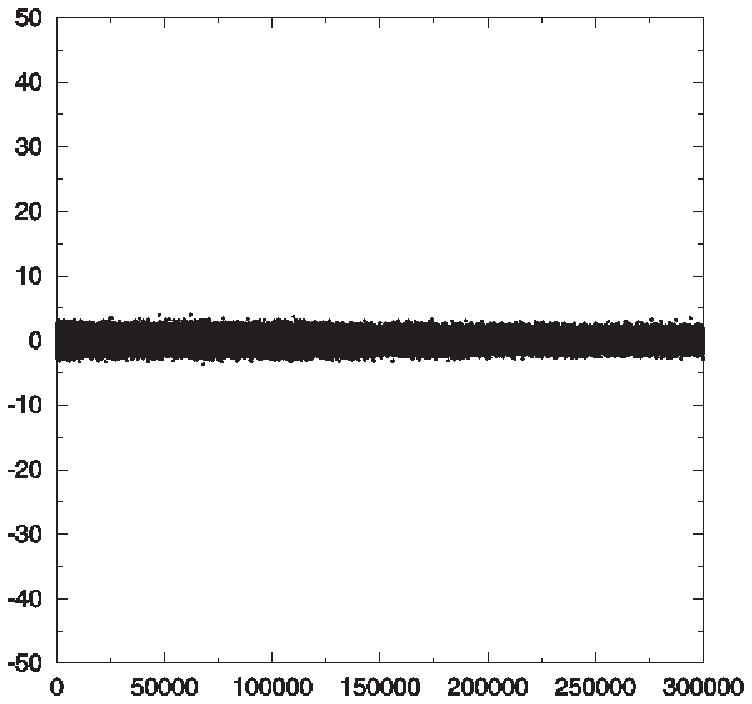}}
   \put(360,0){\includegraphics[height=160\unitlength,width=160\unitlength]{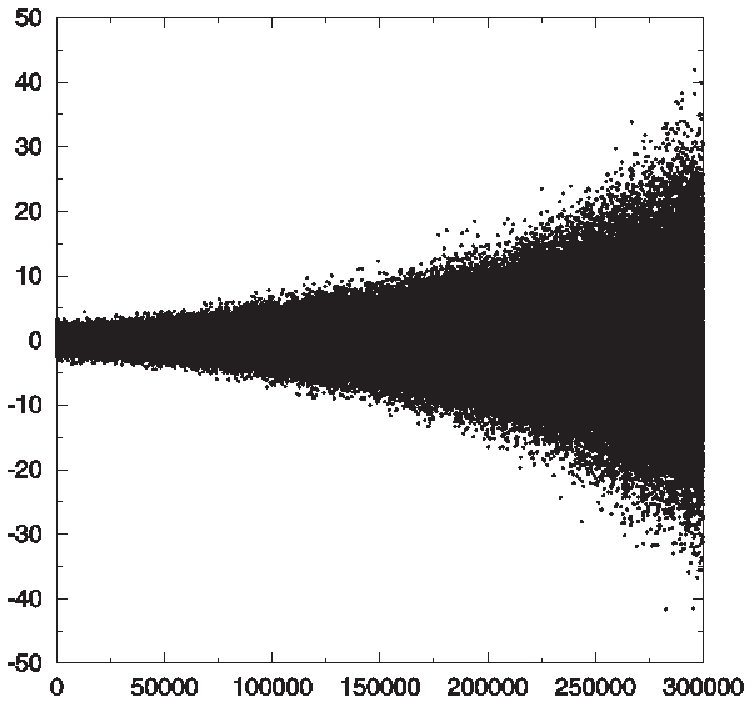}}
   \put(480,0){\includegraphics[height=160\unitlength,width=160\unitlength]{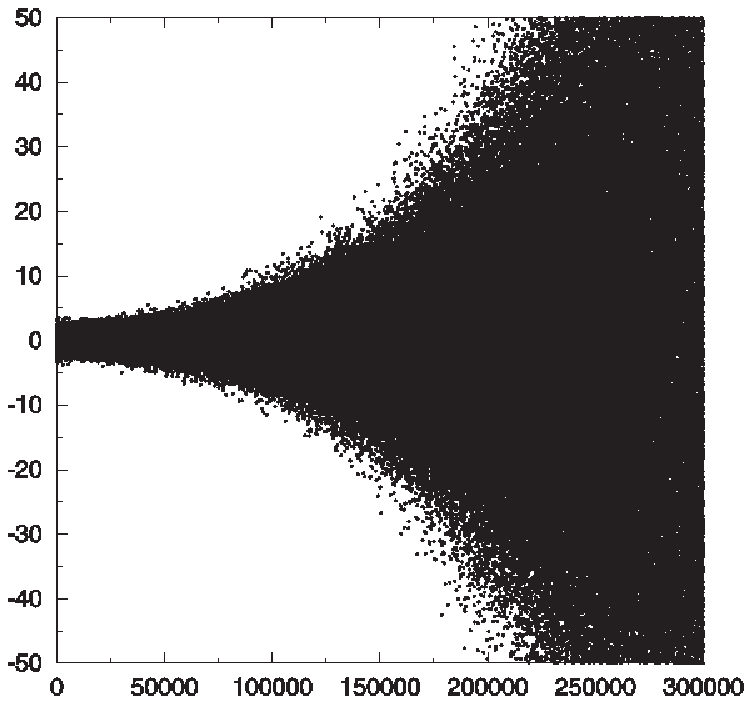}}

  \put(-16,80){\small\here{$A(\ell)$}} \put(-16,250){\small\here{$A(\ell)$}}

 \put(55,-23){\small $\ell$}    \put(175,-23){\small $\ell$}     \put(295,-23){\small $\ell$}
 \put(415,-23){\small $\ell$}  \put(535,-23){\small $\ell$}

 \put(20,135){\small $\alpha=4$}    \put(140,135){\small $\alpha=5$}     \put(260,135){\small $\alpha=6$}
 \put(380,135){\small $\alpha=7$}  \put(500,135){\small $\alpha=8$}

 \put(20,305){\small $\alpha=4$}    \put(140,305){\small $\alpha=5$}     \put(260,305){\small $\alpha=6$}
 \put(380,305){\small $\alpha=7$}  \put(500,305){\small $\alpha=8$}

\end{picture}
 \vspace*{7mm}
\caption{Overall bids $A(\ell)$ versus time $\ell=1\ldots 300,\!000$, as measured in numerical simulations of the inner product MG with $f[A]=A$ and $\zeta=0$ (fully real histories), in systems
 of size $N=1025$. Top row: unbiased initial conditions, $q_i(0)=\order(10^{-5})$ (zero average random); bottom row: biased initial conditions, $q_i(0)=\order(1)$ (zero average random).
The data indicate that there exists a critical value $\alpha^+_c$ such that for $\alpha>\alpha^+_c$, rather then evolve
towards $\sigma=0$, the system is unstable and the bid fluctuations diverge. If the (weak) dependence on initial conditions in the graphs for $\alpha=6$ is not a finite size effect, then there is a parameter regime
where the zero volatility state and the `runaway' solution coexist.
}\label{fig:viewbids_A}
\end{figure}

\begin{figure}[t]
\vspace*{-0mm} \hspace*{1mm} \setlength{\unitlength}{0.27mm}
\begin{picture}(500,170)


    \put(0,0){\includegraphics[height=160\unitlength,width=160\unitlength]{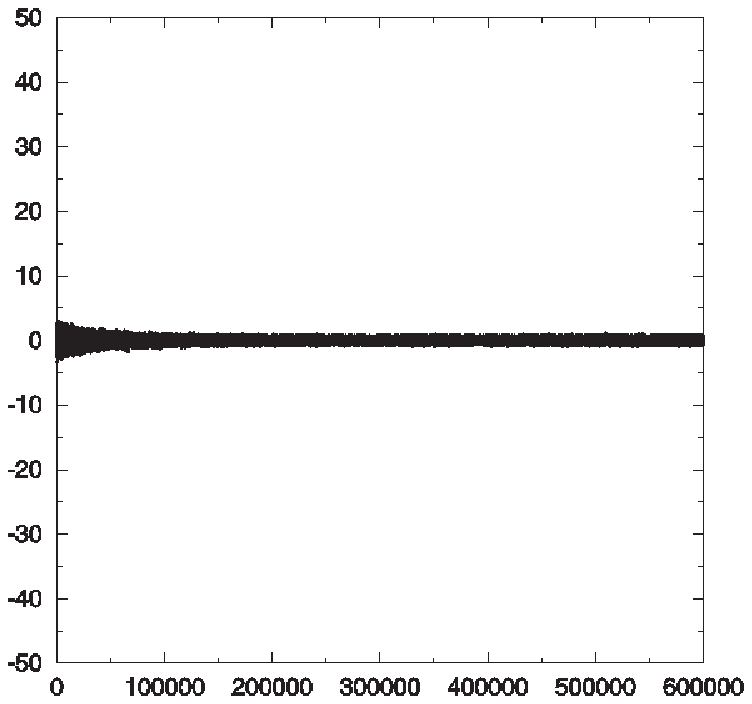}}
   \put(120,0){\includegraphics[height=160\unitlength,width=160\unitlength]{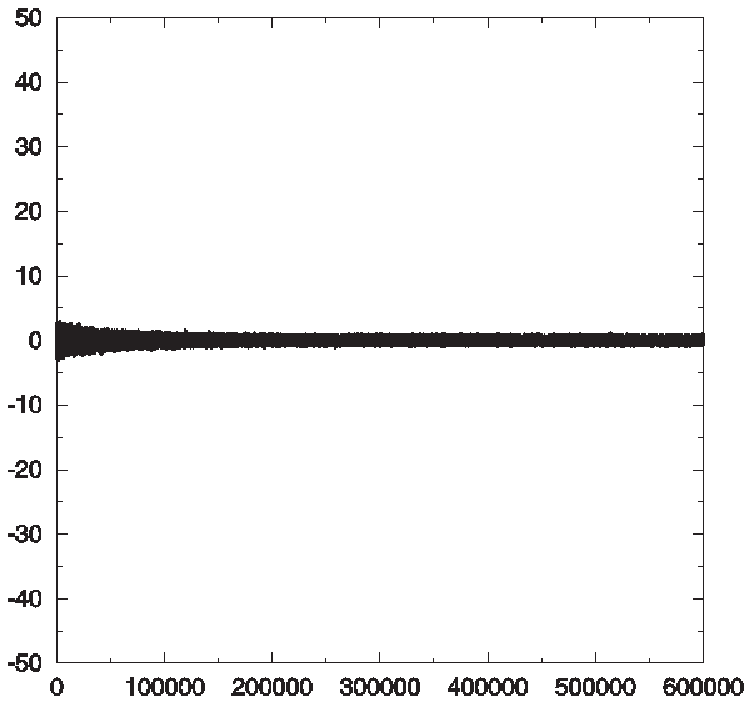}}
   \put(240,0){\includegraphics[height=160\unitlength,width=160\unitlength]{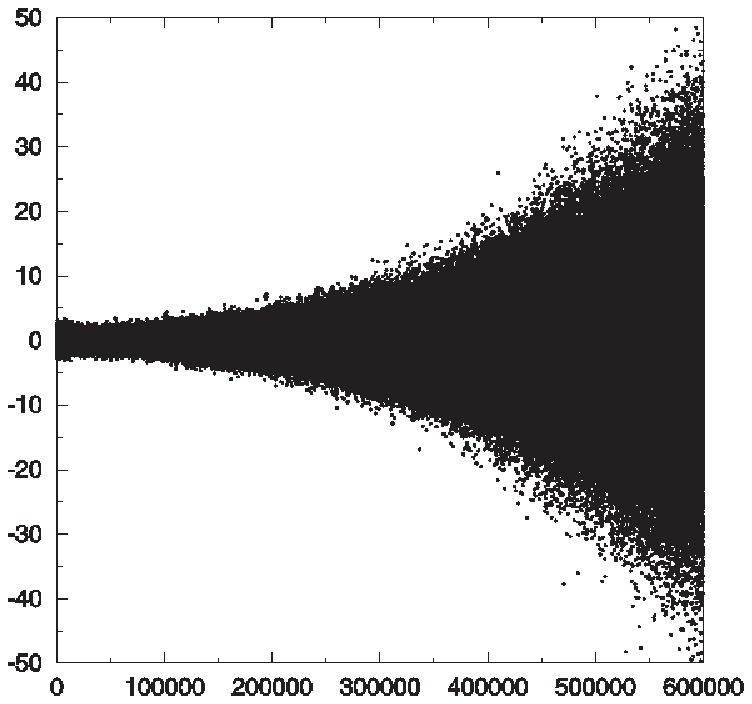}}
   \put(360,0){\includegraphics[height=160\unitlength,width=160\unitlength]{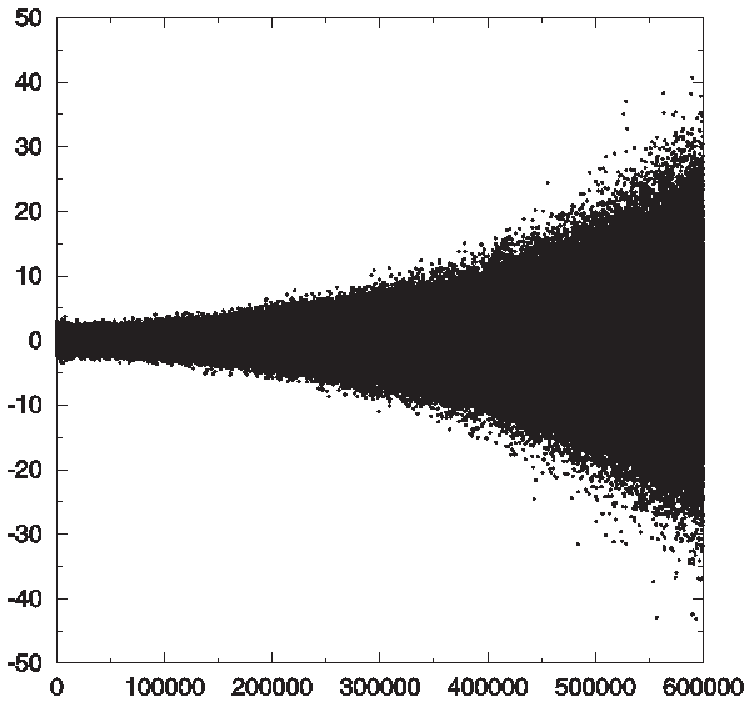}}
   \put(480,0){\includegraphics[height=160\unitlength,width=160\unitlength]{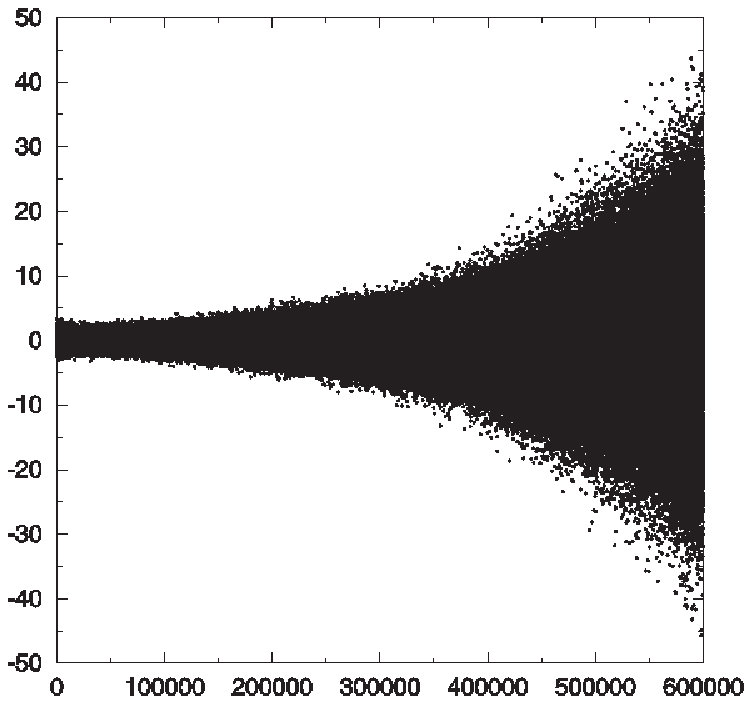}}

  \put(-16,80){\small\here{$A(\ell)$}}

 \put(55,-23){\small $\ell$}    \put(175,-23){\small $\ell$}     \put(295,-23){\small $\ell$}
 \put(415,-23){\small $\ell$}  \put(535,-23){\small $\ell$}

 \put(20,135){\small $\alpha=8$}    \put(140,135){\small $\alpha=9$}     \put(260,135){\small $\alpha=10$}
 \put(380,135){\small $\alpha=11$}  \put(500,135){\small $\alpha=12$}

\end{picture}
 \vspace*{7mm}
\caption{Overall bids $A(\ell)$ versus time $\ell=1\ldots 300,\!000$, as measured in numerical simulations of the inner product MG with $f[A]=A$, in systems
 of size $N=1025$ and following unbiased initial conditions. Here $\zeta=0.1$.
 The unstable state with diverging volatility exists also for $\zeta>0$, with the instability setting in at a critical  value $\alpha<\alpha^+_c(\zeta)$ that increases with $\zeta$. Also the characteristic time scale for the instability to manifest itself appears to increase with $\zeta$ (note the different time scales of figures \ref{fig:viewbids_A} and \ref{fig:viewbids_A_varyzeta}).
}\label{fig:viewbids_A_varyzeta}
\end{figure}

\begin{figure}[t]
\vspace*{5mm} \hspace*{30mm} \setlength{\unitlength}{0.5mm}
\begin{picture}(200,100)

   \put(0,0){\includegraphics[height=100\unitlength,width=140\unitlength]{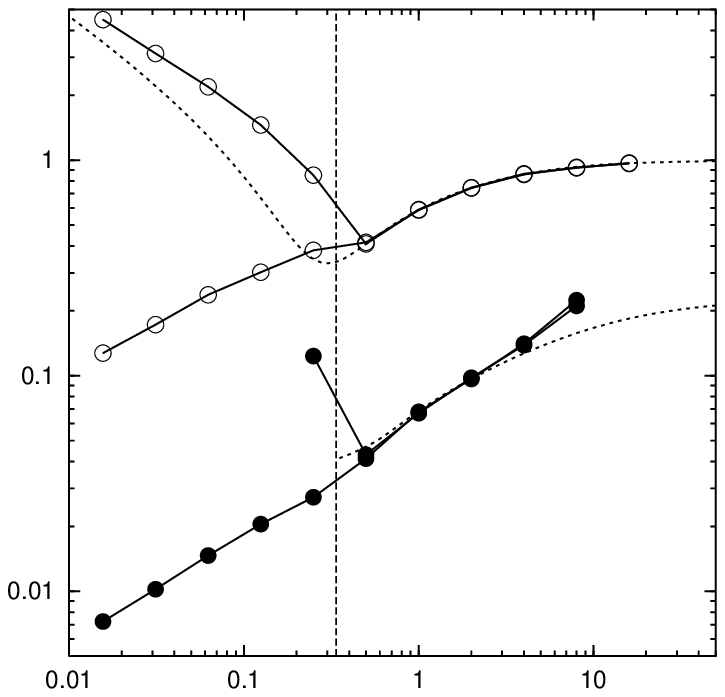}}
     \put(-3,50){\small\here{$\sigma$}}

 \put(120,0){\includegraphics[height=100\unitlength,width=140\unitlength]{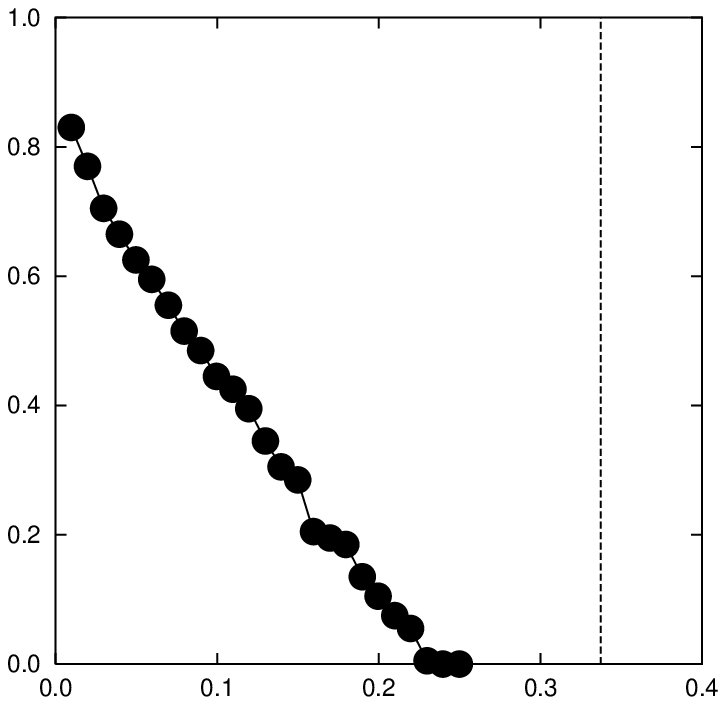}}
 \put(117,50){\small\here{$\zeta$}}
 \put(137,15) {$\sigma=\infty$} \put(170,70){$\sigma<\infty$}

  \put(49,-12){\small $\alpha$}    \put(169,-12){\small $\alpha$}
\end{picture}
 \vspace*{7mm}
\caption{Left: volatility as a function of $\alpha=p/N$, measured in numerical simulations for $f[A]=A$ and $S=1$.
  Measurements are taken over $200.p$ time steps, after $200.p$ equilibration time steps; system sizes: $pN^2=2^{30}$ (except for $\alpha=16$ and $\alpha=32$, where $N=1025$ and $N=513$, respectively).
  We show the results of four runs: $\zeta=0.1$ (predominantly real histories, connected full circles) with biased and unbiased initial
  conditions, and $\zeta=1$
  (strictly fake histories, connected open circles) with biased and unbiased initial conditions; initial conditions are as in previous figures. Dotted curves give the theoretical predictions for $\zeta=0.1$ and $\zeta=1$, upon assuming ergodicity.  The system destabilizes ($\sigma=\infty$) for $\alpha>\alpha^+_c(\zeta)$ (any initial conditions, terminating branches at $\alpha\approx 9$) and for $\alpha<\alpha^-_c$ (unbiased initial conditions only, terminating branch at $\alpha\approx 0.2$).
  Right: estimated critical line $\alpha_c^-(\zeta)$ in the low $\alpha$ region of the $(\alpha,\zeta)$ plane, marking the onset of the $\sigma=\infty$ instability, obtained by numerical simulations with $N=2049$ ($\zeta<0.77$) and $N=4097$ ($\zeta>0.77$). The marker sizes indicate the error bars.
  The vertical dashed lines in both figures mark the $\chi\to\infty$ transition value $\alpha_c$ below which the ergodic theory no longer applies.
}\label{fig:observables_v_alpha_A}
\end{figure}

\begin{figure}[t]
\vspace*{-0mm} \hspace*{31mm} \setlength{\unitlength}{0.65mm}
\begin{picture}(100,100)

   \put(0,0){\includegraphics[height=100\unitlength,width=140\unitlength]{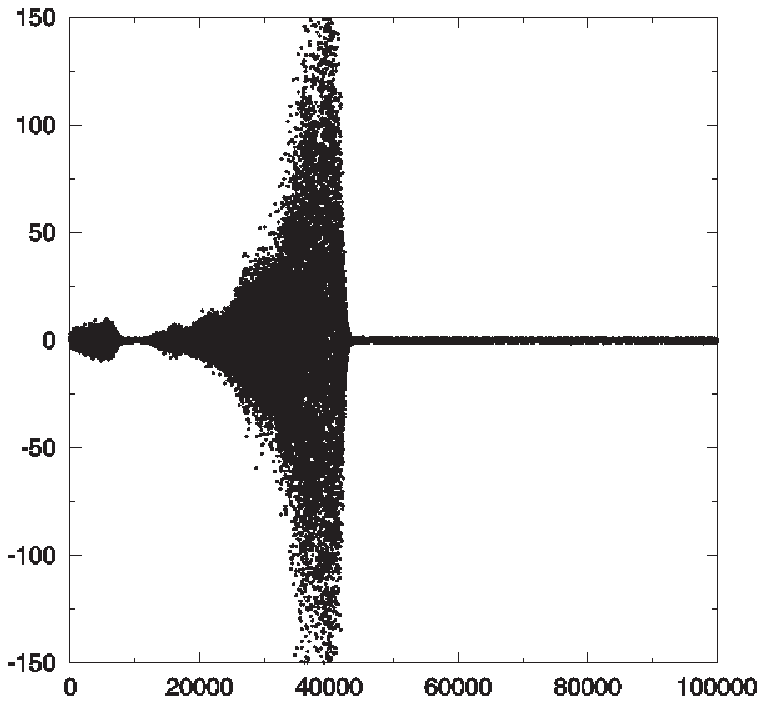}}

     \put(-3,50){\small\here{$A$}}
  \put(55,-10){\small $t$}
\end{picture}
 \vspace*{7mm}
\caption{Typical example of the bid evolution close to the critical line  that marks
the onset of the $\sigma=\infty$ instability, for $f[A]=A$ (see previous figure for the estimated location of this line).
The large intermittent fluctuations exhibited by the system limit the reliability with which the transition can be located via numerical simulations.
The present
example corresponds to $(\alpha,\zeta)=(0.14,0.32)$, with $N=2048$ and unbiased
initial conditions.
}\label{fig:Acritical}
\end{figure}

In contrast to the inner product MG with $f[A]=\sgn(A)$, for $f[A]=A$ the theory predicts a strong dependence of the volatility on the parameter $\zeta$,
with a degeneracy at $\zeta=0$. This prediction was already confirmed qualitatively by the data in figure \ref{fig:covariances}. In figure
\ref{fig:observables_v_zeta_A} we inspect this dependence in more detail for $\alpha=2$, and find an excellent agreement between theory and simulations, even for the volatility (in contrast to the situation for $f[A]=\sgn(A)$, where the predicted volatility was somewhat lower than the observed one). According to (\ref{eq:zero_sigma}) we should expect that the only stationary solution for $\zeta=0$ will have $\sigma=0$. In the limit $\zeta\to 0$ (where solving our closed equations becomes nontrivial as a result of having $\sigma\to 0$ and $\chi\to\infty$) figure
\ref{fig:observables_v_zeta_A} indeed confirms the predicted stationary state with vanishing volatility, and the numerical solution
 of the order parameter equations confirms that $\chi\to\infty$. However, if the experiment
is repeated for larger values of $\alpha$ one finds that beyond a critical value $\alpha^+_c\approx 6$
this zero volatility stationary state is no longer stable, and the fluctuations diverge. This is illustrated in figure \ref{fig:viewbids_A}. In fact, this
runaway solution is found to exist also  for $\zeta>0$, see figure \ref{fig:viewbids_A_varyzeta}, suggesting a $\zeta$-dependent criticality $\alpha^+_c(\zeta)$. Also at the lower end of the $\alpha$-scale, below the conventional transition point, one finds the system generally entering the runaway state, see e.g. the volatility data in figure \ref{fig:observables_v_alpha_A}, below a second critical point $\alpha_c^-(\zeta)$, provided one chooses unbiased initial conditions. For intermediate $\alpha$ values, where the volatility remains finite, our theory is found to predict the volatility $\sigma$ correctly (within the accuracy limitations imposed by finite size effects). In the right panel of figure \ref{fig:observables_v_alpha_A} we show the result of estimating
from numerical simulations the location of $\alpha_c^-(\zeta)$ in the $(\alpha,\zeta)$ plane, which appears to approach $\zeta=1$ for $\alpha\to 0$;
 determining this line experimentally is not entirely trivial in view of the (expected) nontrivial bid fluctuations close to this line, see eg. figure \ref{fig:Acritical}.
Carrying out a similar exercise for the location of the instability $\alpha^+_c(\zeta)$ in the high $\alpha$ region is unfortunately ruled out, due to the extreme relation times required for large $\alpha$.
 In summary: for sufficiently small $\zeta$ (i.e. sufficiently `real' histories),
the usual high volatility state of conventional MGs in the non-ergodic regime becomes a runaway solution in inner product MGs with linearly sampled histories (a true instability with infinite volatility), which for $\zeta=0$ vanishes slightly below
the standard MG's $\chi=\infty$ transition $\alpha=\alpha_c$ if one increases $\alpha$, but then re-appears at a larger value $\alpha^+_c(\zeta)$ and there becomes the {\em only} solution.
It must in principle be possible to calculate the critical values $\alpha^\pm_c(\zeta)$ from our order parameter equations; in practice, however, this requires solving our equations for the correlation and response functions (formulated in terms of both the effective agent process and the effective overall bid
process) at finite times, which at present we are unable to do.

\section{Discussion}

In this paper we have generalized the theory of look-up table Minority Games with real market history \cite{Coolen2005}
to a larger family of models, and applied it to MGs with so-called inner product strategy definitions. The latter MG versions
had, surprisingly, never been solved; not even in their simplest Markovian version where the histories are fake.
At a mathematical level it was not a priori clear which form the inner-product MG theory would take:
in \cite{Coolen2005} it was found that the key object in the theory of look-up table MGs with real market history
was the so-called history frequency distribution, but already on simple scaling grounds it is clear that this object cannot be defined for inner
product MGs. It is satisfactory to find in the present study how the generating functional formalism resolves the issue:
a more general quantity takes over the role of the history frequency
distribution, viz. the eigenvalue spectrum of the history covariance matrix, which reduces to the former for look-up table MGs but is
also well-defined for inner product MGs. This resolution involves a generalization of the short history correlation time ansatz that led in \cite{Coolen2005} to closed stationary state order parameter  equations, resulting (after a random matrix spectrum calculation)
also for the present inner-product MGs in closed equations for observables such as the persistent correlations, the fraction of `frozen'
agents, the integrated response, and the market volatility (although the formula for the latter as always involves further approximations).

We find, provided all the relevant integrals and averages exists, that the phase diagrams of the two MG versions
are identical. However,  we encounter interesting differences in terms of the models' static and dynamic phenomenology, dependent upon
which functions $f[A]$ of the past overall market bids $A$ are being sampled in the inner product MGs. For bounded functions such as $f[A]=\sgn(A)$ one continues to
find behaviour that, although quantitatively different, is qualitatively similar to that of look-up table MGs. If the histories are strictly
real, the theory even predicts complete universality of the static observables, independent of the choice made for $f[A]$.
 In contrast, for unbounded sampling functions such as $f[A]=A$ the situation is quite different. There can now be a
 profound dependence of the volatility on the degree to which the histories are real. Especially when the histories are strictly real and
 $f[A]=A$ we
find that the inner product MG has only one stationary solution, where the volatility $\sigma$ is zero and the model is structurally critical
(i.e. $\chi=\infty$).
In terms of the usual control parameter $\alpha=p/N$, we find
that in the regime $\alpha>\alpha_c$ (the standard non-ergodicity transition of the MG)
this  state $(\sigma,\chi)=(0,\infty)$ can only be reached for $\alpha<\alpha^+_c(\zeta)$, with
 $\alpha^+_c(0)\approx 6$ and $\alpha^+_c(\zeta)$ increasing with $\zeta$. For $\alpha>\alpha^+_c(\zeta)$  the system destabilizes and the fluctuations diverge.
 In the low $\alpha$ regime one finds a similar critical line $\alpha^-_c(\zeta)$ marking a destabilization transition as one reduces $\alpha$ further, with $\sigma=\infty$ for $\alpha<\alpha_c^-(\zeta)$, but here
 this destabilization occurs only for unbiased initial conditions.

Those technical questions that remain open in the present study are the familiar ones that are similarly answered for
all previous MG models. They relate mostly to our inability so far to extract from our order parameter equations exact solutions for time-dependent observables (beyond just a few time steps), or exact stationary state solutions in the non-ergodic regime.
In addition one would like to explore further the possibility of finding solutions for MG versions in which the effects of having real histories are more dramatic;  both the present study and its predecessor \cite{Coolen2005} rely on expansions that require history correlations to be weak.
Finally, we would like to point out that the observed instability of the $f[A]=A$ inner product MG  is qualitatively different from what has been observed in previous MG versions. Although in its present form this specific model is ill-constructed, it could perhaps form the basis of new models aimed at explaining more robustly the so-called stylized facts in real markets, since
its structural instability is {\em not} a finite size effect and therefore requires no careful tuning of control parameters.
The phenomenology observed for $f[A]=A$ is novel, should be expected to emerge also for other unbounded
 choices for the sampling function $f[A]$, and emphasizes even more the need for progress in solving the GFA order parameter equations for finite times.

\end{document}